\documentclass{elsart3-1}

\usepackage{amssymb,epsfig}

\usepackage[english,francais]{babel}

\newtheorem{e-proposition}[theorem]{Proposition}

\newtheorem{e-definition}[theorem]{Definition\rm}

\renewcommand{\theequation}{\arabic{equation}}
\newcommand{\vct}[1]{{\mbox {\boldmath $#1$}}}

\def\build#1_#2^#3{\mathrel{\mathop{\kern 0pt#1}\limits_{#2}^{#3}}}

\usepackage{epsfig,amsmath,amssymb}

\def\og{\leavevmode\raise.3ex\hbox{$\scriptscriptstyle\langle\!\langle$~}}
\def\fg{\leavevmode\raise.3ex\hbox{~$\!\scriptscriptstyle\,\rangle\!\rangle$}}

\begin{document}

\begin{frontmatter}


\selectlanguage{english}
\title{A phenomenological theory of Eulerian and Lagrangian velocity  fluctuations in turbulent flows}


\selectlanguage{english}
\author{Laurent Chevillard, Bernard Castaing} 
\author{Alain Arneodo, Emmanuel L\'ev\^eque, Jean-Fran\c{c}ois Pinton, St\'ephane Roux}


\address{Laboratoire de Physique de l'ENS Lyon, CNRS, Universit\'e de Lyon, 46 all\'ee d'Italie, 69007 Lyon, France.}


\medskip
\begin{center}
{\small Received *****; accepted after revision +++++}
\end{center}

\begin{abstract}
A phenomenological theory of the fluctuations of velocity occurring in a fully developed homogeneous and isotropic turbulent 
flow is presented. The focus is made on the fluctuations of the spatial (Eulerian) and temporal (Lagrangian) velocity increments. 
The universal nature of the intermittency phenomenon as observed in experimental measurements and numerical simulations 
is shown to be fully taken into account by the multi scale picture proposed by the multifractal formalism, 
and its extensions to the dissipative scales and to the Lagrangian framework. 
The article is devoted to the presentation of these arguments and to their comparisons against empirical data. In particular, 
explicit predictions of the statistics, such as probability density functions and high order moments, of the velocity gradients 
and acceleration are derived. In the Eulerian framework, at a given Reynolds number, they are shown to depend on a 
single parameter function called the singularity spectrum and to a universal constant governing the transition between 
the inertial and dissipative ranges. The Lagrangian singularity spectrum compares well with its Eulerian counterpart by a 
transformation based on incompressibility, homogeneity and isotropy and the remaining constant is shown to be difficult 
to estimate on empirical data. It is finally underlined the limitations of the increment to quantify accurately the singular nature of Lagrangian velocity. This is confirmed using higher order increments unbiased by the presence of linear trends, as they are observed on velocity along a trajectory.
\vskip 0.5\baselineskip

\selectlanguage{francais} \noindent{\bf R\'esum\'e} \vskip
0.5\baselineskip \noindent {\bf Une th\'eorie ph\'enom\'enologique des fluctuations de vitesse Eul\'erienne et Lagrangienne dans un \'ecoulement 
turbulent} Nous pr\'esentons une th\'eorie ph\'enom\'enologique des fluctuations de vitesse dans un \'ecoulement turbulent 
pleinement d\'evelopp\'e isotrope et homog\`ene. Nous mettons l'accent sur les fluctuations des incr\'ements spatiaux (Eul\'erien) et temporels 
(Lagrangien) de vitesse. La nature universelle du ph\'enom\`ene d'intermittence observ\'e sur les mesures exp\'erimentales 
et les simulations num\'eriques est compl\`etement pris en compte par les arguments d\'evelopp\'es par le formalisme multifractal, et ses 
extensions aux \'echelles dissipatives et au cadre Lagrangien. Cet article  pr\'esente les pr\'edictions de cette description multifractale et les compare
aux donn\'ees empiriques. En particulier, des pr\'edictions explicites sont obtenues pour des grandeurs statistiques, comme les fonctions 
de densit\'e de probabilit\'e et les moments d'ordres sup\'erieurs, des gradients de vitesse et de l'acc\'el\'eration. Dans le cadre Eul\'erien, 
\`a un nombre de Reynolds donn\'e, nous montrons que ces pr\'edictions ne d\'ependent que d'une fonction \`a param\`etres, appel\'ee spectre de 
singularit\'es, et d'une constante r\'egissant la transition entre les r\'egimes inertiels et dissipatifs. Le spectre des singularit\'es Lagrangien 
est reli\'e \`a son homologue Eul\'erien par une transformation bas\'ee sur la nature incompressible, homog\`ene et isotrope de l'\'ecoulement, alors que la constante restante est difficile \`a estimer \`a partir des donn\'ees. Nous montrons finalement que l'incr\'ement est inadapt\'e \`a quantifier pr\'ecis\'ement  la nature singuli\`ere de la vitesse Lagrangienne. Cela est confirm\'e par l'utilisation d'incr\'ements d'ordres sup\'erieurs non biais\'es par la pr\'esence de comportements lin\'eaires, comme nous l'observons sur la vitesse le long d'une trajectoire. 
{\it Pour citer cet article~: L. Chevillard et al., C. R. Physique
\textbf{volume} (ann\'ee).}
\vskip 0.5\baselineskip \noindent{\small{\it Mots-cl\'es~:} Turbulence;
Intermittence~; Eul\'erien et Lagrangien}}

\end{abstract}
\end{frontmatter}

\selectlanguage{francais}

\selectlanguage{english}
\tableofcontents

\section{Introduction}
\label{sec:Intro}




As a long-standing challenge in 
classical physics \cite{Ric22,Kol41,Bat53,TenLum72,Kra74,MonYag75,Fri95,Pop00,Tsi01}, fully developed turbulence is an archetypical non linear and non local phenomenon. 
When a flow is stirred at a large scale $L$, typically the mesh size of a grid in a wind tunnel or the width of the blades 
in a von Karman washing machine, the input energy cascades towards the small scales without being dissipated 
according to the classical picture of Richardson and Kolmogorov \cite{Ric22,Kol41}. 
One of the most important objectives in turbulence research is to understand the processes that lead to this very 
peculiar cascading structure of energy. As far as we know, recent theoretical progress made in this long lasting 
field comes from the systematic analysis and description of empirical (experimental and numerical) velocity data. 

\begin{figure}[t]
\center{\epsfig{file=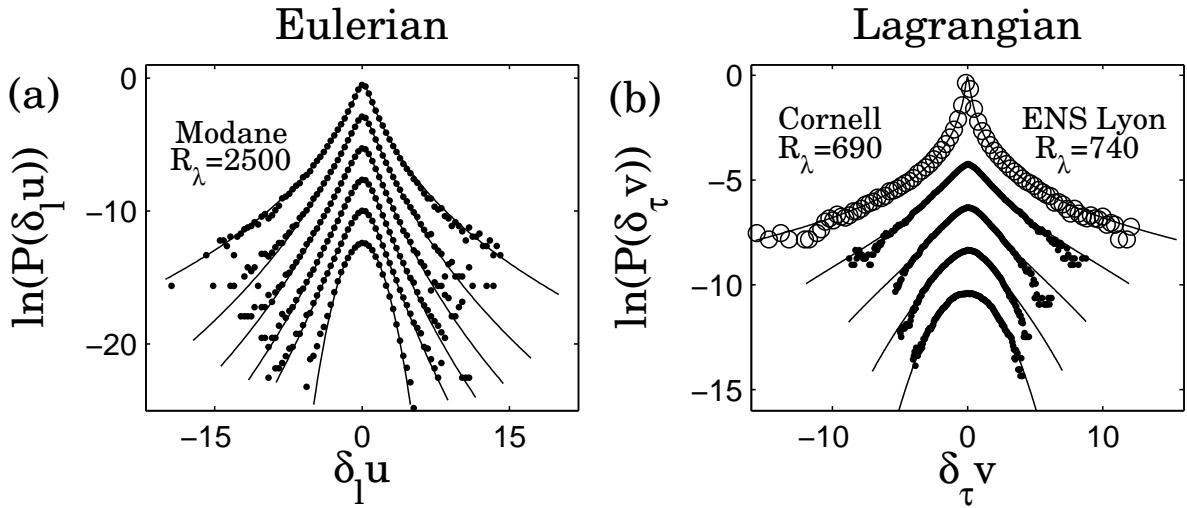,width=16cm}}
\caption{(a) PDFs of signed longitudinal velocity increments in Modane data \cite{KahMal98}. 
Represented scales (from top to bottom): $\ln(\ell/L)$ = -6.4137, -5.6028, -4.6645, -3.6411, -2.7501, -1.8598,
-0.8685, 0.1226. All curves are arbitrarily vertically
shifted for the sake of clarity and their variance is set to unity. The solid curves correspond to our theoretical
predictions (see text in section \ref{section:PdfAsymTheo} and Ref. \cite{CheCas06}). (b) PDFs of Lagrangian temporal increments from the ENS-Lyon experiment,
for time lags (from bottom to top, symbols $\bullet$) $\tau/T=$  0.07, 0.16, 0.35, 1
and from Cornell acceleration data (symbols $\circ$, from Ref. \cite{MorCro04}). Also, the curves are displayed with an arbitrary
vertical shift for clarity, the variance is set to unity at any scales, and the original axis for the
acceleration PDF ($\circ$) has been shrunk by a factor 4.
Solid lines correspond to theoretical predictions (see section \ref{sec:fulllagdescrip} and Ref. \cite{CheRou03}).} \label{fig:PdfEulLag}
\end{figure}

The first experimental measurements of turbulent velocity were performed in the Eulerian framework and
focussed on the longitudinal velocity profile. In this context, the Taylor hypothesis allowed to interpret the time 
dependence of the measurements obtained in a wind tunnel behind a grid or in an air jet as a spatial dependence \cite{Fri95}. 
These experiments gave access to the longitudinal velocity increments, i.e.
\begin{equation}
\delta_\ell u (\textbf{x})  \equiv \left(\textbf{u}(\textbf{x}+\vct{\ell})- \textbf{u}(\textbf{x})\right).\frac{\vct{\ell}}{\ell}\mbox{ ,}
\end{equation}
where $\textbf{u}$ is the velocity vector and $\vct{\ell}$ a vector of norm $\ell$, and the respective structure functions are given by
\begin{equation}
M_n(\ell) = \langle |\delta_\ell u|^n \rangle\mbox{ .}
\end{equation}
For such flows, the Reynolds number defined as
\begin{equation} 
\mathcal R_e = \frac{\sigma L}{\nu}\mbox{ ,}
\end{equation}
where $\sigma = \sqrt{\langle (\delta_Lu)^2\rangle}$ is a characteristic velocity at the so-called integral length scale $L$ and 
$\nu$ the kinematic viscosity, can be considered as very large compared to unity.
At these high Reynolds numbers, Kolmogorov showed, in a first seminal article \cite{Kol41} using a dimensional analysis, 
that the second order structure function $M_2(\ell)$ behaves as a power law, i.e.
\begin{equation}\label{eq:PredM2K41}
M_2(\ell)=\langle (\delta_\ell u)^2\rangle = \sigma^2\left(\frac{\ell}{L}\right)^{2/3} = c_K \langle \epsilon\rangle^{2/3} \ell^{2/3}\mbox{ ,}
\end{equation}
where $c_K$ is the Kolmogorov constant of order unity \cite{YeuZho97,DonSre10} and $\langle \epsilon \rangle$ the averaged dissipation 
that will be properly defined latterly. Equivalently, the power spectrum, i.e. the Fourier Transform of the velocity 
autocorrelation function, follows the celebrated Kolmogorov law,
\begin{equation}\label{eq:PredSpecK41}
E(k) \propto  c_K \langle \epsilon\rangle^{2/3} k^{-5/3}\mbox{ ,}
\end{equation}
where the proportionality constant can be calculated \cite{YeuZho97,DonSre10}.
As stated by Kolmogorov himself \cite{Kol41}, the laws predicted (Eqs. \ref{eq:PredM2K41} and \ref{eq:PredSpecK41}) are only 
valid in a range of scales, called the inertial range $\eta_K \ll \ell \ll L$ (or equivalently $L^{-1} \ll k \ll \eta_K^{-1}$) 
delimitated by the Kolmogorov length scale $\eta_K \sim L \mathcal R_e ^{-3/4}$ under which dissipative effects dominate the physics. 
These laws have been successfully compared to empirical data \cite{Fri95}. If velocity fluctuations were Gaussian, these predictions 
could be easily generalized to higher order structure functions. As observed experimentally, the probability density 
functions (PDFs) of velocity increments undergo a continuous shape deformation, starting from the integral length scale 
$L$ at which statistics can be considered as Gaussian, down to the dissipative scales where the PDF is highly non Gaussian
\cite{CasGag90,BenBif91,KaiSre92}. 
This phenomenon is a manifestation of the intermittent nature of turbulence, as first underlined by Kolmogorov and Obukhov 
\cite{Obo62,Kol62}. We show in Fig. \ref{fig:PdfEulLag}(a) the estimation of the PDFs of longitudinal velocity increments 
obtained in the giant wind tunnel of Modane \cite{KahMal98} at a high Taylor-based Reynolds number $\mathcal R_\lambda=2500$ 
(Eq. \ref{eq:DefRlambdaRstar} provides a link between the large-scale Reynolds number $\mathcal R_e$ and $\mathcal R_\lambda$). 
The curves are arbitraly shifted vertically for the sake of clarity. 
This continuous shape deformation of these PDFs motivated Castaing et al. to build a statistical description 
of these longitudinal velocity fluctuations \cite{CasGag90}. Moreover, by a simple visual inspection, we can remark that the PDFs are not symmetric. 
This is related to the Skewness phenomenon associated with the mean energy transfer of energy towards the small scales that takes 
place in the inertial range. We will come back to this point latterly.

The first part of this article is devoted to review the predictions that can be made using both 
the so-called multifractal formalism \cite{Fri95} and the propagator approach \cite{CasGag90}. This description, that will 
be shown to depend only on a parameter function $\mathcal D(h)$ and a universal constant $\mathcal R^*$ 
(independent on the flow geometry and the Reynolds number), accurately reproduces the non Gaussian features 
formerly presented.

More recently, experimental \cite{OttMan00,PorVot01,VotPor02,MorMet01,MorDel02,MorDel03,CheRou03,MorCro04,XuBou06,BerOtt09}) 
and numerical \cite{MorMet01,MorDel02,CheRou03,YeuPop89,Yeu01,BifBof04} data have revealed a similar phenomenon in the Lagrangian 
framework (see recent review articles \cite{Yeu02,ArnICTR08,TosBod09}). Lagrangian velocity is defined as the Eulerian velocity of a fluid particle at the position $\textbf{X}(t)$, 
initially at the position $\textbf{X}(t_0)$  via the following identity
\begin{equation}\label{eq:DefLagVelocity}
\textbf{v}(\textbf{X}(t_0),t) = \textbf{u}(\textbf{X}(t),t)\mbox{ .}
\end{equation}
In this framework, the study of the Lagrangian velocity fluctuations focusses on the Lagrangian time increment defined as
\begin{equation}\label{eq:VelTimIncr}
\delta_\tau v (t) = v(t+\tau)-v(t)\mbox{ ,} 
\end{equation}
with $v$ a component of the Lagrangian velocity vector $\textbf{v}$ (Eq. \ref{eq:DefLagVelocity}). A similar dimensional analysis \textit{\`a la} Kolmogorov would give a linear dependence of the second order structure function
\begin{equation}\label{eq:DimAnalM2Lag}
M_2(\tau)=\langle (\delta_\tau v)^2\rangle = \sigma^2\left(\frac{\tau}{T}\right) = c^L_K \langle \epsilon\rangle \tau\mbox{ ,}
\end{equation}
where $T=L/\sigma$ is the integral time scale and $c^L_K$ the respective Lagrangian Kolmogorov constant \cite{DuSaw95,LieAsa02}. 
This corresponds to a Lagrangian power spectrum of the form
\begin{equation}\label{eq:DimAnalSpecLag}
E(\omega) \propto c_K^L \langle \epsilon\rangle \omega^{-2}\mbox{ .}
\end{equation}
Once again, these laws are valid only in the respective inertial range $\tau_{\eta_K} \ll \tau \ll T$ 
(or  $T^{-1} \ll \omega \ll \tau_{\eta_K}^{-1}$). Unfortunately, these laws cannot be easily generalized 
to higher order structure functions because of the fundamental non Gaussian nature of the velocity fluctuations. 
We show in Fig. \ref{fig:PdfEulLag}(b) the estimation of the experimental velocity increments PDFs at various scales 
obtained at ENS Lyon \cite{MorMet01} and the acceleration PDF obtained at the university of Cornell \cite{MorCro04}. 
Let us first remark that the acceleration can be seen as a Lagrangian velocity increment at a scale $\tau$ much smaller than the 
Kolmogorov dissipative time scale. Again we observe a continuous shape deformation from Gaussian statistics at large scale, 
to long-tail acceleration statistics at vanishing scale. This is again a manifestation of the intermittency phenomenon. 

We will show in this article that a similar statistical description can be developed in a Lagrangian context. 
The free parameters are again the corresponding singularity spectrum and a constant, as in the Eulerian framework. 
The Lagrangian singularity spectrum will be shown consistent with the prediction derived from its Eulerian 
counterpart using a transformation, presented latterly.

This article is devoted to the presentation of a phenomenological theory of the statistics of the Eulerian and Lagrangian 
velocity increments, from the integral length scale $L$ (or from the integral time scale $T=L/\sigma$), 
down to the far dissipative scales based on the so-called multifractal formalism \cite{Fri95}. In this approach, statistical properties in the inertial range are 
assumed and compared to empirical data. This includes the classical K41 predictions ``$k^{-5/3}$" (Eq.  \ref{eq:PredSpecK41}) 
and ``$\omega^{-2}$" (Eq. \ref{eq:DimAnalSpecLag}), but also the intermittent (or multifractal) corrections. 
From this descriptive analysis, we predict the statistical properties of the (Eulerian) velocity gradients and (Lagrangian) 
acceleration as functions of the Reynolds number and of the corresponding singularity spectra. In this sense, we can consider the multifractal formalism, properly generalized to the dissipative range, as a \textit{phenomenological theory}, i.e. statistical properties are \textit{assumed} in the inertial range (given some free parameters fully encoded in the so-called singularity spectrum), and are \textit{predicted} in the dissipative range. This was already recognized in Ref. \cite{FriVer91}. To this regard, multifractal formalism can be viewed as a \textit{standard model} of turbulence, and, as far as we know, it is the only formalism able to reproduce accurately higher order statistics of velocity increments. The most important perspective would be to establish a link between this formalism (in particular the existence of a singularity spectrum) and the equations of motion (i.e. the Navier-Stokes equations). This is out of the scope of the present article. Let us also mention some alternative statistical formalisms that are able to reproduce the non-Gaussian nature of the underlying statistics, such as, among others, ``superstatistics'' \cite{BeckSuper}, continuous-time random walk models \cite{Fried03}, vortex filament calculations \cite{ZybSir10} and kinetic equations approach \cite{DaiFri12}. As an interesting application of this multifractal formalism, we reinterpret 
former measurements by the group of Tabeling \cite{TabZoc96,TabWil02}, concerning  the velocity gradients flatness  as a function of the Reynolds number. 
We end with a general discussion of the skewness phenomenon and of its implications on the modeling of velocity increments in the inertial range. In the Lagrangian framework, we introduce higher order velocity increments designed to quantify accurately the singular nature of velocity and justify their use.

\section{The Eulerian longitudinal velocity fluctuations}

\subsection{Behavior of the flatness of velocity increments from experimental investigation}

 \begin{figure}[t]
\center{\epsfig{file=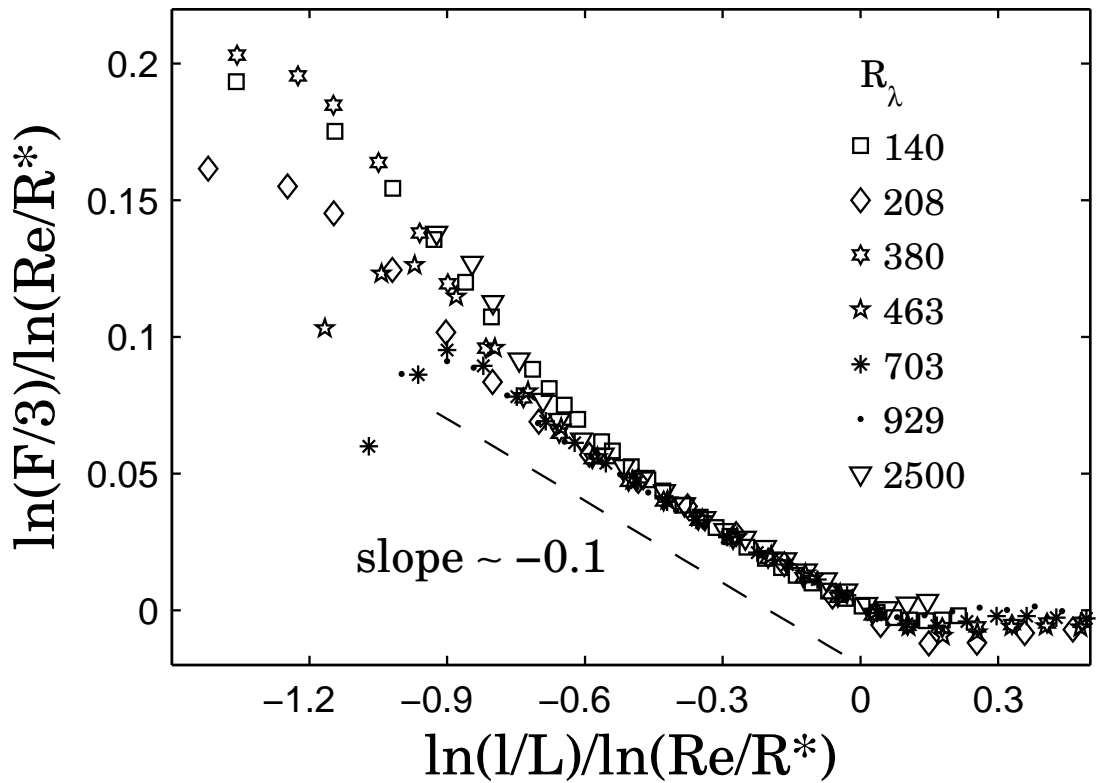,width=15cm}}
\caption{Behavior of the velocity increment flatness as a function of the scales in a normalized logarithmic representation. The universal constant 
entering in the normalization is set to $\mathcal R^*=52$. Different symbols represent different Reynolds numbers: $\mathcal R_\lambda=140$ (DNS by E. L\'ev\^eque \cite{CheCas05}), 
$\mathcal R_\lambda=208, 463, 703, 929$ (Experimental Helium jet \cite{ChaCha00}), 
$\mathcal R_\lambda=380$ (Air jet measurements by C. Baudet and A. Naert), $\mathcal R_\lambda=2500$ (Modane's wind tunnel \cite{KahMal98}). 
A straight line of slope -0.1 is superimposed on the data in the inertial range.} \label{fig:FlatEul}
\end{figure}

We show in Fig. \ref{fig:FlatEul} the estimation of the flatness 
$F = \frac{\langle (\delta_\ell u)^4\rangle}{\langle (\delta_\ell u)^2\rangle^2}$ of the longitudinal velocity increments in various
flow configurations and various Reynolds numbers. The first set of data has been obtained in a Helium jet \cite{ChaCha00} at four different 
Reynolds numbers: $\mathcal R_\lambda = 208, 463, 703, 929$, where the Taylor-based Reynolds number $\mathcal R_\lambda$ is proportional to the square-root of the large-scale Reynolds number $\mathcal R_e$ (see Eq. \ref{eq:DefRlambdaRstar}). Also are displayed the results in an air jet experiment at the ENS Lyon at $\mathcal R_\lambda=380$ of C. Baudet and A. Naert.
The higher Reynolds number ($\mathcal R_\lambda=2500$) comes from the wind tunnel experiment  in Modane \cite{KahMal98}. 
For comparisons are also reported the results coming from a classical direct numerical simulation (DNS) of E. L\'ev\^eque \cite{CheCas05} at a moderate Reynolds number 
$\mathcal R_\lambda=140$ based on $256^3$ grid points in a periodic domain.

The flatness $F$ as a function of the scale $\ell$ is represented in a logarithmic representation. Flatness is divided by 3, i.e. by the value 
of the flatness of a Gaussian random variable. The scales are renormalized by the length scale $L$ estimated such that 
the power-laws that are observed in the inertial range are indistinguishable. Then,  $\ln (\ell/L)$ is itself normalized by the 
Reynolds number, i.e. $\ln (\mathcal R_e/\mathcal R^*)$, where $\mathcal R^*=52$ is a universal constant, that will be shown to be linked to the Kolmogorov 
constant (subsection \ref{subsub:dissipative}). A similar procedure is applied to $\ln(F/3)$.

In this representation, the Kolmogorov length scale can be seen directly as $\ln (\eta_K/L)/\ln (\mathcal R_e/\mathcal R^*) = -3/4$. 
It delimits the inertial range, i.e. $\ln (\ell/L)/\ln (\mathcal R_e/\mathcal R^*) \in [-3/4,0]$, and the dissipative range, i.e.
 $\ln (\ell/L)/\ln (\mathcal R_e/\mathcal R^*)\le -3/4$.

\subsubsection{The inertial range}
\label{sec:EulInert}

Let us first focus on the inertial range. Over this range, we observe a universal (i.e. 
independent on both the Reynolds number and the flow geometry) power law $(\ell/L)^{-0.1}$ of exponent $0.1$. In the same range, 
the second order structure function behaves as a power law, i.e. $\langle (\delta_\ell u)^2\rangle \sim (\ell/L)^{2/3}$ (data not
shown). Furthermore, we can see that the statistics at scales $\ell$ larger than the integral length scale $L$ are consistent 
with a Gaussian process since the Flatness is very close to 3. Indeed the flatness is slightly smaller than 3. 
This can be justified  theoretically from PDF closures \cite{WilFri09,WilFri10}. We will neglect this effect in 
the sequel. At this stage, for $\ell\ge L$, a probabilistic description of the velocity increments is straightforward, i.e. 
$\delta_\ell u \build{=}_{}^{\mbox{law}} \sigma\delta$, where $\delta$ is a zero average unit variance Gaussian noise and
$\sigma^2 = \langle (\delta_L u)^2\rangle$, independent on the scale $\ell$. Henceforth, the symbol $\build{=}_{}^{\mbox{law}}$ 
stands for an equality in law, i.e. in probability. It means the the PDF and all the moments of the random variables on the two sides of the equality are equal.

In the inertial range, we need a probabilistic formulation of this power law behavior. This is clear at this stage that a simple 
Gaussian modeling is not enough since it would predict a scale-independent flatness of constant value 3. Furthermore, the parameters 
that we will use must be universal since the observed power-law is universal. The main idea to build up a probabilistic description 
is to mix Gaussian variables and, for example, to use a Gaussian random variable with a fluctuating variance. 
This was proposed in the so-called propagator approach \cite{CasGag90}. The form of the 
fluctuations of this stochastic variance will be given by the standard arguments of the multifractal formalism \cite{Fri95}. 
The corresponding non-Gaussian modeling consists in writing:
\begin{equation}\label{eq:StochVar}
\delta_\ell u \build{=}_{}^{\mbox{law}} \sigma\left(\frac{\ell}{L}\right)^h\delta\mbox{ ,}
\end{equation}
with $h$ a fluctuating variable, independent on the unit-variance zero-average Gaussian noise $\delta$, and characterized by 
its distribution function
\begin{equation}\label{eq:DistrH}
 \mathcal P_h^{(\ell)} (h) = \frac{\left( \frac{\ell}{L}\right)^{1-\mathcal D^E(h)}}{\int_{h_{\min}}^{h_{\max}}\left( \frac{\ell}{L}\right)^{1-\mathcal D^E(h)}dh}\mbox{ .}
\end{equation}
In the limit of vanishing values of $\ell$, $h$ and $\mathcal D^E(h)$ gain the mathematical status of  the 
H\"{o}lder exponent and the singularity spectrum respectively. Let us stress that the $h$-distribution $\mathcal P_h^{(\ell)} (h)$ (Eq. \ref{eq:DistrH}) is indeed 
normalized, and the range of integration [$h_{\min}$;$h_{\max}$] depends on the precise shape of the singularity spectrum. 
One of the main hypothesis of the multifractal formalism is to assume that $\mathcal D^E(h)$ is independent on the scale $\ell$ \cite{Fri95}.  
In this case, assuming $\min_h[1-\mathcal D^E(h)]=0$, a steepest-descent calculation (see Appendix \ref{ann:SDC}) 
shows that (recall that $h$ and $\delta$ are assumed independent)
\begin{equation}\label{eq:MomPInertialEuler}
\langle|\delta_\ell u|^p \rangle = \sigma^p\langle |\delta|^p \rangle\int_{h_{\min}}^{h_{\max}} \left(\frac{\ell}{L}\right)^{ph} 
\mathcal P_h^{(\ell)}(h)dh \build{\approx}_{\ell\rightarrow 0}^{}
\sigma^p \langle |\delta|^p \rangle\left( \frac{\ell}{L}\right)^{\min_h\left[ph+1-\mathcal D^E(h)\right]} \mbox{ ,}
\end{equation}
where $\langle |\delta|^p\rangle = \Gamma\left( \frac{p+1}{2}\right)/\sqrt{2^p\pi}$ and $\Gamma$ is the Gamma function. Hence, the form 
of the density $\mathcal P_h^{(\ell)}(h)$ (Eq. \ref{eq:DistrH}) implies that the structure functions behave as power-laws, with a set of exponents $\langle|\delta_\ell u|^p \rangle\sim \ell^{\zeta_p}$
linked to $\mathcal D^E(h)$ via a Legendre transform \cite{Fri95}
\begin{equation}\label{eq:Legendre}
 \zeta_p = \min_h \left[ph+1-\mathcal D^E(h)\right]\mbox{ .}
\end{equation}
It gives in a straightforward manner the behavior of the flatness in the inertial range, i.e. $F= 3(\ell/L)^{\zeta_4-2\zeta_2}$.
Let us first mention that in a K41 framework, without 
any intermittency corrections, the singularity spectrum is equal to $\mathcal D^{\mbox{K41}}(1/3)=1$ and  
$\mathcal D^{\mbox{K41}}(h)=-\infty$ for $h\ne 1/3$. In this case, using Eq. \ref{eq:Legendre}, we can easily show that $\zeta_p^{\mbox{K41}} = p/3$, 
and because of the linearity of the $\zeta_p$ function, the flatness is independent on the scale $\ell$. Clearly, the universal 
power-law behavior of the flatness in the inertial range as shown in Fig.  \ref{fig:FlatEul}, is the signature of  the presence of intermittency 
in turbulence.

In the literature, several models for $\mathcal D^E(h)$ have been proposed. One of the most widely used is the lognormal approximation, 
giving the simplest quadratic form of the $\mathcal D^E(h)$ spectrum including intermittency corrections, as proposed by Kolmogorov and Oboukhov 
\cite{Obo62,Kol62}:
\begin{equation}\label{eq:DHLN}
\mathcal D^E(h) = 1-\frac{(h-c_1)^2}{2c_2}\mbox{ .}
\end{equation}
In this case, the proposed stochastic modeling of the velocity increments (Eq. \ref{eq:StochVar}) has a simple probabilistic 
interpretation: with $h_{\min} = -\infty$ and $h_{\max} = +\infty$, the velocity increments are modeled as a Gaussian noise 
$\delta$  multiplied by a lognormal multiplicator $\sigma(\ell/L)^h$. 
This was the proposition made in the propagator approach \cite{CasGag90}. It is easily seen that $\zeta_p = c_1p-c_2\frac{p^2}{2}$, leading 
to $F= 3(\ell/L)^{-4c_2}$. In the sequel, we will choose 
$h_{\min} = 0$ and $h_{\max} = 1$ because (i) a numerical integration of Eq. \ref{eq:MomPInertialEuler} shows no difference with the 
indefinite case, (ii) as we will see, the extension to the dissipative range implies $h_{\min}>-1$ (see section \ref{subsub:dissipative})
and (iii) rigorously, only H\"{o}lder exponents greater than 0 and smaller than 1 are accessible when using increments. Experimental 
and numerical data as displayed in Fig. \ref{fig:FlatEul} show that the flatness behaves 
as a universal power-law of exponent $-0.10 \pm 0.01$. This corresponds to $c_2 = 0.025$, and $c_2$ is called 
the intermittency coefficient. 
Also seen on empirical data (data not shown), $\zeta_3 \approx 1$ \cite{Arneodoetal96}. Thus, in the sequel, we will take 
$c_1=1/3+3c_2/2\approx 0.37$, very close to its K41 prediction $1/3$. This defines completely the quadratic singularity spectrum 
(Eq. \ref{eq:DHLN}). 

Another widely used singularity spectrum is the She-L\'ev\^eque spectrum \cite{SheLev94},
\begin{equation}\label{eq:DHSL}
 \mathcal D^E(h) = -1+3\left[\frac{1+\ln(\ln(3/2))}{\ln(3/2)}-1\right](h-1/9)-\frac{3}{\ln(3/2)}(h-1/9)\ln(h-1/9)\mbox{ .}
\end{equation}
This spectrum is based on log-Poisson statistics 
\cite{Dub94} and yields $\zeta_p^{\mbox{SL}} = p/9+2[1-(2/3)^{p/3}]$. This implies that $F\approx 3(\ell/L)^{-0.11}$, 
a power-law behavior in good agreement with the empirical data shown in Fig. \ref{fig:FlatEul}. More sophisticated methods based on 
wavelets \cite{MuzBac93,WenAbr07} agree on the difficulty to discriminate between log-normal and 
log-Poisson statistics, and none of these approximations, for the moment, have been derived rigorously from first principles (i.e. 
the Navier-Stokes equations). For these reasons, we will use, in the sequel, the simplest lognormal approximation 
(Eq.  \ref{eq:DHLN}) that compares well with empirical data, at least for low order moments $\zeta_p$ with $p\le 6$ and that 
gives a Gaussian distribution for the $h$-exponent, that is easy to manipulate (see in particular in the stochastic modeling 
proposed in section \ref{sec:stochmodel}).  

Finally, the proposed stochastic modeling (Eqs. \ref{eq:StochVar} and \ref{eq:DistrH}) includes a functional 
form for the probability density functions of the velocity increments. A formal derivation 
of these PDFs, starting from the product of two independent random variables is reported in Appendix \ref{ann:PDF}. In simple words, we can see that the random variable $\delta_\ell u$ (Eq. \ref{eq:StochVar}) is fully defined when is given the law of $h$ (Eq. \ref{eq:DistrH}) and $\delta$ (let us say $\mathcal P_\delta$), that we assume, at this stage, independent.  To simplify the derivation of the PDF  $\mathcal P_{\delta_\ell u} (\delta_\ell u)$ of $\delta_\ell u$ (see Appendix \ref{ann:PDF} for a more rigorous derivation), let us consider only the absolute value of the velocity increment. We get
$$ \ln |\delta_\ell u| = \ln\sigma + h\ln\frac{\ell}{L}+\ln|\delta| \mbox{.} $$
Given that the random variables $h$ and $\ln|\delta|$ are independent, the PDF of $ \ln |\delta_\ell u|$ is the convolution product of the PDFs of  $h\ln\frac{\ell}{L}$ and $\ln|\delta|$, as it was noticed in Ref. \cite{CasGag90}, namely
\begin{align*} 
\mathcal P_{\ln \frac{ |\delta_\ell u|}{\sigma}}(\ln \frac{ |\delta_\ell u|}{\sigma}) &= \int \mathcal P_{\ln |\delta|}(\ln \frac{ |\delta_\ell u|}{\sigma}-h\ln\frac{\ell}{L})\mathcal P_{h\ln\frac{\ell}{L}}(h\ln\frac{\ell}{L})d(h\ln\frac{\ell}{L})\\
&= \int \mathcal P_{\ln |\delta|}(\ln \frac{ |\delta_\ell u|}{\sigma}-h\ln\frac{\ell}{L})\mathcal P_{h}(h)dh\mbox{ .}
\end{align*}
Noticing that $\mathcal P_{\ln |\delta|}(\ln |\delta|)= |\delta|\mathcal P_{|\delta|}(|\delta|)$ and $\mathcal P_{\ln \frac{ |\delta_\ell u|}{\sigma}}(\ln \frac{ |\delta_\ell u|}{\sigma}) =  |\delta_\ell u|\mathcal P_{|\delta_\ell u|}(|\delta_\ell u|)$, we finally get
\begin{equation}\label{eq:PDFAbsIncr}\mathcal P_{|\delta_\ell u|}(|\delta_\ell u|) =\int_{h_{\min}}^{h_{\max}}\frac{1}{\sigma}\left(\frac{\ell}{L}\right)^{-h}
\mathcal P_{|\delta|} \left[\frac{|\delta_\ell u|}{\sigma}\left(\frac{\ell}{L}\right)^{-h}\right]\mathcal P_h^{(\ell)} (h)dh\mbox{ ,}
\end{equation}
which shows the expression of the PDF of the absolute value of the velocity increment. Without further assumptions, this derivation cannot be generalized to derive the PDF of the signed velocity increment, since we are taking at one point a logarithm, but it clarifies the fact that defining in probability a random variable (such as in Eq. \ref{eq:StochVar}) allows to derive the associated PDF.

Actually, Eq. \ref{eq:PDFAbsIncr} is also true for the signed velocity increment, as shown in Appendix \ref{ann:PDF}, and we get for the velocity increment PDF 
$\mathcal P_{\delta_\ell u} (\delta_\ell u)$ at the scale 
$\ell$ the following form:
\begin{equation}\label{eq:PdfEulInert}
 \mathcal P_{\delta_\ell u} (\delta_\ell u) = \int_{h_{\min}}^{h_{\max}}\frac{1}{\sigma}\left(\frac{\ell}{L}\right)^{-h}
\mathcal P_\delta \left[\frac{\delta_\ell u}{\sigma}\left(\frac{\ell}{L}\right)^{-h}\right]\mathcal P_h^{(\ell)} (h)dh\mbox{ ,}
\end{equation}
where  $\mathcal P_\delta(x) = \exp(-x^2/2)/\sqrt{2\pi}$ is the PDF of a unit-variance zero-mean Gaussian variable. 
A numerical investigation (data not shown) of the PDF (Eq. \ref{eq:PdfEulInert}) shows the characteristic continuous shape deformation associated 
to the intermittency phenomenon, in a similar way than observed in Fig. \ref{fig:PdfEulLag}(a). Once again, this PDF is symmetric, i.e.
 $\mathcal P_{\delta_\ell u} (\delta_\ell u) =\mathcal P_{\delta_\ell u} (-\delta_\ell u)$ and thus fails to describe the 
skewness phenomenon. We invite the reader to have a look at the sections \ref{sec:fullskew} and \ref{sec:stochmodel} that 
provide a formalism able to reproduce the asymmetry of the PDFs, as observed in empirical data.  
We will see in the following section how to introduce the dissipative effects in order to obtain predictions on the statistics of the 
velocity gradients.

\subsubsection{The dissipative range}
\label{subsub:dissipative}

The flatness of velocity increments, shown in Fig. \ref{fig:FlatEul}, behaves in a very different way in the dissipative range, 
i.e. for scales  $\ln (\ell/L)/\ln (\mathcal R_e/\mathcal R^*)\le -3/4$. First of all, no power-law is observed. Then, we 
can see a strong Reynolds-number dependence. This is very different from the behavior observed in the inertial range.
In the limit of vanishing scales, the velocity increments can be Taylor expanded and we obtain 
$\delta_\ell u(x) = \ell\partial_xu(x)$, i.e. a linear behavior as a function of the scale. This implies that the flatness tends to a 
Reynolds-number dependent function, independent on the scale $\ell$, i.e. 
$F(\ell)\build{\rightarrow}_{\ell\rightarrow 0}^{}\frac{\langle (\partial_xu)^4\rangle}{\langle (\partial_xu)^2\rangle^2}$. The aim 
of this section is to understand and model the transition from the observed power-law behavior of the flatness in the inertial range, 
to this scale-independent behavior in the dissipative range. But first of all, let us have a look at the empirical data.

In Fig. \ref{fig:FlatEul}, we see that not all of the measurements exhibit at very small scales a scale-independent flatness. Only 
the DNS ($\mathcal R_\lambda=140$), the air-jet ($\mathcal R_\lambda=380$) and the Helium-jet ($\mathcal R_\lambda=208$) have
succedeed in resolving scales smaller than the Kolmogorov length scale. Indeed, the measurements are difficult 
because the Hot-wire size is usually of the order of the Kolmogorov length scale. If the Reynolds number is too high, 
dissipative scales are too small to be resolved (as for $\mathcal R_\lambda=463$, $703$ and $929$). In the Modane's wind tunnel, 
the integral length scale is very large, and despite the large Reynolds number ($\mathcal R_\lambda=2500$), $\eta_K$ should have 
been accessible (which is of the order of the millimeter). Unfortunately, the noisy local environment of the giant wind tunnel and the high temperatures reached ($\sim 60^\circ$ Celsius) implied by the strong level of turbulence, prevented electronics from working properly at high frequency.

Given the experimental limitations, we see that the flatness underlies a rapid increase. This was the subject of 
Ref. \cite{CheCas05}. The main underlying idea is the differential action of the viscosity. The multiplicator $\sigma(\ell/L)^h$
fluctuates in the inertial range. In the dissipative range, up to a (random) coefficient, the multiplicator becomes 
linear with respect to the scale in order to be consistent with the Taylor's expansion. The viscosity will regularize this 
inertial-range singular behavior at a scale that depends on the strength $h$ of the singularity: the bigger is the 
multiplicator, corresponding to smaller $h$-exponents, the smaller is the scale of regularization. 
Purely kinematic arguments, proposed in Ref. \cite{CheCas05}, show that the width of the so-called intermediate dissipative range 
$[\eta^-,\eta^+]$ \cite{FriVer91}, where $\eta^-$ and  $\eta^+$ are respectively the smallest and biggest dissipative scales, indeed
depends weakly on the Reynolds number
\begin{equation}\label{eq:predkinematicintdiss}
 \ln\left( \frac{\eta^+}{\eta^-}\right) \sim \sqrt{\ln \mathcal R_e}\mbox{ ,}
\end{equation}
the Kolmogorov length scale $\eta_K$ lying in between. In the representation chosen in Fig. \ref{fig:FlatEul}, this prediction
(Eq. \ref{eq:predkinematicintdiss}) implies that the observed rapid increase occuring in the intermediate dissipative range 
should be steeper and steeper as the Reynolds number increases. This is what is qualitatively observed as long 
as experimental technics were able to reach these very small scales.

This can be fully modeled in the context of the multifractal formalism. To do so, one has to come up with a dissipative 
scale that depends explicitly on the strength of the multiplicator, or in a more straightforward manner, on the exponent $h$. 
Paladin and Vulpiani \cite{PalVul87} first proposed such a $h$-dependent cut-off. Their reasoning was based on the local 
Reynolds number. 
For a scale lying in the inertial range, a fluctuating Reynolds number can be defined using a fluctuating 
characteristic velocity $v_\ell = \sigma (\ell/L)^h$, leading to the local Reynolds number $\mathcal R_\ell = v_\ell \ell/\nu$. 
We can see that at the integral length scale, the associated Reynolds number $\mathcal R_L$ is unique, does not fluctuate, 
and is given by $\mathcal R_L=\mathcal R_e = \sigma L/\nu$. This allows to define unambiguously a dissipative length scale: 
a scale $\eta$ at which 
the local Reynolds number is of order unity $\mathcal R_\eta = \mathcal R^* = O(1)$. The order one constant $\mathcal R^*$, is a
free parameter of the formalism, a priori universal. It has been introduced phenomenologically in Refs. 
\cite{ChaCha00,MalAur00,GagCas04}. We will take it to be $\mathcal R^*=52$ and we will show that it is related to the Kolmogorov 
constant $c_K$. From there, one obtains directly the main result of Ref. \cite{PalVul87}, namely 
\begin{equation}\label{eq:etah}
 \eta(h) = L \left(\frac{\mathcal R_e}{\mathcal R^*}\right)^{-\frac{1}{h+1}}\mbox{ .}
\end{equation}
Based on this result, Nelkin \cite{Nel90} showed the implication of a fluctuating dissipative scale (Eq. \ref{eq:etah}) on the 
modeling of the velocity increments for scales lying in the dissipative range. Within our approach summarized by Eq. 
\ref{eq:StochVar}, the velocity increment at a scale smaller than the dissipative scale $\eta(h)$ will be modeled in the
following way:
\begin{equation}\label{eq:StochVardiss}
\delta_\ell u \build{=}_{\ell\le\eta(h)}^{\mbox{law}} \sigma\frac{\ell}{L}\left(\frac{\eta(h)}{L}\right)^{h-1}\delta\mbox{ ,}
\end{equation}
where $\delta$, as in Eq. \ref{eq:StochVar}, is a zero-mean unit variance Gaussian random variable,  
$\sigma^2=\langle (\delta_Lu)^2\rangle$, independent of the random exponent $h$. This probabilistic modeling 
(Eq. \ref{eq:StochVardiss}) is consistent with the Taylor
expansion of the velocity increment. It implies a model for the velocity gradient \cite{Nel90}, namely:
\begin{equation}\label{eq:StochVardiffdiss}
\partial_x u \build{=}_{}^{\mbox{law}}\frac{\sigma}{L}\left(\frac{\eta(h)}{L}\right)^{h-1}\delta\mbox{ .}
\end{equation}
Also we see that the proposed modeling for $\ell\ge\eta(h)$ (Eq. \ref{eq:StochVar}) and for 
$\ell\le\eta(h)$ (Eq. \ref{eq:StochVardiss}) is continuous at the dissipative length scale $\ell=\eta(h)$. 

As in Eqs. \ref{eq:StochVar} and \ref{eq:DistrH}, to fully characterize in a probabilistic manner the random variable 
$\delta_\ell u$ (Eq.  \ref{eq:StochVardiss}), we need to define the distribution of the exponent $h$. The proposition of Ref. 
\cite{Nel90} is to take for the distribution of $h$ at the dissipative scales $\ell\le\eta(h)$ a scale-independent distribution (up to a normalizing function $\mathcal Z(\ell)$)
that also has to be continuous at the transition $\ell=\eta(h)$, namely
 \begin{equation}\label{eq:DistrHdiss}
 \mathcal P_h^{(\ell)} (h)  \build{=}_{\ell\le\eta(h)}^{} \frac{1}{\mathcal Z(\ell)}\left( \frac{\eta(h)}{L}\right)^{1-\mathcal D^E(h)}\mbox{ ,}
\end{equation}
where  $\mathcal D^E(h)$ is the same universal 
function entering in the $h$-distribution of the inertial range (Eq. \ref{eq:DistrH}), that we will approximate to be quadratic 
(Eq. \ref{eq:DHLN}).

As a general remark, as previously underlined in Ref. \cite{FriVer91}, the multifractal formalism becomes predictive. Indeed, given the 
description of the inertial range (Eqs. \ref{eq:StochVar} and \ref{eq:DistrH}), and in particular given the singularity spectrum 
$\mathcal D^E(h)$, we can predict the behavior of the velocity increments in the dissipative range 
(Eqs. \ref{eq:StochVardiss} and \ref{eq:DistrHdiss}) by a simple continuity argument. As a consequence, using Eqs. \ref{eq:etah}, \ref{eq:StochVardiffdiss}, and \ref{eq:DistrHdiss}, the $2p^{\mbox{th}}$-order 
moment of the velocity gradients is given by following 
function of the Reynolds number:
\begin{equation}\label{eq:predgradeul}
 \langle (\partial_x u)^{2p}\rangle = \langle\delta^{2p}\rangle \left(\frac{\sigma}{L}\right)^{2p}\frac{1}{\mathcal Z(0)}\int_{h_{\min}}^{h_{\max}} 
\left(\frac{\mathcal R_e}{\mathcal R^*}\right)^{-\frac{2p(h-1)+1-\mathcal D^E(h)}{h+1}}dh\mbox{ ,}
\end{equation}
where $\langle\delta^{2p}\rangle = \frac{(2p)!}{p!2^p}$, and $\mathcal Z(0) = \int_{h_{\min}}^{h_{\max}} 
\left(\frac{\mathcal R_e}{\mathcal R^*}\right)^{-\frac{1-\mathcal D(h)}{h+1}}dh$. The distribution of a  Gaussian variable $\delta$ is even, so all the odd moments of both velocity increments (Eqs. \ref{eq:StochVar},  
\ref{eq:StochVardiss}) and gradients (Eq. \ref{eq:StochVardiffdiss}) are vanishing at this stage. 
To include the skewness phenomenon, that requires further probabilistic modeling, we invite the reader to take a look at 
section \ref{sec:fullskew}.

The explicit form of even-order moments of velocity gradients (Eq. (\ref{eq:predgradeul})) allows us to derive two important 
predictions: the computation of the average dissipation and the dependence on the Reynolds number of the flatness of the 
velocity derivatives. The dissipation $\epsilon$ is a key quantity in turbulence theory \cite{Fri95}. In an isotropic and homogeneous 
flow, the average dissipation is related to the second order moment of the velocity gradients as 
$\langle\epsilon\rangle = 15\nu\langle (\partial_xu)^2\rangle$. Using Eq. (\ref{eq:predgradeul}), 
one obtains
\begin{equation}\label{eq:ComputMultiEpsilon}
 \langle\epsilon\rangle = 15\nu\sigma^2\frac{1}{L^2\mathcal Z(0)}\int_{h_{\min}}^{h_{\max}} 
\left(\frac{\mathcal R_e}{\mathcal R^*}\right)^{-\frac{2(h-1)+1-\mathcal D^E(h)}{h+1}}dh \mbox{ .}
\end{equation}
In the limit of large Reynolds number, a steepest-descent calculation (see Appendix \ref{ann:SDC}) shows that, up to an order one multiplicative constant, the former integral, 
 is dominated by the term $(\mathcal R_e/\mathcal R^*)^{\chi_2}$, where $\chi_2=
-\min_h\left(\frac{2(h-1)+1-\mathcal D^E(h)}{h+1}\right)$. Because by construction, and for a wide class of singularity spectrum 
such that (via a Legendre transform, c.f. Eq. \ref{eq:Legendre}) $\zeta_3=1$, which is the case with the quadratic approximation
 (Eq. \ref{eq:DHLN}) with $c_1=1/3+3c_2/2$, we can show \cite{Fri95} that $\chi_2=1$. This implies that $\langle\epsilon\rangle$ is 
independent on the Reynolds number. This verifies a basic hypothesis of Kolmogorov, namely the finiteness of  
$\langle\epsilon\rangle$ at infinite Reynolds number. A precise estimation of the integrals entering in Eq. 
\ref{eq:ComputMultiEpsilon}  (see Appendix \ref{ann:SDC}) shows that 
\begin{equation}\label{eq:ComputMultiEpsilonSimp}
 \langle\epsilon\rangle \build{\approx}_{\mathcal R_e \rightarrow +\infty}^{}\frac{15}{\mathcal R^*}\frac{\sigma^3}{L} \mbox{ .}
\end{equation}
As shown in Refs. \cite{CheCas05,CheCas06}, Eq. \ref{eq:ComputMultiEpsilonSimp} gives a prediction of the Kolmogorov constant $c_K$. Indeed, 
in the inertial range, neglecting intermittent corrections, the structure function is given by 
$\langle (\delta_\ell u)^2\rangle = \sigma^2(\ell/L)^{2/3} = c_K (\langle\epsilon\rangle \ell)^{2/3}$. Using Eq. 
\ref{eq:ComputMultiEpsilonSimp}, it gives $c_K = \left(\frac{\mathcal R^*}{15}\right)^{2/3}$, showing indeed that this
constant $\mathcal R^*$ is related to the Kolmogorov constant $c_K$. Using $\mathcal R^*=52$, one finds $c_K = 2.33$, which is 
slightly bigger than what has be measured on empirical data $c_K\approx 2$ \cite{YeuZho97,DonSre10}. One of the reasons of this apparent discrepancy 
is related to the fact that the constant $\mathcal R^*$ is clearly linked to the presence 
of intermittency and the implied extension of the intermediate dissipative range, whereas the definition of the $c_K$ assumes 
the absence of intermittency corrections. The prediction of the 
second order moment of the gradients allows also to link precisely the (large-scale) Reynolds number $\mathcal R_e$ to the Taylor 
based Reynolds number $\mathcal R_\lambda$ in the following way:
\begin{equation}\label{eq:DefRlambdaRstar}
 \mathcal R_e = \frac{4}{\mathcal R^*}\mathcal R_\lambda^2\mbox{ .}
\end{equation}

Another important prediction is the dependence of the flatness of derivatives on the 
Reynolds number \cite{Nel90}. From Eq. \ref{eq:predgradeul}, using a steepest-descent argument, we can show that, up to multiplicative constants of order unity (given in Appendix \ref{ann:SDC}): 
\begin{equation}
 \frac{ \langle (\partial_x u)^{4}\rangle}{ \langle (\partial_x u)^{2}\rangle^2}\build{\approx}_{\mathcal R_e \rightarrow +\infty}^{}
3\left(\frac{\mathcal R_e}{\mathcal R^*}\right)^{\chi_4-2\chi_2}\mbox{ with }\chi_p = \build{\min}_{h}^{} \left[ -\frac{p(h-1)+1-\mathcal D^E(h)}{h+1}\right]\mbox{ .}
\end{equation}
Using the quadratic spectrum (Eq. \ref{eq:DHLN}), $\mathcal R^*=52$, and Eq. \ref{eq:DefRlambdaRstar} 
for the Taylor-based Reynolds number dependence, we obtain:
\begin{equation}\label{eq:predflatderiveul}
 \frac{ \langle (\partial_x u)^{4}\rangle}{ \langle (\partial_x u)^{2}\rangle^2}\build{\approx}_{\mathcal R_e \rightarrow +\infty}^{}
3\left(\frac{\mathcal R_e}{\mathcal R^*}\right)^{0.18} = 1.47 \mathcal R_e^{0.18} = 0.93\mathcal R_\lambda^{0.36} \mbox{ ,}
\end{equation}
which is consistent with empirical observations \cite{VanAnt80,SreAnt97,GylAyy04,IshKan07}. More precisely, we compare in 
Fig. \ref{fig:FlatEulTheo} the multifractal prediction (Eq. \ref{eq:predflatderiveul}) with various experimental measurements and 
numerical simulations as compiled in Ref. \cite{IshKan07}. The present theoretical prediction reproduces quantitatively the 
empirical observations. Let us stress that 
the quantitative dependence of the flatness of velocity derivatives on the Reynolds number (Eq. \ref{eq:predflatderiveul}) 
is a genuine consequence of the intermittent nature of turbulence. In a K41 framework, this quantity would be Reynolds number 
independent.

\subsubsection{Full multi scale description}
\label{sec:fulleuldescrip}

 \begin{figure}[t]
\center{\epsfig{file=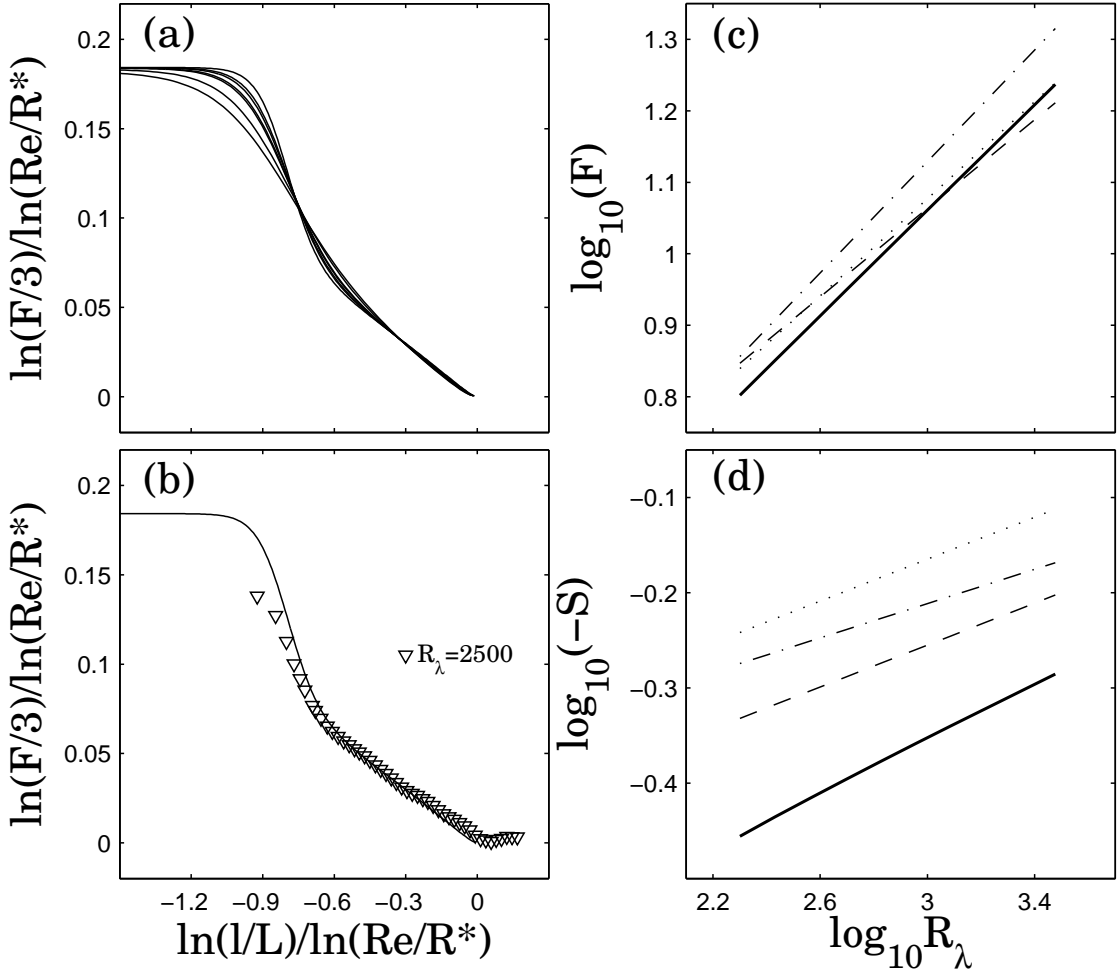,width=15cm}}
\caption{(a) Theoretical predictions of the velocity increment flatness 
obtained using Eqs. \ref{eq:betah} and \ref{eq:phmeneeul} with a quadratic singularity spectrum (see text) 
at different Reynolds numbers, as proposed in Fig. \ref{fig:FlatEul}. The higher is the Reynolds number, the steeper the  
increase in flatness that takes place in the intermediate dissipative range. (b) Direct comparison of the velocity increment flatness obtained 
at Modane $\mathcal R_\lambda=2500$ (symbols) and from the model (solid line). (c) Comparison of various observations of the Reynolds number dependence of the flatness of the velocity derivatives 
with the present theoretical predictions (Eq. \ref{eq:predflatderiveul}): $F=1.36\mathcal R_\lambda^{0.31}$ \cite{AntCha81} 
(Athmospheric measurements, dashed line), $F=0.91\mathcal R_\lambda^{0.39}$ \cite{GylAyy04} (wind-tunnel measurements, 
dash-dotted line), $F=1.14\mathcal R_\lambda^{0.34}$ \cite{IshKan07} (DNS, dotted line) and $F=0.93\mathcal R_\lambda^{0.36}$ for 
the present prediction Eq. \ref{eq:predflatderiveul} (solid line). (d) Comparison of various observations of the Reynolds
number dependence of the skewness of the velocity derivatives 
with the present theoretical predictions (Eq. \ref{eq:PredMom3Grad}): $S=-0.26\mathcal R_\lambda^{0.11}$ \cite{AntCha81} 
(Athmospheric measurements, dashed line), $S=-0.33\mathcal R_\lambda^{0.09}$ \cite{GylAyy04} (wind-tunnel measurements, 
dash-dotted line), $S=-0.32\mathcal R_\lambda^{0.11}$ \cite{IshKan07} (DNS, dotted line) and $S=-0.175\mathcal R_\lambda^{0.134}$ for 
the present prediction Eqs. \ref{eq:PredMom3Grad} and \ref{eq:PredSkewNumEul} (solid line).} \label{fig:FlatEulTheo}
\end{figure}

In the preceding sections (\ref{sec:EulInert} and \ref{subsub:dissipative}), we have seen how to model velocity fluctuations 
in respectively the inertial and far-dissipative ranges. It remains to write down a formalism, based on these two well known 
limiting ranges, that reproduces velocity statistics in the entire range of scales, including the intermediate dissipative 
range.

A first naive idea to gather both the inertial and dissipative ranges in a unified description is to consider, at a given scale $\ell$ and 
Reynolds number $\mathcal R_e$, the $h$-exponents. Within the range $[h_{\min};h_{\max}]$, if $h\le h^*$, then the velocity 
increment lies in the inertial range; On the opposite, if $h\ge h^*$, then the velocity increment lies in the dissipative range. The transition 
occurs at the exponent $h^*$ defined by $\ell = \eta(h^*)$, namely
\begin{equation}
 h^*(\ell, \mathcal R_e) = -\left(1+\frac{\ln(\mathcal R_e/\mathcal R^*)}{\ln (\ell/L)} \right)\mbox{ .}
\end{equation}
Using both the laws of the velocity increments in the inertial (Eqs. \ref{eq:StochVar} and \ref{eq:DistrH}) and 
dissipative (Eqs. \ref{eq:StochVardiss} and \ref{eq:DistrHdiss}) ranges, we then obtain 
\begin{equation}
 \langle |\delta_\ell u|^p\rangle =  \frac{\langle |\delta|^p\rangle}{\mathcal Z(\ell)} \left[ \int_{h_{\min}}^{h^*}
\left(\frac{\mathcal R_e}{\mathcal R^*}\right)^{-\frac{p(h-1)+1-\mathcal D(h)}{h+1}}dh + \int^{h_{\max}}_{h^*}
\left(\frac{\ell}{L}\right)^{ph+1-\mathcal D(h)}dh\right]\mbox{ ,}
\end{equation}
where $\mathcal Z(\ell)$ normalizes the probability densities, namely
\begin{equation}
\mathcal Z(\ell) = \int_{h_{\min}}^{h^*}
\left(\frac{\mathcal R_e}{\mathcal R^*}\right)^{-\frac{1-\mathcal D(h)}{h+1}}dh + \int^{h_{\max}}_{h^*}
\left(\frac{\ell}{L}\right)^{1-\mathcal D(h)}dh\mbox{ .}
\end{equation}
Unfortunately, this model does not compare well with empirical data because the transition is too sharp (data not shown, 
see \cite{ChePhD}). Furthermore, the present modeling gives a continuous but not differentiable modeling of the velocity increment.

A continuous and differentiable transition, inspired by the interpolation function of Batchelor 
to model the second-order structure function \cite{Bat51}, has been proposed in Ref. \cite{Men96} as the entire range of scales. In Ref. \cite{CheCas06}, a slight modification of the proposition of Ref. \cite{Men96} has been introduced in order to make the transition compatible with the far-dissipative 
predictions of Ref. \cite{Nel90}. It reads
\begin{equation}\label{eq:betah}
 \delta_\ell u \build{=}_{}^{\mbox{law}} \sigma\beta_\ell \delta \mbox{ with } \beta_\ell  =
\frac{\left(\frac{\ell}{L}\right)^h}{\left[
1+\left(\frac{\ell}{\eta(h)}\right)^{-2}\right]^{(1-h)/2}} \mbox{
,}
\end{equation}
where the random variable $\delta$ is again a zero-mean unit-variance Gaussian noise, $\sigma^2 = \langle (\delta_Lu)^2\rangle$, and
\begin{equation}\label{eq:phmeneeul}
\mathcal P_h^{(\ell)}(h)
=\frac{1}{\mathcal Z(\ell)}
\frac{\left(\frac{\ell}{L}\right)^{1-\mathcal D(h)}}{\left[
1+\left(\frac{\ell}{\eta(h)}\right)^{-2}\right]^{(\mathcal
D(h)-1)/2}}\mbox{ .}
\end{equation}
The normalizing constant is again such that $\int_{h_{\min}}^{h_{\max}}\mathcal P_h^{(\ell)}(h)dh=1$, namely
\begin{equation}\label{eq:ZoflMene}
 \mathcal Z(\ell)=\int_{h_{\min}}^{h_{\max}}\frac{\left(\frac{\ell}{L}\right)^{1-\mathcal D(h)}}{\left[
1+\left(\frac{\ell}{\eta(h)}\right)^{-2}\right]^{(\mathcal
D(h)-1)/2}}dh\mbox{ .}
\end{equation}
Note that at a fixed given scale $\ell$, the velocity increment 
(Eq. \ref{eq:betah}) and the $h$-distribution  (Eq. \ref{eq:phmeneeul}) tend to the inertial description (given by 
Eqs. \ref{eq:StochVar} and \ref{eq:DistrH}) in the limit of infinite Reynolds number. In the same manner, at a given finite 
Reynolds number, this proposed description tends to the dissipative predictions (Eqs. \ref{eq:StochVardiss} and \ref{eq:DistrHdiss}))
in the limit of vanishing scales $\ell\rightarrow 0$.

We show in Fig. \ref{fig:FlatEulTheo}(a) the theoretical predictions of the flatness as a function of the scales $\ell$ for 
the different Reynolds numbers previously investigated in Fig. \ref{fig:FlatEul}. To do so,
we integrated numerically Eqs. \ref{eq:betah} and \ref{eq:phmeneeul} using a quadratic singularity spectrum, with $c_2=0.025$ and 
$c_1=1/3+3c_2/2$, $h_{\min}=0$ and $h_{\max}=1$ and $\mathcal R^* =52$. The Reynolds number $\mathcal R_e$ is obtained from the 
Taylor-based Reynolds number using Eq. \ref{eq:DefRlambdaRstar}.
The proposed description based on  Eqs. \ref{eq:betah} and \ref{eq:phmeneeul}, reproduces the main characteristics shown in 
Fig. \ref{fig:FlatEul}, namely, the universal power-law behavior in the inertial range and the rapid increase of the flatness 
in the dissipative range. Obviously, the theoretical predictions do not suffer from a lack of resolution and so with the 
proposed renormalization of the scales and of the flatness, all the curves tend to a universal plateau given by Eq. 
\ref{eq:predflatderiveul} when $\ell\rightarrow 0$. The rapid increase that takes place in the intermediate dissipative range 
is consistent with the kinematic prediction given in Eq. \ref{eq:predkinematicintdiss}, namely the slope of this increase behaves 
as $1/\sqrt{\ln (\mathcal R_e/\mathcal R^*)}$ in this representation. 
We can see indeed that the higher is the Reynolds number, the steepest is the increase.

In  Fig. \ref{fig:FlatEulTheo}(b), we compare more precisely the velocity increment flatness obtained in Modane's wind tunnel 
\cite{KahMal98} to the multifractal prediction. The model does reproduce both the inertial and intermediate dissipative ranges, given the experimental limitations to reach the far dissipative range.

\subsubsection{Reinterpretation of the Tabeling's data as a non trivial effect of the dissipative physics}
\label{sec:tabelingexp}

 \begin{figure}[t]
\center{\epsfig{file=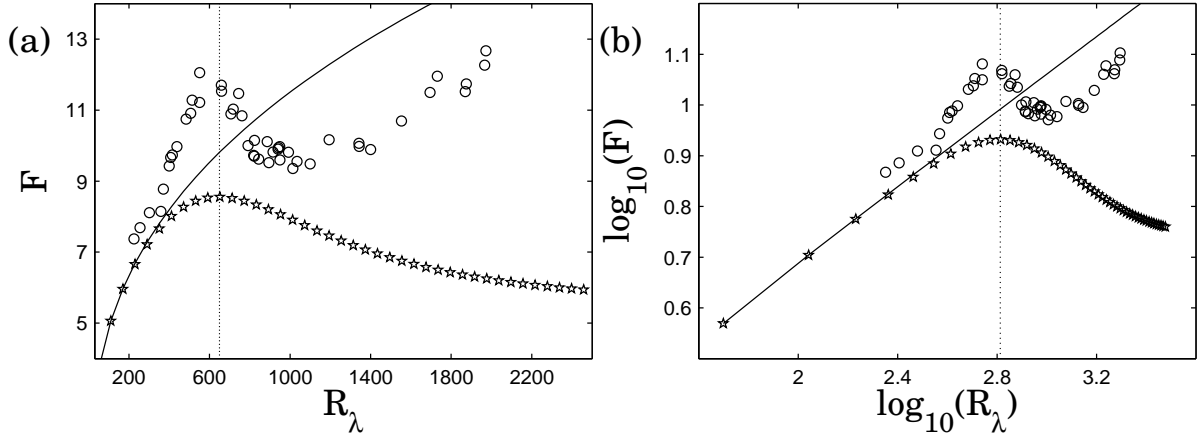,width=16cm}}
\caption{Flatness of the velocity derivatives as a function of the Taylor-based Reynolds number in a (a) linear and (b) logarithmic representation: ($\circ$) experimental data from a Von-Karman flow in gazeous helium at low temperature \cite{TabWil02}, (solid line) Multifractal prediction for the velocity derivative flatness (Eq. \ref{eq:predflatderiveul}), (stars) Multifractal prediction for the velocity increment flatness at a fixed scale $\ell_o/L=2.5 \: 10^{-3}$.
 } \label{fig:FlatDerivTab}
\end{figure}

As an example of the implications of the present theory, we reexamine in this section the observations of Tabeling and Willaime  \cite{TabWil02}. These authors investigated fully developed turbulence in a Von-Karman flow in gazeous helium at low temperature. Varying the pressure of the gas, they could span an unusually large range of Taylor scale based Reynolds numbers $200<R_{\lambda}<2200$. They measured the local velocity using a hot wire probe. Through the Taylor ``frozen turbulence'' hypothesis, they could access to the longitudinal derivative of the velocity, and its flatness $F$ (Eq. \ref{eq:predflatderiveul}).
Up to $R_{\lambda}=600$, their results are in good agreement with previous literature and the present prediction (Eq. \ref{eq:predflatderiveul}). Surprisingly, at $R_{\lambda} \simeq 650$, $F$ presents a maximum, and goes down up to $R_{\lambda} \simeq 1000$, then raises again slowly (Fig. \ref{fig:FlatDerivTab}).Tabeling and Willaime interpreted their results as some evidence of a transition in turbulent flows. Some comments suggested that this behavior of $F$ could be due to the finite size of the probe. However, as remarked by Tabeling and Willaime, this size (about 10$\mu$m) is much smaller than the Kolmogorov dissipation length $\eta$ at $R_{\lambda} = 650$. Moreover, such a limitation was expected to yield a saturation of $F$ at a constant value, not a well pronounced maximum.

Indeed, following the present multifractal theory, identifying the ``velocity derivative'' with a finite difference at a constant length $\ell_o$ gives a maximum for the flatness as shown in Fig. \ref{fig:FlatDerivTab}. This is due to the rapid rise of the longitudinal velocity difference flatness in the intermediate dissipative range. This maximum occurs when the length $\ell_o$ coincides with the lowest scale of this intermediate range, which is much smaller than $\eta$. Using Eqs. \ref{eq:betah} and \ref{eq:phmeneeul} with a quadratic singularity spectrum (Eq. \ref{eq:DHLN}), $\mathcal R^*=52$, stars in Fig. (\ref{fig:FlatDerivTab}) show the behavior predicted for $F$ if $\ell_o/L=2.5 \: 10^{-3}$, in reasonable agreement with the size of the sensor.

However, the present theory cannot predict a further rise of $F$ as observed experimentally. Also, the width of the predicted peak is much wider than observed. The present theory can explain some of the surprising features observed, not all.

\subsection{Consistent description of the skewness phenomenon}
\subsubsection{General discussion on the skewness phenomenon}
\label{sec:fullskew}

This section is devoted to the modeling of the skewness of the velocity increments. As we just saw, modeling the velocity increment as 
a Gaussian random variable multiplied by a random amplitude (see Eq. \ref{eq:betah}) cannot reproduce the asymmetric nature of the 
distribution of velocity increments since the Gaussian random variable $\delta$, and its independence on the amplitude $\beta_\ell$
lead to vanishing odd-order moments, i.e. $\forall p\in \mathbb N$, $\langle (\delta_\ell u)^{2p+1}\rangle =0$. Nevertheless, 
keeping the same probabilistic description as in Eqs. \ref{eq:betah} and \ref{eq:phmeneeul} for the second order moment of velocity increments, allows us to predict in a consistent 
way the third-order moment of velocity increments if we use the Karman-Howarth-Kolmogorov equation:
\begin{equation}\label{eq:KarmanHowarth}
 \langle (\delta_\ell u)^3\rangle = -\frac{4}{5}\langle \epsilon \rangle\ell + 6\nu \frac{d\langle (\delta_\ell u)^2\rangle}{d\ell}\mbox{ .}
\end{equation}
In Ref. \cite{CheCas06} (see also Ref. \cite{BosChe10}), we compared experimental data to the predictions obtained for 
$\langle (\delta_\ell u)^3\rangle$ using the Karman-Howarth-Kolmogorov equation \ref{eq:KarmanHowarth} and the second order 
structure function $\langle (\delta_\ell u)^2\rangle$ obtained from Eqs. \ref{eq:betah} and \ref{eq:phmeneeul}. Predictions 
and emprical data compares well in the whole range of scales \cite{CheCas06} without additional free parameters. In particular, 
the level of skewness in the inertial range is well reproduced and can be shown to be related to the universal constant $\mathcal R^*$. 
If we neglect dissipative effects in the exact relation Eq. \ref{eq:KarmanHowarth}, and the intermittent corrections on the 
second order structure function, it is easy to see that $S(\ell)=\langle (\delta_\ell u)^3\rangle/\langle (\delta_\ell u)^2\rangle^{3/2}$ 
is independent on the scale $\ell$ and can be further approximated to $S=-12/\mathcal R^* = -0.23$, in excellent agreement with 
empirical data \cite{CheCas06}. Moreover, applying a Taylor development on both Eq. \ref{eq:KarmanHowarth} and on the 
multifractal predictions (Eqs. \ref{eq:betah} and \ref{eq:phmeneeul}) for the second order structure function, in the limit of 
vanishing scales, we get the following multifractal prediction for the third order moment of the derivatives:\begin{equation}\label{eq:PredMom3Grad}
\langle (\partial_x u)^3\rangle =   -\frac{6\nu\sigma^2}{L^4} \left[
\frac{2}{\mathcal Z(0)}
\int_{h_{\min}}^{h_{\max}} \left[2h-1-\mathcal D^E(h) \right]\left(
\frac{\mathcal R_e}{\mathcal R^*}\right)^{-\frac{2(h-2)+1-\mathcal D^E(h)}{h+1}}dh + \mathcal F
\right]\mbox{ ,}
\end{equation}
where $\mathcal F$ is a negligible additive term, coming from the
Taylor's development of the normalizing factor $\mathcal
Z(\ell)$ (Eq. \ref{eq:ZoflMene}) \cite{CheCas06}. This prediction of the third order moment 
of the velocity gradient (Eq. \ref{eq:PredMom3Grad})  using a Batchelor-Meneveau 
type of transition between the inertial and dissipative ranges (Eqs. \ref{eq:betah} and \ref{eq:phmeneeul}), 
depends on the singularity spectrum $\mathcal D^E$, measured in the inertial range on empirical data, and on the universal constant 
$\mathcal R^*$. We show in Fig. \ref{fig:FlatEulTheo}(d) the numerical estimation of the multifractal prediction of the 
Skewness of derivatives (using Eqs. \ref{eq:betah}, \ref{eq:phmeneeul} and \ref{eq:PredMom3Grad}) as a function of the 
Taylor-based Reynolds number, using the quadratic singularity spectrum (Eq. \ref{eq:DHLN}) and $\mathcal R^*=52$. 
This prediction is compared to empirical data \cite{GylAyy04,IshKan07,AntCha81} as described in the figure 
caption and compiled in Ref. \cite{IshKan07}. We observe some dispersion between the three different empirical skewnesses, although the dependence on the Reynolds 
number seems to be universal. The difference in amplitude could be due to a lack of 
experimental and numerical resolution, and/or to a lack of statistical convergence. Thus, if the multifractal approach fails to predict 
the value of the skewness, it does reproduce accurately the Reynolds number dependence.  Indeed, using Eq. \ref{eq:PredMom3Grad}, 
a steepest-descent calculation shows that the skewness of the derivatives behaves as a power law of the Reynolds number, i.e. 
\begin{equation}
\ln(-S(0))/\ln(\mathcal R_e) \sim \chi_S-1\mbox{ ,}
\end{equation} 
with
\begin{equation}
\chi_S =  \min_h \left[ -\frac{2(h-2)+1-\mathcal D^E(h)}{h+1}\right]-\frac{3}{2}\min_h \left[ -\frac{2(h-1)+1-\mathcal D^E(h)}{h+1}\right] \mbox{ .}
\end{equation}
Using a quadratic approximation for the parameter function $\mathcal D^E$ (Eq. \ref{eq:DHLN}), we get $\chi_S-1 = 0.067$.
This power-law dependence on the Reynolds number has been already obtained by Nelkin \cite{Nel90} using a different, 
although related, approach based on the asymptotically exact relationship $\langle (\partial_xu)^3\rangle = -2\nu \langle (\partial_x^2u)^2\rangle$.
As shown is Fig. \ref{fig:FlatEulTheo}(d), further numerical estimations of relation  (\ref{eq:PredMom3Grad}), once rephrased in terms 
of Taylor-based Reynolds numbers using Eq. \ref{eq:DefRlambdaRstar}, leads to the following dependence of the velocity derivative skewness 
on $\mathcal R_\lambda$:
\begin{equation}\label{eq:PredSkewNumEul}
 \frac{\langle (\partial_x u)^3\rangle}{\langle (\partial_x u)^2\rangle^{3/2}} \build{\approx}_{\mathcal R_e\rightarrow +\infty}^{}
-0.175\mathcal R_\lambda^{0.134}\mbox{ .}
\end{equation}

\subsubsection{Modeling the velocity increments probability density function}

\label{section:PdfAsymTheo}

Indeed, consistent predictions for 
higher odd order structure functions are needed to predict the shape of the full velocity increment probability density function (PDF).
Several propositions were made in the literature to account for the asymmetry of the PDF linked to the skewness 
phenomenon \cite{CasGag90,CheCas06}. To do so, we must modify the noise $\delta$ entering in the probabilistic formulation 
Eq. \ref{eq:betah}, and/or correlate this noise $\delta$ with the exponent $h$. 
Asymmetric PDFs can be obtained if we change 
the Gaussian random variable $\delta$ to a non-Gaussian noise, still independent on the scale $\ell$ and on the multiplicator $\beta_\ell$. More 
precisely, it was proposed in Ref. \cite{CasGag90} to consider the random variable $\delta$ as being a variable of density 
$\mathcal P_\delta(\delta)$ that now reads
\begin{equation}\label{eq:BruitCastain90}
 \mathcal P_\delta(\delta) \propto \exp\left[ -\frac{\delta^2}{2}\left(1+a_S\frac{\delta}{\sqrt{1+\delta^2}} \right)\right]\mbox{ ,}
\end{equation}
where $a_S\approx 0.18$ is a universal constant, independent on both Reynolds number and scales. The main problem using this peculiar noise (Eq. \ref{eq:BruitCastain90}) is that it leads to non zero average velocity increments. This could be fixed by introducing 
a  scale dependent free parameter that centers the whole velocity increment PDF. Furthermore,  to reproduce the 
non trivial behavior of the skewness in the dissipative range, we need to modify the parameter $a_S$, and to make it dependent on 
both scale and Reynolds number. In this spirit, still based on the hypothesis of independence of the two random variables $h$ and 
$\delta$, a general development of the PDF of $\delta$ 
on a basis made of the successive derivatives of a Gaussian, called the Edgeworth's development, was proposed in Ref. \cite{CheCas06}: 
\begin{equation}\label{eq:Edgeworth}
 \mathcal P_\delta(\delta) = \frac{1}{\sqrt{2\pi}}\sum_{n=0}^{+\infty}\lambda_n(\ell)\frac{d^n}{d\delta^n}e^{-\frac{\delta^2}{2}}\mbox{ ,}
\end{equation}
 where the coefficients  $\lambda_n(\ell)$ are functions of the scale $\ell$. As previously shown, the symmetric part  (even terms) is well described by a Gaussian noise, which means that $\lambda_0(\ell)=1$ and 
$\lambda_{2n}(\ell)=0$ for $n\ge 1$. The coefficient $\lambda_1(\ell)$ is set to zero since, from Eq. \ref{eq:Edgeworth}, 
$\langle\delta_\ell u\rangle = -\sigma\langle\beta_\ell\rangle \lambda_1(\ell) = 0$. Under these hypotheses, the third order moment 
is then given by $\langle(\delta_\ell u)^3\rangle = -6\sigma^3\langle\beta_\ell^3\rangle \lambda_3(\ell)$, which fully 
determines the coefficient $\lambda_3(\ell)$ thanks to the Karman-Howarth-Kolmogorov equation (Eq. \ref{eq:KarmanHowarth}). Importantly, $\lambda_3(\ell)$ does depend on scale and Reynolds number. 
As Eq. \ref{eq:KarmanHowarth} is the only available constraint on $\lambda_n$,
it is tempting (as a first approximation) to restrict the
expansion to $\lambda_3$: $\lambda_{2n+1}(\ell) = 0$ for $n \ge 2$.
Additional statistical equations involving higher order odd moments of $\delta_\ell u$
would be needed to give the next $\lambda_{2n+1}(\ell)$. This would require
further modeling (primarily to get ride of pressure terms),
which is outside the scope of the present work. Unfortunately,
this crude approximation for the odd terms $\lambda_{2n+1}$ leads
to severe pathologies, such as negative probability for rare large events, and is not consistent with higher order statistics
such as hyperskewness $\langle (\delta_\ell u)^5\rangle/\langle (\delta_\ell u)^2\rangle^{5/2}$ (data not shown). To remedy for this weakness, Ref. \cite{CheCas06} proposed to modify the variance of the Gaussian associated to the third term in the development 
(Eq. \ref{eq:Edgeworth}), in the following way:
\begin{equation}\label{eq:PDeltaEdge}
\mathcal P_\delta(\delta) = \frac{1}{\sqrt{2\pi}}\left[ e^{-\delta^2/2}-\lambda_3(\ell)\delta(\delta^2-1)e^{-\delta^2/(2a^2)}\right]\mbox{ ,}
\end{equation}
where $\lambda_3$ is fully determined by the exact relation \ref{eq:KarmanHowarth}, and $a$ an add-hoc free parameter, close to 
unity \cite{CheCas06}, aimed at describing higher order odd statistics. Then, from Eqs. \ref{eq:betah}, 
\ref{eq:phmeneeul} and \ref{eq:PDeltaEdge}, the velocity increment PDF can be written, 
under the hypothesis of independence of $\delta$ and $h$ as (see Appendix \ref{ann:PDF}): 
\begin{equation}\label{eq:PredPDFEulEdgeworth}
 \mathcal P_{\delta_\ell u}(\delta_\ell u) = \int_{h_{\min}}^{h_{\max}}\frac{dh}{\sigma\beta_\ell(h)}\mathcal P_h^{(\ell)} (h)\mathcal P_\delta\left[\frac{\delta_\ell u}{\sigma\beta_\ell(h)}\right]\mbox{ .}
\end{equation}
Then, using a quadratic singularity spectrum (Eq. \ref{eq:DHLN}), $\mathcal R^*=52$ and the add-hoc coefficient $a=\sqrt{0.9}$ (see 
Ref. \cite{CheCas06}), the predicted PDF (Eq. \ref{eq:PDeltaEdge})  successfully compares to empirical data 
\cite{CheCas06} as shown in Fig. \ref{fig:PdfEulLag}(a) for various scales. The shape of the experimental velocity increment PDFs, from the inertial, to the intermediate dissipative and far dissipative ranges 
is well captured by the present theoretical prediction (Eq. \ref{eq:PDeltaEdge}), consistently with the behaviors of the flatness 
and skewness of the velocity increments. Unfortunately, at this stage, it is not possible to motivate the choice of the additional
free parameter $a$. To avoid having recourse to this parameter, we are forced to abandon the hypothesis of independence of the 
singularity exponent $h$ and the noise $\delta$. A formalism that takes into account possible correlations between these two random variables 
is presented in the following section.

 \begin{figure}[t]
\center{\epsfig{file=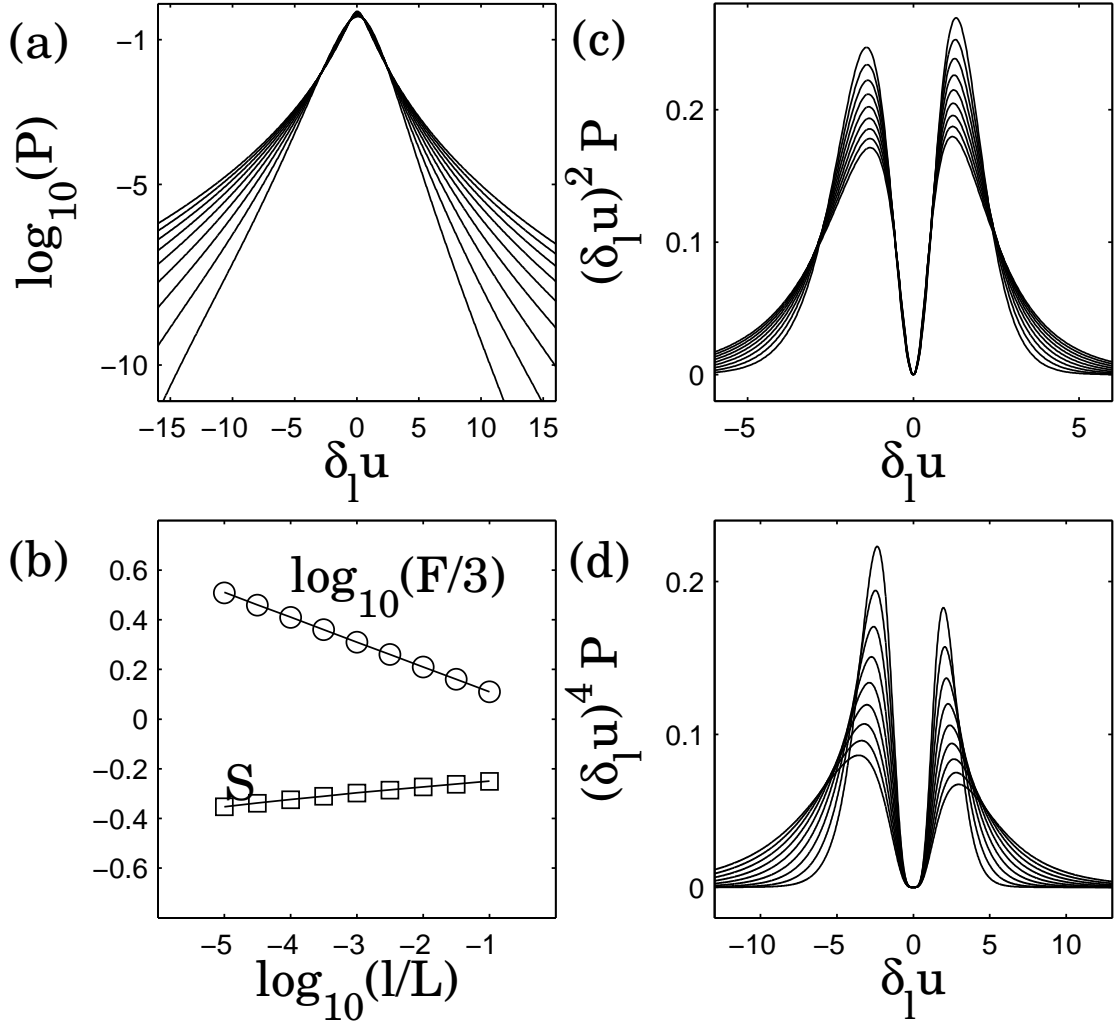,width=15cm}}
\caption{(a) Numerical estimation of the theoretical prediction of the velocity increment PDFs at various scales in the inertial 
range (Eq. \ref{eq:DistrDeltalu}). All PDFs are of unit variance and the scales used are (from bottom to top) $\ell/L = 10^{-1}, 10^{-1.5}, 
10^{-2}, 10^{-2.5}, 10^{-3}, 10^{-3.5}, 
10^{-4}, 10^{-4.5}, 10^{-5}$. (b) Numerical estimation of the velocity increments flatness $F(\ell)$ ($\circ$) and the skewness 
$S(\ell)$ ($\square$) 
from Eq. \ref{eq:DistrDeltalu}. Solid lines correspond to theoretical calulations (Eq. \ref{eq:SkewFlatJoint}). (c) and (d) 
Pre-multiplied PDFs $(\delta_\ell u)^2\mathcal P_{\delta_\ell u}(\delta_\ell u)$ and  
$(\delta_\ell u)^4\mathcal P_{\delta_\ell u}(\delta_\ell u)$, absissa are renormalized by $\sqrt{\langle (\delta_\ell u)^2\rangle}$ 
and ordinates are renormalized such that the integrals are unity. 
 } \label{fig:PdfAsymTheo}
\end{figure}

\subsubsection{Probabilistic modeling of the skewness in the inertial range}
\label{sec:stochmodel}

It remains to give a consistent description of the asymmetric part of velocity increment PDFs in both the inertial and dissipative 
ranges without invoking an additional free parameter $a$ in the PDF of the noise $\delta$ (Eq. \ref{eq:PDeltaEdge}). 
In this section, we propose such a formalism for the inertial range, and leave the extension to the dissipative range to 
future investigations. The main idea  is to correlate the exponent $h$ and the noise $\delta$   
in the multifractal description provided by Eq. \ref{eq:betah}.

Let us first assume that the velocity increment can be written as a product of a Gaussian noise $\delta$ and an amplitude 
$(\ell/L)^h$: $\delta_\ell u \build{=}_{}^{\mbox{law}} \sigma(\ell/L)^h \delta$, as in Eq. \ref{eq:betah}. The main 
difference with former assumptions is to let possible a correlation between $h$ and $\delta$. The simplest way to deal with such a 
probabilistic formalism is to assume $h$ and $\delta$ jointly Gaussian. In this case, see Appendix \ref{ann:PDF}, the velocity 
increment PDF $\mathcal P_{\delta_\ell u}(\delta_\ell u)$ can be written as
\begin{equation}\label{eq:DistrDeltalu}
 \mathcal P_{\delta_\ell u}(\delta_\ell u) = \int_{-\infty}^{\infty} \frac{1}{\sigma}\left(\frac{\ell}{L}\right)^{-h}
\mathcal P_{\delta,h}\left[ \frac{\delta_\ell u}{\sigma} \left(\frac{\ell}{L}\right)^{-h},h\right]dh\mbox{ ,}
\end{equation}
where $\mathcal P _{\delta,h}(\delta,h)$ is the joint probability of the random variables $\delta$ and $h$ given by:
\begin{equation}\label{eq:JointGaussianPdf}
 \mathcal P_{\delta,h} (\delta,h) = \frac{1}{2\pi\sigma_\delta\sigma_h\sqrt{1-\rho^2}}\exp\left\{ -\frac{1}{2(1-\rho^2)}
\left[\frac{(\delta-m_\delta)^2}{\sigma_\delta^2} + \frac{(h-m_h)^2}{\sigma_h^2}-\frac{2\rho(\delta-m_\delta)(h-m_h)}{\sigma_\delta\sigma_h}\right]\right\}\mbox{ .}
\end{equation}
In Eq. \ref{eq:JointGaussianPdf}, $m_\delta$ and $\sigma_\delta^2$ (resp. $m_h$ and $\sigma_h^2$) stand for the mean and variance 
of the random variable $\delta$ (resp. $h$). The correlation coefficient $\rho(\ell) = \frac{\langle h\delta \rangle}{\sigma_h\sigma_\delta}$
lies in the range $[-1,1]$. At large scale $\ell=L$ one has to recover $\sigma^2 = \langle (\delta_L u)^2\rangle$. This implies 
$\sigma_\delta=1$.

Consistently with Eqs. \ref{eq:DistrH} and \ref{eq:DHLN}, the multifractal formalism sets the mean and variance of the random variable $h$ to:
\begin{equation}
 m_h = \frac{1}{3} + \frac{3c_2}{2} \mbox{ and } \sigma_h^2 = \frac{c_2}{\ln(L/\ell)}\mbox{ ,}
\end{equation}
where $c_2$ remains the intermittency coefficient ($c_2=0.025$ in empirical data, c.f. Fig. \ref{fig:FlatEul}). Using the Legendre 
transform (Eq. \ref{eq:Legendre}), the $h$-distribution leads to a set of structure function exponent 
\begin{equation}
\zeta_p = m_hp-c_2p^2/2\mbox{ .}
\end{equation}
Using Eq. \ref{eq:DistrDeltalu}, we can show that
\begin{equation}
 \langle \delta_\ell u\rangle = \int_{-\infty}^{\infty}x\mathcal P_{\delta_\ell u}(x)dx = \sigma
\left(\frac{\ell}{L}\right)^{\zeta_1}\left[ m_\delta -\rho(\ell)\sqrt{c_2\ln(L/\ell)}\right]\mbox{ .}
\end{equation}
The velocity increment statistics are of zero-mean  if and only if
\begin{equation}
 m_\delta = \rho(\ell)\sqrt{c_2\ln(L/\ell)}\mbox{ .}
\end{equation}
For such parameters, we obtain the following velocity increment moments
\begin{align}\label{eq:MomentsTemp}
 \langle \delta_\ell u\rangle &= 0 \notag \\
 \langle (\delta_\ell u)^2\rangle &= \sigma^2\left(\frac{\ell}{L}\right)^{\zeta_2}\left[ 1+\rho^2{c_2\ln(L/\ell)}\right] \notag \\
 \langle (\delta_\ell u)^3\rangle &= -2\sigma^3\left(\frac{\ell}{L}\right)\rho\sqrt{c_2\ln(L/\ell)}\left[ 4\rho^2{c_2\ln(L/\ell)}+3\right] \mbox{ .}
\end{align}
The remaining free parameter is the correlation coefficient $\rho(\ell)$ that will be fully determine by the Karman-Howarth-Kolmogorov 
equation (Eq. \ref{eq:KarmanHowarth}) and the Kolmogorov constant $c_K$ (or equivalently by $\mathcal R^*$) . Indeed, neglecting dissipative effects, Eq. 
\ref{eq:KarmanHowarth} gives $\langle (\delta_\ell u)^3\rangle = -\frac{4}{5}\langle \epsilon \rangle \ell$. The average dissipation 
is given as $\langle \epsilon \rangle = \alpha \sigma^3/L$, where $\alpha=15/\mathcal R^*=0.2885$ (Eq. \ref{eq:ComputMultiEpsilonSimp}). Using the expression of 
$ \langle (\delta_\ell u)^3\rangle$ given in Eq. \ref{eq:MomentsTemp}, we obtain for $\rho(\ell)$, after solving a polynomial of third 
order, the following scale dependence
\begin{equation}\label{eq:CorrCoeff}
 \rho(\ell) = \frac{A(\alpha)}{\sqrt{c_2\ln(L/\ell)}}\mbox{ ,}
\end{equation}
with
\begin{equation}
A(\alpha) =\frac{1}{10}\left[ 50\alpha +25\sqrt{25+4\alpha^2}\right]^{1/3}-\frac{5}{2\left[ 50\alpha +25\sqrt{25+4\alpha^2}\right]^{1/3}}\mbox{ .} 
\end{equation}
Using the 
scale dependence of the correlation coefficient $\rho(\ell)$ (Eq. \ref{eq:CorrCoeff}), we can see that the proposed description is 
consistent as long as the coefficient $\rho(\ell)$ remains bounded in between -1 and 1. It corresponds to scales 
$\ell/L \le \exp(-A^2(\alpha)/c_2) = 0.94$, using $\alpha=15/\mathcal R^*$ and $c_2=0.025$. We finaly obtain for the Skewness and Flatness factors
\begin{align}\label{eq:SkewFlatJoint}
 S = \frac{\langle (\delta_\ell u)^3\rangle}{\langle (\delta_\ell u)^2\rangle^{3/2}} &= \left(\frac{\ell}{L}\right)^{\zeta_3-\frac{3}{2}\zeta_2}
\frac{2A(\alpha)\left[ 4A^2(\alpha)-3\right]}{\left[ 1+A^2(\alpha)\right]^{3/2}} \build{=}_{}^{\alpha=15/\mathcal R^*} -0.23\left(\frac{\ell}{L}\right)^{\zeta_3-\frac{3}{2}\zeta_2}\notag \\
F = \frac{\langle (\delta_\ell u)^4\rangle}{\langle (\delta_\ell u)^2\rangle^{2}} &= 3\left(\frac{\ell}{L}\right)^{\zeta_4-2\zeta_2}
\frac{ 27A^4(\alpha)+18A^2(\alpha)+1}{\left[ 1+A^2(\alpha)\right]^{2}} \build{=}_{}^{\alpha=15/\mathcal R^*}3.07 \left(\frac{\ell}{L}\right)^{\zeta_4-2\zeta_2} \mbox{ .}
\end{align}
We see that the proposed description (Eqs. \ref{eq:DistrDeltalu} and \ref{eq:JointGaussianPdf}), fully determined by 
the intermittent coefficient $c_2$, the Karman-Howarth-Kolmogorov equation (Eq. \ref{eq:KarmanHowarth}, setting $\nu=0$) and 
the Kolmogorov constant $c_K$ (or equivalently $\mathcal R^*$), gives consistent predictions (Eq. \ref{eq:SkewFlatJoint}). Indeed, if intermittency is 
neglected, $\zeta_3\approx \frac{3}{2}\zeta_2$, the skewness is constant $S\approx -0.23$ and compares well with empirical skewness 
in the inertial range \cite{CheCas06}. Let us also notice that the flatness does not tend to its Gaussian value $3$ when 
$\ell\rightarrow L$, but to $3.07$. This could be corrected if a large-scale cut-off is introduced in an ad-hoc way.

In Fig. \ref{fig:PdfAsymTheo} are shown the results of the numerical integration of Eqs. 
\ref{eq:DistrDeltalu} and \ref{eq:JointGaussianPdf}, using $c_2 = 0.025$ and $\alpha=15/\mathcal R^*$. In Fig. 
\ref{fig:PdfAsymTheo}(a) we recover the characteristic shape deformation of the velocity increments PDFs. In Fig. \ref{fig:PdfAsymTheo}(b), 
the velocity increments Skewness and Flatness are estimated and compared succesfully to the analytical calculations given in 
Eq. \ref{eq:SkewFlatJoint}. We also display in Fig. \ref{fig:PdfAsymTheo}(c) and (d) the second and fourth order 
pre-multiplied PDFs. We can see that the core of velocity increments PDF dominates in the second pre-multiplied PDFs, 
whereas the tails of the velocity increments PDFs contribute significantly to the fourth-order one.

\section{The Lagrangian framework}

Recently, several experimental technics \cite{OttMan00,PorVot01,VotPor02,MorMet01,MorDel02,MorDel03,CheRou03,MorCro04,XuBou06,BifBod08,BerOtt09}
and massive numerical computations \cite{YeuPop89,MorMet01,MorDel02,CheRou03,BifBof04,BifBod08,BenBif10}  have been developed aiming following fluid particles 
along their trajectory in a fully developed turbulent flow. As stated in the introduction, a phenomenology similar to the
Eulerian one can be developed in a Lagrangian context. The goal of this section is to present such a phenomenology 
introduced in Ref. \cite{CheRou03}, that has been compared to  a compilation of empirical data in \cite{CheRou03,ArnICTR08}.

\subsection{Probabilistic formalism of the inertial and dissipative ranges}
\subsubsection{The inertial range}

In the same spirit as in the section \ref{sec:EulInert} devoted to the Eulerian framework, a probabilistic formulation of the Lagrangian velocity 
time fluctuations can be written down \cite{CheRou03}. As observed experimentally, at the large integral time scale $T$, related to
the integral length scale  $L=\sigma T$, the statistics of both velocity increments and velocity are very close to Gaussian statistics. 
Indeed, in stationary, incompressible, homogeneous and isotropic turbulent flows, the Lagrangian velocity 
statistics can be related 
to their Eulerian counterparts \cite{TenLum72} (see also Section \ref{sec:ErgoEulLag}). The statistics of the velocity time 
increments at a time scale $\tau$ (Eq. 
\ref{eq:VelTimIncr}) can be modeled once again as the product of two independent random variables:
\begin{equation}\label{eq:StochVarLag}
 \delta_\tau v \build{=}_{}^{\mbox{law}}\sigma\left( \frac{\tau}{T}\right)^h \delta\mbox{ ,}
\end{equation}
where $\delta$ is a zero-mean unit variance Gaussian random variable and $\sigma^2 = \langle (\delta_T v)^2\rangle= 
\langle (\delta_L u)^2\rangle=2 \langle u^2\rangle$. The exponent $h$ is independent on $\delta$ and fluctuates according to the distribution law:
\begin{equation}\label{eq:DistrHLag}
 \mathcal P_h^{(\tau)}(h)= \frac{\left( \frac{\tau}{T}\right)^{1-\mathcal D^L(h)}}{\int_{h_{\min}}^{h_{\max}}\left( \frac{\tau}{T}\right)^{1-\mathcal D^L(h)}dh}\mbox{ ,}
\end{equation}
where, according to the multifractal formalism assumptions, the Lagrangian singularity spectrum $\mathcal D^L(h)$ 
is universal (i.e. Reynolds number independent) and independent on 
the time scale $\tau$. As with Eqs. \ref{eq:StochVar}, \ref{eq:DistrH} and \ref{eq:Legendre}, this probabilistic modeling 
(Eqs. \ref{eq:StochVarLag} and \ref{eq:DistrHLag}) is consistent with a power-law behavior of the Lagrangian structure 
functions $\langle |\delta_\tau v|^p\rangle \sim \tau^{\xi_p}$, the  exponents $\xi_p$ are related 
to the Lagrangian singularity spectrum $\mathcal D^L(h)$ via a Legendre transform:
\begin{equation}\label{eq:SteepLag}
 \xi_p = \build{\min}_{h}^{} \left[ ph+1-\mathcal D^L(h)\right]\mbox{ .}
\end{equation}
Dimensional analysis (Eqs. \ref{eq:DimAnalM2Lag} and \ref{eq:DimAnalSpecLag}) leads to a K41 description of Lagrangian turbulence neglecting
intermittency. In this case, $\mathcal D^L(1/2) = 1$ and $\mathcal D^L(h) = -\infty$ for $h\ne 1/2$. 
Experimental and numerical data actually revealed the presence of intermittency \cite{MorMet01,CheRou03,YeuPop89,BifBof04,ArnICTR08} that can be characterized by a  quadratic  singularity spectrum \cite{CheRou03}:
\begin{equation}\label{eq:DHLNLag}
 \mathcal D^L(h) = 1-\frac{(h-c_1^L)^2}{2c_2^L} \mbox{ , with }c^L_1=\frac{1}{2}+c^L_2 \mbox{ and }c^L_2=0.085\mbox{ .}
\end{equation}
Note that the obtained value of $c_1$ is consistent with empirical data and with the dimensional prediction $\xi_2=1$. As in the Eulerian framework, 
the bounds of the integration domain in Eq. \ref{eq:DistrHLag} are $h_{\min}=0$ and $h_{\max}=1$ (see Section \ref{sec:CommentBounds}). 
We will see in the following (Section \ref{sec:StatAcce})
that the quadratic approximation (Eq. \ref{eq:DHLNLag}) is too crude to describe the statistics of the acceleration 
 and is not compatible with a quadratic Eulerian singularity spectrum via the 
Borgas' transformation (see section \ref{sec:Borgas}). Nevertheless, it gives a consistent description of the Lagrangian velocity statistics in the inertial range as well as in the intermediate and far-dissipative range \cite{CheRou03}.

\subsubsection{The dissipative range}

As previously reported, the quadratic Eulerian spectrum (Eq. \ref{eq:DHLN}) has been extensively compared to a large set of empirical data,  
some of them displaying a large inertial range $[\eta_K,L]$ (in particular the Modane wind tunnel data, $\mathcal R_\lambda=2500$). 
Sophisticated signal 
analysis procedures \cite{WenAbr07,DelMuz01} concluded that this quadratic approximation cannot be distinguished from the data and this for 
both the increasing part  (positive order structure functions) and the decreasing part (negative order 
structure functions) of the $\mathcal D^E(h)$ spectrum. In the Lagrangian framework, such an analysis is much more difficult because the experimental technologies 
(silicon strip detectors \cite{PorVot01}, acoustic scatering \cite{MorMet01}, rapid cameras \cite{XuBou06,BerOtt09}, etc.) are not 
as efficient as a well known hot-wire probe under the Taylor's hypothesis \cite{Fri95}. More fundamentaly, at a same Reynolds 
number, the 
width of the inertial range is much greater in the Eulerian framework than in the Lagrangian counterpart. Indeed, dimensional analysis predicts a Kolmogorov dissipative scale $\tau_{\eta_K}$ proportional to $L\mathcal R_e^{-1/2}$, and in turn
$\sigma\tau_{\eta_K}/\eta_K\sim \mathcal R_e^{1/4}\gg 1$. 

Let us now adapt the arguments justifying a fluctuating dissipative 
length scale in Ref. \cite{PalVul87} to the dissipative time scales \cite{Bor93}. A local Reynolds number 
can be defined as $\mathcal R_\tau = v_\tau^2\tau/\nu$, where $v_\tau$ is a 
characteristic fluctuating velocity at the time scale $\tau$. From  Eq. \ref{eq:StochVarLag}, we obtain:
\begin{equation}\label{eq:loclagReynolds}
 \mathcal R_\tau = \left( \frac{\tau}{T}\right)^{2h+1}\mathcal R_e\mbox{ .}
\end{equation}
Reproducing the argumentation developed in Refs. \cite{PalVul87,Bor93}, a h-dependent dissipative time scale can be defined as the scale such that the local Reynolds number 
(Eq. \ref{eq:loclagReynolds}) is of order unity:  $\mathcal R_{\tau_\eta} = \mathcal R^\dag = O(1)$. A priori, the universal constant 
$ \mathcal R^\dag$ is different from the corresponding Eulerian one $\mathcal R^*$. We thus obtain \cite{Bor93},
\begin{equation}\label{eq:TauEtaH}
 \tau_\eta(h) = T\left( \frac{\mathcal R_e}{\mathcal R^\dag}\right)^{-\frac{1}{2h+1}}\mbox{ .}
\end{equation}
If we neglect the intermittency corrections (in a K41 framework), then the exponent $h=1/2$ is unique, and we recover the Kolmogorov's 
dimensional prediction
\begin{equation}\label{eq:TauEtaKK41}
 \tau_{\eta_K} = T\left( \frac{\mathcal R_e}{\mathcal R^\dag}\right)^{-\frac{1}{2}}\mbox{ .}
\end{equation}
 For a time scale $\tau$ smaller than the dissipative time scale $\tau_\eta(h)$, the velocity time increment can be Taylor expanded, 
i.e. $\delta_\tau v = \tau a +o(\tau^2)$, $a$ being the acceleration. Like for the Eulerian velocity increments 
(Eq. \ref{eq:StochVardiss}), we obtain the following stochastic modeling of the velocity time increments in the far dissipative range:
\begin{equation}\label{eq:StochVarLagdiss}
 \delta_\tau v \build{=}_{\tau\le\tau_\eta (h)}^{\mbox{law}}\sigma\frac{\tau}{T}\left( \frac{\tau_{\eta}(h)}{T}\right)^{h-1} \delta\mbox{ .}
\end{equation}
In analogy with the Eulerian framework (Eq. \ref{eq:DistrHdiss}) \cite{Nel90,Bor93}, the distribution of the 
$h$-exponents does not depend on the scale $\tau$ (up to the normalizing function $\mathcal Z(\tau)$) and is a function of the Reynolds number:
\begin{equation}\label{eq:DistrHlagdiss}
 \mathcal P_h^{(\tau)}(h) \build{=}_{\tau\le\tau_\eta (h)}^{}\frac{1}{\mathcal Z(\tau)} \left( \frac{\mathcal R_e}{\mathcal R^\dag}\right)^{-\frac{1-\mathcal D^L(h)}{2h+1}}\mbox{ .}
\end{equation}
From Eqs. \ref{eq:TauEtaH}, \ref{eq:StochVarLagdiss} and \ref{eq:DistrHlagdiss}, we can derive the expression of the high order 
moments of the acceleration, in the same way we derived the prediction of the moments of the velocity gradients in the Eulerian framework (Eq. \ref{eq:predgradeul}): 
\begin{equation}\label{eq:HighMomAcce}
 \langle a^{2p}\rangle = \langle\delta^{2p}\rangle\left(\frac{\sigma}{T}\right)^{2p}\frac{1}{\mathcal Z(0)}\int_{h_{\min}}^{h_{\max}} \left(\frac{\mathcal R_e}{\mathcal R^\dag}\right)
^{-\frac{2p(h-1)+1-\mathcal D^L(h)}{2h+1}}dh \mbox{ ,}
\end{equation}
where $\mathcal Z(0) =\int_{h_{\min}}^{h_{\max}} \left(\frac{\mathcal R_e}{\mathcal R^\dag}\right)
^{-\frac{1-\mathcal D^L(h)}{2h+1}}dh $ and $\langle\delta^{2p}\rangle = \frac{(2p)!}{p!2^p}$. Let us stress that the acceleration odd-order moments are predicted to vanish:  
$\langle a^{2p+1}\rangle = 0$ as observed in data. Finally, acceleration PDF can be expressed as a function of the singularity spectrum $\mathcal D^L(h)$ and the constant $\mathcal R^\dag$:
\begin{equation}\label{eq:PredMultiAcceleration}
\mathcal P_a(a) = \frac{T}{\sigma}\frac{1}{\mathcal Z(0)}\int_{h_{\min}}^{h_{\max}} \left(\frac{\mathcal R_e}{\mathcal R^\dag}\right)
^{\frac{h-1}{2h+1}-\frac{1-\mathcal D^L(h)}{2h+1}}\mathcal P_\delta\left[ \frac{aT}{\sigma}\left(\frac{\mathcal R_e}{\mathcal R^\dag}\right)
^{\frac{h-1}{2h+1}}\right]dh\mbox{ ,}
\end{equation}
where $\mathcal P_\delta(x) =e^{-x^2/2}/\sqrt{2\pi}$. A detailed 
discussion of the Reynolds number dependence of acceleration variance and flatness is provided in section 
\ref{sec:StatAcce}.

\subsubsection{Full multi scale description}
\label{sec:fulllagdescrip}

As in section \ref{sec:fulleuldescrip}, we need an interpolation formula linking the velocity increments statistics 
in the inertial (Eqs. \ref{eq:StochVarLag} and \ref{eq:DistrHLag}) and dissipative 
(Eqs. \ref{eq:StochVarLagdiss} and \ref{eq:DistrHlagdiss}) ranges. An adapted 
Batchelor-Meneveau form inspired from the Eulerian framework (Eqs. \ref{eq:betah} and \ref{eq:phmeneeul}) was proposed in Ref. \cite{CheRou03}:
\begin{equation}\label{eq:betahlag}
 \delta_\tau v \build{=}_{}^{\mbox{law}} \sigma \beta_\tau \delta \: \mbox{ with } \: \beta_\tau  =
\frac{\left(\frac{\tau}{T}\right)^h}{\left[
1+\left(\frac{\tau}{\tau_{\eta}(h)}\right)^{-\gamma}\right]^{(1-h)/\gamma}} \mbox{
,}
\end{equation}
where again the random variable $\delta$ is a Gaussian noise of zero-mean and unit-variance, and
\begin{equation}\label{eq:phmenelag}
\mathcal P_h^{(\tau)}(h)
=\frac{1}{\mathcal Z(\tau)}
\frac{\left(\frac{\tau}{T}\right)^{1-\mathcal D^L(h)}}{\left[
1+\left(\frac{\tau}{\tau_{\eta}(h)}\right)^{-\gamma}\right]^{(\mathcal
D^L(h)-1)/\gamma}}\mbox{ ,}
\end{equation}
with
\begin{equation}
\mathcal Z(\tau)
=\int_{h_{\min}}^{h_{\max}} 
\frac{\left(\frac{\tau}{T}\right)^{1-\mathcal D^L(h)}}{\left[
1+\left(\frac{\tau}{\tau_{\eta}(h)}\right)^{-\gamma}\right]^{(\mathcal
D^L(h)-1)/\gamma}}\mbox{ .}
\end{equation}

As in Eq. \ref{eq:phmeneeul}, the normalizing function $\mathcal Z(\tau)$ is determined by imposing 
$\int_{h_{\min}}^{h_{\max}}\mathcal P_h^{(\tau)}(h)dh=1$. The free parameter $\gamma$ entering in 
Eqs. \ref{eq:betahlag} and \ref{eq:phmenelag} actually controls the transition from inertial to dissipative physics: the bigger 
$\gamma$ is, the steeper the transition. In the sequel, we will use the Batchelor value $\gamma=2$, as in the Eulerian 
framework. Let us point out that we have found $\gamma<2$ when dealing with experimental signals and focussing on the second order 
log-cumulant \cite{CheRou03} whereas the value $\gamma=4$ was used in Refs. \cite{BenBif10,ArnICTR08} to describe the 
logarithmic local slope. Following Appendix \ref{ann:PDF}, the velocity increment PDF then reads
(Eqs. \ref{eq:betahlag} and \ref{eq:phmenelag}):
\begin{equation}\label{eq:VIPDFLag}
 \mathcal P_{\delta_\tau v}(\delta_\tau v) = \int_{h_{\min}}^{h_{\max}} \frac{dh}{\sigma \beta_\tau (h)}\mathcal P_h^{(\tau)}(h) \mathcal P_\delta \left[ \frac{\delta_\tau v}{\sigma\beta_\tau}\right]\mbox{ .}
\end{equation}

We show in Fig. \ref{fig:PdfEulLag}(b) the numerical estimation of the Lagrangian velocity increment PDF (Eq. \ref{eq:VIPDFLag}) at different  scales using 
a quadratic singularity spectrum (Eq. \ref{eq:DHLNLag})  and for two Reynolds numbers corresponding to the two sets of experiments ($\mathcal R_\lambda=740$ for the acoustic scattering measurements at the ENS 
Lyon \cite{MorMet01} and $\mathcal R_\lambda = 690$ \cite{MorCro04} for the Cornell's sillicon strip detectors). We can 
see that the multifractal model predictions for the PDFs compared well to both sets of experimental velocity measurements \cite{CheRou03}. 
We have used $c_2^L = 0.0753$, $c_1^L=1/2+c_2^L$, $\gamma=1.08$ and $\mathcal R^\dag = 30$ to describe ENS Lyon PDFs. For the 
Cornell acceleration data, we have used Eq. \ref{eq:VIPDFLag} with $c_2^L = 0.079$, $c_1^L=1/2+c_2^L$, $\gamma=1.3$ and $\mathcal R^\dag = 30$. As far as ENS Lyon data are considered, to obtain these free parameters, we have fitted the second order cumulant, i.e. the variance of $\ln|\delta_\tau v|$, over the whole range of scales (inertial and dissipative), $\mathcal R_e$, $\mathcal R^\dag$ and $T$ assumed known, and defining $c_2^L$ and $\gamma$ as the  minimizers of the quadratic error of the theoretical and empirical second order cumulant \cite{CheRou03}. We have further shown \cite{CheRou03} that a similar fitting procedure 
on DNS Data ($\mathcal R_\lambda=140$) leads to $c_2^L = 0.086$ and $\gamma=1.98$, a value of the transition parameter $\gamma$ closer 
to its Eulerian counterpart (i.e. $\gamma=2$, Eqs. \ref{eq:betah} and \ref{eq:phmeneeul}). The low value of the parameter $\gamma$ 
found in experiments can be interpreted as resulting from a low-pass filtering induced by the finitesize of the particule tracers.

As for Cornell's data, we did not use the true acceleration PDF (Eq. \ref{eq:PredMultiAcceleration}) because the measured flatness is much lower than (i)
the one predicted by the present multifractal formalism and (ii) what is obtained in DNS (see the discussion in section \ref{sec:StatAcce}). 
We chose  in Ref. \cite{CheRou03} to consider the Cornell's PDF shown in Fig. \ref{fig:PdfEulLag}(b) as a velocity increment 
PDF at a small time scale $\tau = 0.0029T$, corresponding to $\tau = \tau_{\eta_K}/8.62$ in units of the Kolmogorov's dissipative time scale 
(Eq. \ref{eq:TauEtaKK41}). Then, the respective intermittency coefficient $c_2^L$ and $\gamma$ are extracted from Cornell's acceleration data while minimizing the quadratic error of the theoretical and experimental PDF.

\begin{figure}[t]
\center{\epsfig{file=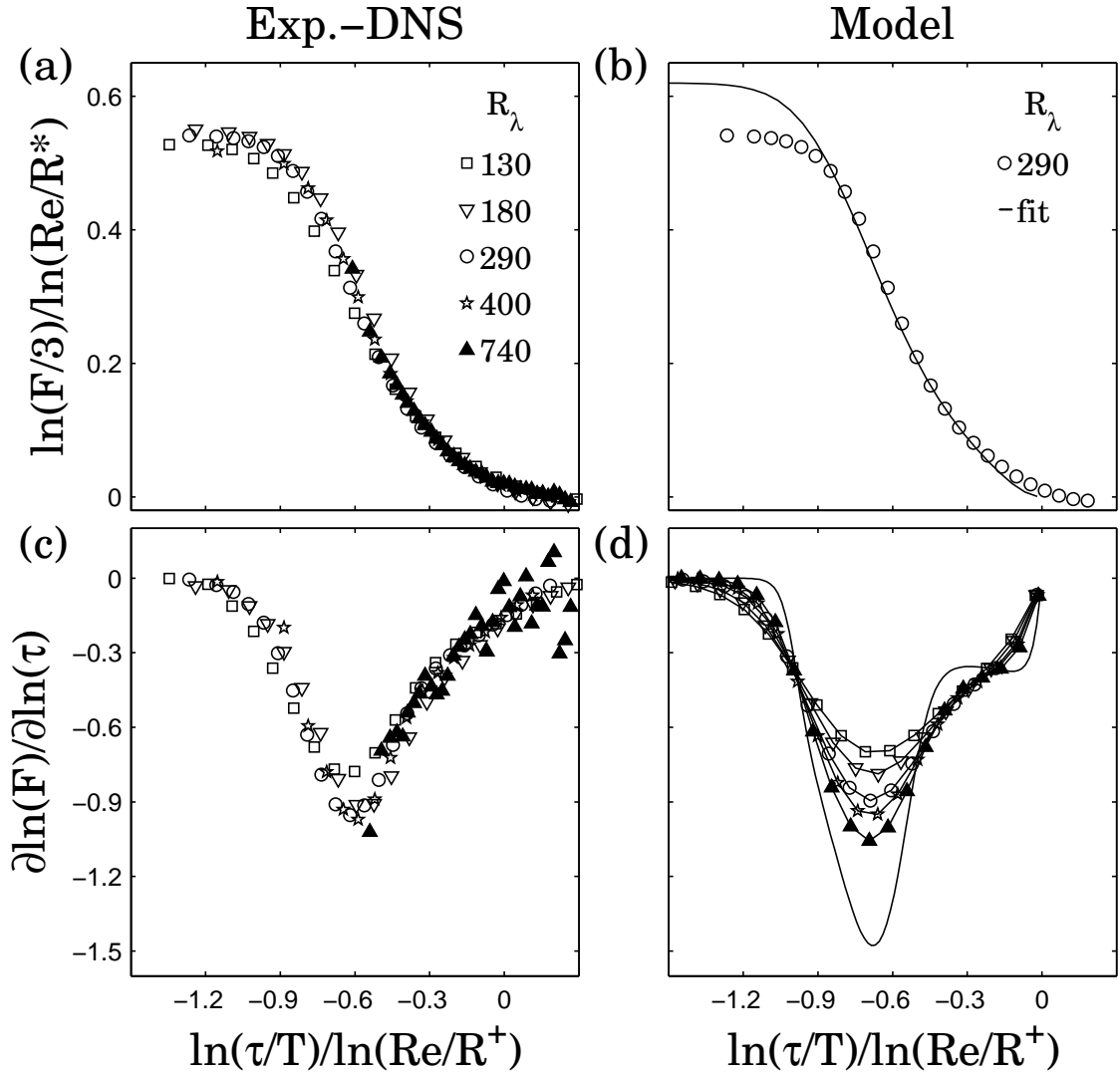,width=15cm}}
\caption{(a) Flatness of Lagrangian velocity time increments as a function of the normalized scales $\tau/T$. Lower Reynolds numbers
correspond to numerical data ($\mathcal R_\lambda = 130$, $180$, $290$ from E. L\'ev\^eque \cite{MorDel03,CheRou03} and 
$\mathcal R_\lambda=400$ from the Roma group \cite{BifBof04,BenBif10}). The highest Reynolds number $\mathcal R_\lambda=740$ has 
been achieved experimentally at the ENS Lyon \cite{MorMet01}. Time scales are renormalized by the integral time scale $T$ and we  
use $\mathcal R^\dag=30$. (b) Comparison of the proposed predictions (using Eqs. \ref{eq:betahlag} and \ref{eq:phmenelag}), using 
a quadratic singularity spectrum (Eq. \ref{eq:DHLNLag}) with 
$c_2=0.085$, $c_1=1/2+c_2$, $\gamma=2$ and $\mathcal R^\dag=30$ (solid line) to the numerical data at $\mathcal R_\lambda = 290$. (c) Logarithmic local slope of the Flatness of the data displayed in Fig. \ref{fig:FlatLagFit}(a). (d) Theoretical logarithmic local slope of the Flatness obtained from a numerical integration of Eqs. \ref{eq:betahlag} and \ref{eq:phmenelag}, using 
a quadratic singularity spectrum (Eq. \ref{eq:DHLNLag}) with 
$c_2=0.085$, $c_1=1/2+c_2$, $\gamma=2$ and $\mathcal R^\dag=30$, and for the various Reynolds numbers as shown in Fig. \ref{fig:FlatLagFit}(a). We furthermore display (solid line, without symbols) a theoretical logarithmic local slope at a very high Reynolds number ($\mathcal R_e = 10^{10}$). } 
\label{fig:FlatLagFit}
\end{figure}

In Fig. \ref{fig:FlatLagFit}(a) is shown the behavior of the flatness of Lagrangian velocity time increments as a function of the 
scale $\tau$, for various Reynolds numbers (from $R_\lambda=130$ to $\mathcal R_\lambda=740$) and flow configurations 
(DNS and experimental von Karman flows), as previously done in the Eulerian framework (Fig. \ref{fig:FlatEul}). 
As expected for scales $\tau$ greater than the integral time scale $T$, the statistics of velocity increments are close to Gaussianity. This can be deduced from  the Gaussianity of Eulerian velocity increments for $\ell\ge L$,  using an ergodicity argument (see section \ref{sec:Borgas}). 

In the inertial range, 
for time scales in the range $-1/2 \le \ln(\tau/T)/\ln(\mathcal R_e/\mathcal R^\dag) \le 0$, an universal Reynolds-number independent behavior is observed. Recall that in the Lagrangian framework, at a given Reynolds number, the width of the inertial range $T/\tau_{\eta_K}\sim \mathcal R_e^{1/2}$ is expected smaller than its Eulerian counterpart $L/\eta_K\sim \mathcal R_e^{3/4}$. Therefore, we may think that observing clear power-laws for Lagrangian velocity fluctuations asks for higher Reynolds number than in the Eulerian framework. Indeed, for the range of Reynolds number available, no clear power laws are observed. This is confirmed when we display in Fig. \ref{fig:FlatLagFit}(c) the local logarithmic slope where no plateau is observed in the inertial range at any Reynolds number. Thus, as far as Flatness is concerned, no clear power laws are observed. 

Scales  $\ln(\tau/T)/\ln(\mathcal R_e/\mathcal R^\dag) \le -1/2$ correspond to the intermediate dissipative scales. 
Note that the extension of the intermediate dissipative range in the Lagrangian framework is much wider than in the Eulerian 
framework. This is related to the fact that, as we will see, Lagrangian velocity is more intermittent than the Eulerian one, implying stronger 
fluctuations of the local dissipative time scale $\tau_{\eta}(h)$ defined in Eq. \ref{eq:TauEtaH}. In this range of scales, we can 
also notice a Reynolds number dependence, that could be interpreted, as previously done in the Eulerian framework (Section \ref{subsub:dissipative}), as a 
direct consequence of the fluctuating nature of the dissipative time scale. When the scale $\tau$ tends to zero, we 
observe a saturation of the velocity increments flatness to the corresponding acceleration flatness. This 
representation shows clearly the Reynolds number dependence of the acceleration flatness that will be further discussed in 
Section \ref{sec:StatAcce}.

As we see, estimating the intermittency coefficient is difficult since, contrary to the Eulerian framework, no strict power laws are observed. Nevertheless, using Eqs. \ref{eq:betahlag} and \ref{eq:phmenelag}, we can predict the behavior of Flatness over the entire range of scales, not only the inertial range. In Fig. \ref{fig:FlatLagFit}(b), we compare the DNS data ($\mathcal R_\lambda=290$) against the proposed formalism 
(Eqs. \ref{eq:betahlag} and \ref{eq:phmenelag}), using a quadratic singularity spectrum $\mathcal D^L(h)$, an intermittency 
coefficient $c_2^L=0.085$ (and $c_1^L = 1/2+c_2^L$), the Batchelor value $\gamma=2$ for the transition and $\mathcal R^\dag=30$.  Thus, we find that Lagrangian turbulence is more intermittent than its Eulerian counterpart.
The present formalism reproduces quantitatively the behavior of the flatness in the inertial and intermediate dissipative ranges. In the far-dissipative range (when $\tau\rightarrow 0$), this formalism seems to overpredict the value of the 
acceleration flatness (see the discussion in section \ref{sec:StatAcce}). Given this limitation, the comparison 
between theory and empirical data is very satisfactory.

We can see that, even if no clear power laws are observed, the present formalism gives a realistic picture of the Flatness at any scale. We can also see that, even if the multifractal formalism assumes the existence of power-laws in the asymptotic limit of very high Reynolds numbers, the predicted Flatness does not exhibit a clear power law at the finite Reynolds numbers under investigation. This can be clearly seen in  Fig. \ref{fig:FlatLagFit}(d) where the predicted logarithmic local slopes of Flatness, for the various Reynolds numbers given in Fig. \ref{fig:FlatLagFit}(a), are shown. Theoretically speaking, the fact that no power laws are obtained in the model is mostly related to the wide extension of the dissipative range implied by the strong level of intermittency that prevents from getting an extended inertial range. Another reason that explains why no clear inertial range power-laws are obtained in the model is that, for the largest scales of the inertial range for which the ratio $\tau/T$ cannot be considered as small, the steepest-descent calculation (Eq. \ref{eq:SteepLag}), that predicts power-laws, is not a good approximation of Eqs. \ref{eq:betahlag} and \ref{eq:phmenelag}. To this regard, studying the behavior of structure functions in a relative way \cite{BifBof04,BifBod08,ArnICTR08}, in the spirit of the extended self similarity, allows to weaken large-scale anisotropic effects and shows more clearly power-law behaviors. Nevertheless, when working at a very high Reynolds number, the model indeed exhibits a clear power-law for the Flatness. This can been seen in Fig. \ref{fig:FlatLagFit}(d) where we superimpose (solid line with no symbols), as an illustration, the predicted logarithmic local slope of Flatness for $\mathcal R_e=10^{10}$. We see indeed the presence of a plateau in the inertial range.

\subsection{The Borgas' argument: linking Eulerian and Lagrangian intermittencies}
\label{sec:Borgas}

This section is devoted to establish a link between the Eulerian singularity spectrum $\mathcal D^E(h)$ and its Lagrangian 
counterpart $\mathcal D^L(h)$ \cite{CheRou03}. We will mainly recall the work of Borgas \cite{Bor93} and invite the reader to have a look at this 
reference for a detailed derivation. An alternative dimensional derivation of an equivalent relationship has been also proposed in the literature \cite{BofLil02}. 
Let us 
also mention more general arguments developed on kinematic bases \cite{HomKam09,KamFri09}.

\subsubsection{Ergodicity principle}
\label{sec:ErgoEulLag}
Establishing a relationship between Eulerian and Lagrangian fluctuations requires as basic statement some principle of 
ergodicity. In simple words, we will admit that in an isotropic, homogeneous, incompressible and stationnary flow, the Eulerian 
average of a physical variable, obtained from summing up its realizations over space 
is equal to its Lagrangian average, obtained from summing up its values along the trajectory of a particle. This was first recognized 
and formalized by Tennekes and Lumley \cite{TenLum72}.

More formally, let us consider a one-point physical variable $\mathcal F$ (e.g. velocity, dissipation, pressure and its derivatives, etc.) 
that does not depend on the scale. In the Eulerian 
framework, this variable depends on the spatial coordinates and time $\mathcal F^E (x,y,z,t)$. 
In a Lagrangian description of the flow, this variable can be written as a function  of the initial positions 
of the particles and time $\mathcal F^L (x_0,y_0,z_0,t)$. The assumption of incompressibility allows the following relationship 
\begin{equation}\label{eq:TenLumErgo}
 \lim_{V\rightarrow +\infty}\frac{1}{V}\iiint_V \mathcal F^E(x,y,z,t)dxdydz =  \lim_{V\rightarrow +\infty}\frac{1}{V}\iiint_V \mathcal F^L(x_0,y_0,z_0,t)dx_0dy_0dz_0\mbox{ ,}
\end{equation}
that expresses the fact that an incompressible fluid \textit{continues to fill the box as it moves around} \cite{TenLum72}. The 
next step requires   the assumptions of isotropy and homogenity. When ensemble averaging  the equality, this expectation can be taken inside the integrals by linearity. Then, homogeneity and isotropy allow us to remove 
these expectations from these integrals because of the independence over the space. Thus we are left with 
\begin{equation}
\langle \mathcal F^E(x,y,z,t)\rangle =\langle \mathcal F^L(x_0,y_0,z_0,t)\rangle\mbox{ .}
\end{equation} 

The particular case of the observables 
$\mathcal F^E(x,y,z,t) = \exp[i\vec{k}.\vec{u}(x,y,z,t)]$  and $\mathcal F^L(x_0,y_0,z_0,t) = \exp[i\vec{k}.\vec{v}(x_0,y_0,z_0,t)]$ was treated in  Ref.  \cite{TenLum72}. The 
expectation of these variables are the characteristic function the 
Eulerian $\vec{u}$ and Lagrangian $\vec{v}$ velocities. The equality of these characteristic functions implies the equality of the distribution and of all the moments of each 
velocity components, namely $\forall q$, $\langle u_i^q\rangle = \langle v_i^q\rangle$. As observed in data, 
the Eulerian velocity has statistics close to Gaussian, which implies from the ergodicity principle the Gaussianity of the Lagrangian velocity. This 
has been checked in experiments as well as in simulations (see the Gaussian values of the Eulerian and Lagrangian velocity increments flatness
shown in respectively Figs. \ref{fig:FlatEul} and \ref{fig:FlatLagFit}).

The idea of Borgas was to apply this ergodicity principle to the (scalar) observable dissipation $\epsilon$, i.e. 
$\mathcal F^E(x,y,z,t) = \exp[ik\epsilon(x,y,z,t)]$ and $\mathcal F^L(x_0,y_0,z_0,t) = \exp[ik\epsilon(x_0,y_0,z_0,t)]$, with as a main outcome the equality of the moments of the Eulerian dissipation and of the moments of the dissipation as seen by the particle along 
its trajectory. The next subsection 
is devoted to recall the Eulerian multifractal predictions for the moments of dissipation, and to extend these predictions to the Lagrangian 
framework \cite{Bor93}.

\subsubsection{Multifractal description of dissipation fluctuations}

For the sake of completeness, we repeat here the arguments developed in  Ref. \cite{Bor93}. Multifractal predictions have been 
historically developed \cite{Fri95,Obo62,Kol62,Meneveauetal90,MenSre91,SreMen88} for the coarse-grained dissipation over a ball of size $\ell$ centered on the position 
$\vec{r}$: 
\begin{equation}
 \epsilon_\ell (\vec{r},t) = \frac{1}{\frac{4}{3}\pi\ell^3}\int_{|\vec{r}-\vec{r}'|\le \ell} \epsilon(\vec{r}',t)d\vec{r}'\mbox{ .}
\end{equation}
For a scale $\ell$ lying in the inertial range, a formalism similar to the one developed for the velocity increment can be written 
(Eqs. \ref{eq:StochVar} and \ref{eq:DistrH}) using an exponent $\alpha$ and a singularity spectrum $f^E(\alpha)$. Using the notations of Ref. \cite{Bor93}, we can write 
$\epsilon_\ell = \langle \epsilon \rangle (\ell/L)^{\alpha-1}$ (equality in probability law), and the probability to get an exponent $\alpha$ at scale $\ell$ being given by $\mathcal P_\alpha^{(\ell)}(\alpha) \sim (\ell/L)^{1-f^E(\alpha)}$. When the 
scale $\ell$ enters the dissipative range, we must take into account the fluctuating nature of the dissipative scale 
(Eq. \ref{eq:etah}) parametrized by the exponent $\alpha$, i.e. $\eta(\alpha) \sim \mathcal R_e^{-3/(3+\alpha)}$. The choice of Borgas to apply the ergodicity principle to the dissipation
was influenced by the fact that, in the limit of vanishing scales, the coarse-grained dissipation leads to predictions for the 
point wise moments of dissipation. We get, neglecting multiplicative (Reynolds number independent, but $q$ dependent) constants,
\begin{equation}\label{eq:PrediMomDissEul}
 \lim_{\ell\rightarrow 0}\langle \epsilon_\ell^q\rangle = \langle \epsilon^q\rangle \sim \langle \epsilon\rangle^q\int_{\alpha_{\min}}^{\alpha_{\max}}
\mathcal R_e ^{-3\frac{q(\alpha-1)+1-f^E(\alpha)}{\alpha+3}}d\alpha\sim \langle \epsilon\rangle^q \mathcal R_e ^{-\min_\alpha \left[3\frac{q(\alpha-1)+1-f^E(\alpha)}{\alpha+3}\right]}\mbox{ .}
\end{equation}
In a Lagrangian formulation, we will consider the average of dissipation during a time scale $\tau$:
\begin{equation}
 \epsilon_\tau (\vec{r}_0,t) = \frac{1}{\tau}\int_{|t-t'|\le \tau} \epsilon(\vec{r}_0,t')dt'\mbox{ .}
\end{equation}
In a similar fashion as developed for the Eulerian description, the multifractal description of the statistical properties of $\epsilon_\tau$ can be written down using the exponent $\kappa$ and 
the respective Lagrangian singularity spectrum $f^L(\kappa)$ \cite{Bor93}, i.e. $\epsilon_\tau = \langle \epsilon \rangle (\tau/T)^{\kappa-1}$ (equality in probability law) and the probability to get an exponent $\kappa$ at scale $\tau$ being given by $\mathcal P_\kappa^{(\tau)}(\kappa) \sim (\tau/T)^{1-f^L(\kappa)}$. When taking into account the fluctuating nature 
of the dissipative time scale $\tau_{\eta_K}(\kappa) \sim \mathcal R_e^{-1/(1+\kappa)}$, we get
\begin{equation}\label{eq:PrediMomDissLag}
 \lim_{\tau\rightarrow 0}\langle \epsilon_\tau^q\rangle = \langle \epsilon^q\rangle \sim \langle \epsilon\rangle^q\int_{\kappa_{\min}}^{\kappa_{\max}}
\mathcal R_e ^{-\frac{q(\kappa-1)+1-f^L(\kappa)}{\kappa+1}}d\kappa\sim \langle \epsilon\rangle^q \mathcal R_e ^{-\min_\kappa \left[\frac{q(\kappa-1)+1-f^L(\kappa)}{\kappa+1}\right]}\mbox{ .}
\end{equation}
Identifying the leading order Reynolds number power laws exponents entering in the Eulerian and Lagrangian predictions of the pointwise moments of dissipation 
(Eqs. \ref{eq:PrediMomDissEul} and \ref{eq:PrediMomDissLag}), we get a relationship between the Eulerian $f^E(\alpha)$ and 
Lagrangian $f^L(\kappa)$ singularity spectra:
\begin{equation}\label{eq:PredBorgasf}
 f^L(\kappa) = -\frac{1}{2}\kappa +\left( 1 + \frac{1}{2}\kappa\right)f^E\left( \frac{3\kappa}{\kappa+2}\right)\mbox{ .}
\end{equation}
This relationship is the main result (Eq. 5.6) of Ref.  \cite{Bor93}.

\subsubsection{Refined similarity hypotheses}

To investigate the implications of the Borgas' relationship between the singularity spectra $f^{E,L}$ of dissipation and the 
singularity spectra $\mathcal D^{E,L}$ of velocity, we need to use a \textit{dictionnary} \cite{Fri95} between the statistical 
properties of the coarse-grained dissipation and of the velocity increments. 
This is provided by the Refined Similarity Hypothesis (RSH) of Kolmogorov and Oboukhov \cite{Obo62,Kol62}. In the Eulerian 
framework in an $d$-dimensional Euclidian space, this hypothesis reads:
\begin{equation}\label{eq:RSHEul}
 \langle (\delta_\ell u)^q\rangle \sim \langle \epsilon_\ell^{q/3}\rangle\ell^{q/3} \: \leftrightarrow\: h=\frac{\alpha}{3} \mbox{ and }
\mathcal D^E(h) = f^E (\alpha)+d-1 \mbox{ .}
\end{equation}
Similarly, in the Lagrangian framework, the RSH hypothesis reads
\begin{equation}\label{eq:RSHLag}
 \langle (\delta_\tau v)^q\rangle \sim \langle \epsilon_\tau^{q/2}\rangle\tau^{q/2}\: \leftrightarrow\: h=\frac{\kappa}{2} \mbox{ and }
\mathcal D^L(h) = f^L (\kappa)+d-1 \mbox{ .}
\end{equation}
Only recently the Lagrangian RSH hypothesis (Eq. \ref{eq:RSHLag}) has been verified in numerical simulations 
\cite{BenBifCal10,YuMen10}. Using  Eqs. \ref{eq:RSHEul} and \ref{eq:RSHLag} to reinterpret Eq. \ref{eq:PredBorgasf}, we easily 
derive the following relationship 
between the Eulerian $\mathcal D^E(h)$ and Lagrangian $\mathcal D^L(h)$ velocity singularity spectra (with $d=1$) \cite{CheRou03}:
\begin{equation}\label{eq:BorTransf}
 \mathcal D^L(h) = -h+(1+h)\mathcal D^E\left( \frac{h}{1+h}\right)\mbox{ .}
\end{equation}
This relation Eq. \ref{eq:BorTransf} can be inverted
\begin{equation}\label{eq:InvBorTransf}
 \mathcal D^E(h) = h+(1-h)\mathcal D^L\left( \frac{h}{1-h}\right)\mbox{ .}
\end{equation}
To end this section, let us emphasize that the relation \ref{eq:BorTransf} is consistent with a non intermittent (K41) 
picture of turbulence. In this case, the Eulerian singularity spectrum is such that $\mathcal D^E(1/3) = 1$ and 
$\mathcal D^E(h) = -\infty$ if $h\ne 1/3$. Using relation \ref{eq:BorTransf}, we obtain $\mathcal D^L(1/2) = 1$ and 
$\mathcal D^L(h) = -\infty$ if $h\ne 1/2$. This is consistent with $ \langle (\delta_\tau v)^2\rangle\propto \tau$ and more generally 
with Eq. \ref{eq:RSHLag}.

\begin{figure}[t]
\center{\epsfig{file=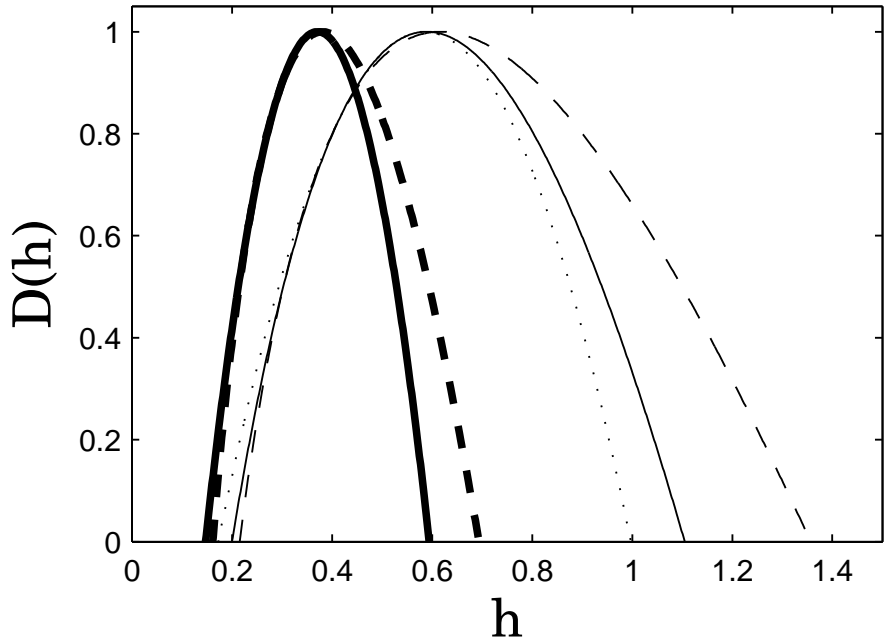,width=12cm}}
\caption{Comparing Eulerian and Lagrangian  singularity spectra: Eulerian quadratic singularity spectrum (Eq. \ref{eq:DHLN}, thick 
solid line), Eulerian She-L\'ev\^eque spectrum (Eq. \ref{eq:DHSL}, thick dashed line), the thin solid and dashed lines correspond 
to their Lagrangian counterpart using the Borgas' transformation (Eq. \ref{eq:BorTransf}), dotted thin line correspond to the Lagrangian 
quadratic spectrum (Eq. \ref{eq:DHLNLag}).}  \label{fig:CompDHFlat}
\end{figure}

We represent in Fig. \ref{fig:CompDHFlat} various singularity spectra used in the present article, in both the Eulerian 
and Lagrangian framework, as it was done in Ref. \cite{CheRou03,ChePhD}. First of all, we display the Eulerian quadratic (Eq. 
\ref{eq:DHLN}) and the She-L\'ev\^eque (Eq. \ref{eq:DHSL}) singularity spectra using respectively a thick solid and dashed lines. 
One can see that the increasing part of the spectra superimposes on each others. It means that, in a purely inertial description 
of turbulence, the quadratic and She-L\'ev\^eque spectra are indistinguishable when using positive-order structure functions. 
The decreasing parts are distinct. More sophisticated signal analysis methods than computing velocity increment moments are required to investigate the decreasing part of 
the singularity spectra, such as the wavelet transform modulus maxima (WTMM) \cite{MuzBac93}, the inverse structure functions method 
\cite{BifCen99} or the wavelet leaders \cite{WenAbr07}. The conclusion of these investigations is that, given the statistical 
limitations and linearization effects, experimental signals are consistent with a quadratic singularity spectrum (Eq. \ref{eq:DHLN}) 
for both the increasing and decreasing parts.

We furthermore display on the same plot (Fig. \ref{fig:CompDHFlat}) the three different singularity spectra we are using. This 
includes the quadratic Lagrangian spectrum (Eq. \ref{eq:DHLNLag}) represented using a thin dotted lines, and the 
transformed Eulerian spectra using the Borgas' relation (Eq. \ref{eq:BorTransf}) displayed using thin solid and dashed lines. 
One can see that the increasing parts coincide for the three cases, quantitative differences are shown concerning the decreasing 
part. Let us stress clearly that the Borgas' relation being nonlinear, a quadratic spectrum in the Eulerian spectrum is transformed 
into a non quadratic spectrum in the Lagrangian framework. This explains why the thin dotted and solid lines are distinct. In 
the sequel, we will see that the decreasing part has a strong influence on the statistics of dissipative quantities in Lagrangian 
turbulence such as the acceleration flatness (see section \ref{sec:StatAcce}), giving several arguments to discriminate these 
spectra when compared to empirical data. Another important remark can be made at this stage. The Borgas' transformation
(Eq. \ref{eq:BorTransf}), given the Eulerian spectra formerly introduced, predicts the existence of $h$-exponent greater than unity. 
This has strong implications on the singular nature of Lagrangian velocity fluctuations as quantified with velocity increments 
and the universal character of acceleration. We invite the reader to sections \ref{sec:CommentBounds} and \ref{sec:HOWavelets} 
for further discussions of this important point.

\subsection{Prediction of the variance and flatness of acceleration}

\label{sec:StatAcce}
\begin{figure}[t]
\center{\epsfig{file=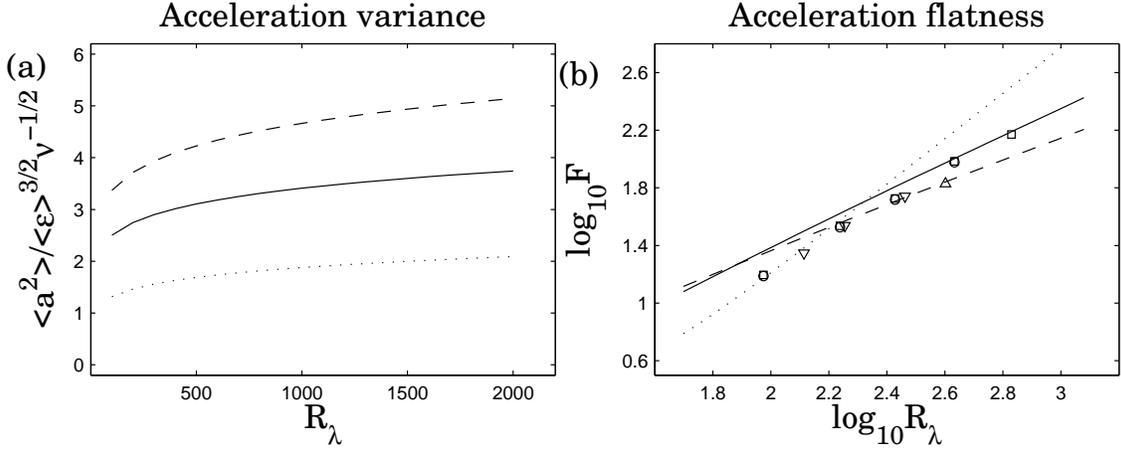,width=15cm}}
\caption{Multifractal predictions for the variance and flatness of acceleration. Different curves correspond to 
different sets of parameters $\mathcal D^L(h)$ and $\mathcal R^\dag$: quadratic Lagrangian spectrum (Eq. \ref{eq:DHLNLag}) and 
$\mathcal R^\dag=30$ (dotted line), quadratic Eulerian spectrum (Eq. \ref{eq:DHLN}) with $\mathcal R^\dag=6$ (solid line), and 
the She-L\'ev\^eque spectrum (Eq. \ref{eq:DHSL}) with $\mathcal R^\dag=2$ (dashed line). In the last two cases, the Borgas' 
transformation has been  used (Eq. \ref{eq:BorTransf}). (a) Acceleration variance given by Eqs. \ref{eq:a0LN}, \ref{eq:a0LNBor} 
and \ref{eq:a0SLBor}. (b) Acceleration flatness based on the numerical integration of Eq. \ref{eq:PredFlatAcceMulti}. 
Symbols correspond to numerical data: $\mathcal R_\lambda = 130$, $180$ and $290$ ($\Delta$) and $\mathcal R_\lambda=400$ 
($\nabla$) (Fig. \ref{fig:FlatLagFit}), symbols $\circ$ and $\square$ correspond respectively to the flatness of acceleration 
and pressure gradient from Ref. \cite{IshKan07}. 
 } \label{fig:VarFlatAcce}
\end{figure}

This section is devoted to the multifractal predictions of acceleration. We will mainly focus on the variance 
and on the flatness. The even order moments of acceleration are given by Eq. \ref{eq:HighMomAcce}. Recall that 
$\langle\delta^{2}\rangle =1$ by definition, we get for the acceleration 
variance:
\begin{equation}
 \langle a^{2}\rangle =  \left(\frac{\sigma}{T}\right)^{2}\frac{1}{\mathcal Z(0)}\int_{h_{\min}}^{h_{\max}} \left(\frac{\mathcal R_e}{\mathcal R^\dag}\right)
^{-\frac{2(h-1)+1-\mathcal D^L(h)}{2h+1}}dh \mbox{ .}
\end{equation}
Using a Gaussian approximation of the former integral (see Appendix \ref{ann:SDC}), we get
\begin{equation}
  \langle a^{2}\rangle \approx \left(\frac{\sigma}{T}\right)^{2}\sqrt{\frac{\left(\frac{\partial^2 \theta^L(h,0)}{\partial h^2}\right)_{h=h^L_0}}{\left(\frac{\partial^2 \theta^L(h,2)}{\partial h^2}\right)_{h=h^L_{2}}}}
\left(\frac{\mathcal R_e}{\mathcal R^\dag}\right)^{-\min_h \left[ \frac{2(h-1)+1-\mathcal D^L(h)}{2h+1}\right]}\mbox{ ,}
\end{equation}
where
\begin{equation}\label{eq:DefFtoLHP}
 \theta^L(h,p) = \frac{p(h-1)+1-\mathcal D^L(h)}{2h+1}\mbox{ ,} 
\end{equation}
and $h_p^L$ is the $h$-exponent for which $f^L(h,p)$ is minimum (in the same spirit as it is presented 
for the Eulerian case in Appendix \ref{ann:SDC}). Once rephrased in terms of mean dissipation (using Eq. 
\ref{eq:ComputMultiEpsilonSimp}, we obtain $\sigma^2/T=\langle \epsilon\rangle\mathcal R^*/15$) and in terms of Taylor-based 
Reynolds number $\mathcal R_\lambda$ (using Eq. \ref{eq:DefRlambdaRstar}), we obtain
\begin{equation}
 \langle a^{2}\rangle =a_0\langle \epsilon\rangle^{3/2}\nu^{-1/2}\mbox{ ,}
\end{equation}
where $a_0$ is a remaining non-dimensional quantity that includes intermittent corrections, tabulated in various flow conditions 
in Refs. \cite{VotPor02,Hil02}
\begin{equation}\label{eq:PredA0}
 a_0 \approx \left( \frac{\mathcal R^*}{15}\right)^{3/2}\frac{1}{\sqrt{\mathcal R^\dag}}
\sqrt{\frac{\left(\frac{\partial^2 \theta^L(h,0)}{\partial h^2}\right)_{h=h^L_0}}{\left(\frac{\partial^2 \theta^L(h,2)}{\partial h^2}\right)_{h=h^L_{2}}}}
\left( \frac{4}{\mathcal R^*}\frac{\mathcal R_\lambda^2}{\mathcal R^\dag}\right)^{-\min_h \left[ \frac{2(h-1)+1-\mathcal D^L(h)}{2h+1}\right]-\frac{1}{2}}\mbox{ .}
\end{equation}
When using the quadratic Lagrangian singularity spectrum (Eq. \ref{eq:DHLNLag}) with $\mathcal R^{\dag}=30$, we get:
\begin{equation}\label{eq:a0LN}
 a_0 = 0.6443 \mathcal R_\lambda^{0.1548}\mbox{ .}
\end{equation}
Alternatively; when introducing the Eulerian quadratic spectrum (Eq. \ref{eq:DHLN}) into the Lagrangian frame using Eq. \ref{eq:BorTransf}, and 
$\mathcal R^\dag=6$, we obtain:
\begin{equation}\label{eq:a0LNBor}
 a_0 = 1.3493 \mathcal R_\lambda^{0.1342}\mbox{ .}
\end{equation}
A very similar result is obtained when plugging the Eulerian She-L\'ev\^eque spectrum (Eq. \ref{eq:DHSL}) into the Lagrangian frame using Eq. \ref{eq:BorTransf}, and 
$\mathcal R^\dag=2$:
\begin{equation}\label{eq:a0SLBor}
 a_0 = 1.7603\mathcal R_\lambda^{0.141}\mbox{ .}
\end{equation}
These three predictions show similar dependence on the Reynolds number \cite{Bor93}, but  
the multiplicative pre factor depends strongly on the constant $\mathcal R^\dag$ (Eq. \ref{eq:PredA0}) that is found itself strongly 
dependent on the shape of the singularity spectrum (see also the following discussion on the acceleration flatness). We show 
in Fig. \ref{fig:VarFlatAcce}(a) the Reynolds number dependence of the factor $a_0$ (Eq. \ref{eq:PredA0}) for the three different 
sets of singularity spectrum $\mathcal D^L(h)$ and constants $\mathcal R^\dag$ (Eqs. \ref{eq:a0LN}, \ref{eq:a0LNBor} 
and \ref{eq:a0SLBor}).

Similar predictions can be derived for the acceleration flatness. This study will underline the limitations of the quadratic 
Lagrangian singularity spectrum (Eq. \ref{eq:DHLNLag}), to describe the fluctuations of the velocity increments.
The acceleration flatness can be expressed as the following function of the Reynolds number 
(Eq. \ref{eq:HighMomAcce})
\begin{equation}\label{eq:PredFlatAcceMulti}
 \mathcal F(a) = \frac{\langle a^4\rangle}{\langle a^2\rangle^2} = 3\mathcal Z(0)\frac{\int_{h_{\min}}^{h_{\max}} \left(\frac{\mathcal R_e}{\mathcal R^\dag}\right)
^{-\frac{4(h-1)+1-\mathcal D^L(h)}{2h+1}}dh}{\left(\int_{h_{\min}}^{h_{\max}} \left(\frac{\mathcal R_e}{\mathcal R^\dag}\right)
^{-\frac{2(h-1)+1-\mathcal D^L(h)}{2h+1}}dh\right)^2}
\mbox{ .}
\end{equation}
The predictions of the acceleration flatnesses obtained by numerical 
integration of the integrals entering Eq. \ref{eq:PredFlatAcceMulti} are shown in Fig. \ref{fig:VarFlatAcce}(b). These predictions 
are compared to 
DNS data, the ones used in Fig. \ref{fig:FlatLagFit} and the ones provided in Ref. \cite{IshKan07}. Some 
quantitative differences are observed for both the Reynolds number dependence and the multiplicative pre factor that 
depends strongly on the value of $\mathcal R^\dag$. An analytical expression of the flatness (Eq. \ref{eq:PredFlatAcceMulti})
can be obtained once again using a Gaussian approximation (c.f. Appendix \ref{ann:SDC}): 
\begin{equation}\label{eq:ApproxGaussFlatAcce}
 \mathcal F(a)\approx 3 \frac{b_4}{b_2^2}
\left( \frac{\mathcal R_e}{\mathcal R^\dag}\right)^{\chi_4^a-2\chi_2^a}\mbox{ ,}
\end{equation}
where
\begin{equation}\label{eq:DefChipAcce}
 \chi_p^a = -\min_h \left[ \frac{p(h-1) + 1-\mathcal D^L(h)}{2h+1}\right]\mbox{  and } b_p = \sqrt{\frac{\left(\frac{\partial^2 \theta^L(h,0)}{\partial h^2}\right)_{h=h^L_0}}{\left(\frac{\partial^2 \theta^L(h,p)}{\partial h^2}\right)_{h=h^L_{p}}}}\mbox{ ,}
\end{equation}
and $\theta^L(h,p)$ is defined in Eq. \ref{eq:DefFtoLHP}. The exponents $\chi_p^a$ control the Reynolds number dependence of 
the acceleration flatness. They depend only of the shape of the singularity spectrum $\mathcal D^L(h)$. In contrast, 
the multiplicative pre factor depends on both the shape of $\mathcal D^L(h)$ and $\mathcal R^\dag$. When 
using the quadratic Lagrangian spectrum (Eq. \ref{eq:DHLNLag}) with $\mathcal R^{\dag}=30$, we get
\begin{equation}\label{eq:FaLN}
 \mathcal F(a) = 0.0115\mathcal R_\lambda^{1.73}\mbox{ .}
\end{equation}
When introducing the Eulerian quadratic spectrum (Eq. \ref{eq:DHLN}) into the Lagrangian framework using Eq. \ref{eq:BorTransf}, 
and $\mathcal R^\dag=6$, we obtain
\begin{equation}\label{eq:FaLNBor}
 \mathcal F(a) = 0.4107\mathcal R_\lambda^{0.9174}\mbox{ ,}
\end{equation}
whereas for the Eulerian She-L\'ev\^eque spectrum (Eq. \ref{eq:DHSL}) and 
$\mathcal R^\dag=2$, we get
\begin{equation}\label{eq:FaSLBor}
 \mathcal F(a) = 1.1192\mathcal R_\lambda^{0.715}\mbox{ .}
\end{equation}
The three predictions (Eqs. \ref{eq:FaLN}, \ref{eq:FaLNBor} and \ref{eq:FaSLBor}) we made using the three different 
sets of parameters ($\mathcal D^L(h)$ and $\mathcal R^\dag$) show different behaviors. First, we can see quantitative differences 
on the dependence on the Reynolds number. When compared to data, the quadratic Lagrangian singularity 
spectrum (Eq. \ref{eq:DHLNLag}) gives an exponent $1.73$ much bigger than for the two other parameter sets. This seems to be consistent with data only 
at the lowest Reynolds numbers, but not at the highest Reynolds numbers where the predicted exponent leads to some overestimate of the flatness. 
On the opposite, the quadratic 
Eulerian singularity spectrum (Eq. \ref{eq:DHLN}) and the log-Poisson one (Eq. \ref{eq:DHSL}), once re-expressed in the Lagrangian 
framework using the Borgas' transformation (Eq. \ref{eq:BorTransf}), give a Reynolds number dependence consistent with 
data, especially at high Reynolds numbers. Second, the multiplicative pre factors depend strongly on the the shape of the respective 
$\mathcal D^L$ and on $\mathcal R^\dag$.

Let us stress that the analytical approximations given in Eqs.  \ref{eq:FaLN}, \ref{eq:FaLNBor} and \ref{eq:FaSLBor} differ from 
a numerical estimation of the integrals entering in Eq. \ref{eq:PredFlatAcceMulti}. In particular, when the quadratic Lagrangian 
spectrum is chosen, differences are quantitatively significant. This is clearly due to the range of integration $[h_{\min};h_{\max}]$ that is finite in the numerical 
integration (i.e. $h_{\min}=0$ and $h_{\max}=1$) whereas it is assumed infinite (i.e. $h_{\min}=-\infty$ and $h_{\max}=+\infty$) 
in the Gaussian approximation in order to get simple analytical formula (thus avoiding any corrections given 
by the error function Erf). Let us point out that if we instead use (i) the range $h_{\min}=0$ and $h_{\max}=1$ is used when working with the Eulerian quadratic 
singularity spectrum further transformed using Eq. \ref{eq:PredBorgasf}, and (ii) the range $h_{\min}=\frac{1}{8}$ and $h_{\max}=1$ 
when working with the She-L\'ev\^eque spectrum (Eq. \ref{eq:DHSL}), then the analytical 
predictions (Eqs. \ref{eq:FaLNBor} and \ref{eq:FaSLBor}) are very close to the numerical estimations of 
Eq. \ref{eq:PredFlatAcceMulti}. This underlines the limitations of the quadratic Lagrangian singularity spectrum 
(Eq. \ref{eq:DHLNLag}) to represent the statistics of acceleration. Indeed, additional analytical work shows that the exponent $h$ that 
minimizes the function $\theta^L(h,4)$ (Eq. \ref{eq:DefFtoLHP}) and leads to $\chi_4^a$ (c.f. Eq. \ref{eq:DefChipAcce}) is 
negative when using the quadratic Lagrangian singularity spectrum (Eq. \ref{eq:DHLNLag}). This would imply the existence 
of unphysical negative $h$-exponents. This drawback does not exist when using the Eulerian quadratic 
(Eq. \ref{eq:DHLN}) or the She-L\'ev\^eque spectra (Eq. \ref{eq:DHSL}). Once again, this tells us that the 
quadratic Lagrangian singularity spectrum (Eq. \ref{eq:DHLNLag}) is not able to reproduce the acceleration 
statistics (see section \ref{sec:CommentBounds} for more detailed discussion about the range of integration $[h_{\min};h_{\max}]$).

\section{Further discussions regarding the singular nature of velocity}

\subsection{Comments on the integration bounds $h_{\min}$ and $h_{\max}$.}
\label{sec:CommentBounds}

This section is devoted to present further discussions on the choice of the integration bounds  $[h_{\min};h_{\max}]$ that has 
been made to define the probabilistic models of Eulerian (c.f. Eqs. \ref{eq:betah} and \ref{eq:phmeneeul}) and 
Lagrangian (c.f. Eqs. \ref{eq:betahlag} and \ref{eq:phmenelag}) velocity increments. This requires a clear definition of the 
H\"{o}lder exponents $h$. Let us stress that the dissipative 
cut-offs, as given by the fluctuating dissipative length scale $\eta(h)$ (Eq. \ref{eq:etah}) and  by the fluctuating dissipative 
time scale $\tau_{\eta}(h)$ (Eq. \ref{eq:TauEtaH}), imply that $h_{\min}\ge -1$ in the Eulerian frame and $h_{\min}\ge -1/2$ in 
the Lagrangian counterpart. In the sequel, we will consider only inertial range fluctuations because the mathematical theory of 
singularities is well established (see Ref. \cite{MuzBac93} and references therein). Our choice of the integration domain will 
be based on these considerations. In Ref. \cite{BenBif10}, in which only the She-L\'ev\^eque spectrum is considered, 
other choices have been made.

The H\"{o}lder exponent $h(x_0)$ at a point $x_0$ of a singular signal $u(x)$ (the Lagrangian velocity $v(t)$ could be chosen without 
loss of generality) is defined as the biggest exponent $h$ such that it exists a polynomial $P_N$ of order $N$ and a positive 
constant C verifying
\begin{equation}
 |u(x) - P_N(x-x_0)| \le C|x-x_0|^{h(x_0)}\mbox{ .}
\end{equation}
Let us stress that $P_N$ is the Taylor's development of order $N$ of the signal $u$ and typically, $N\le h(x_0) < N+1$.
In other words, $h(x_0)$ is the exponent that defines the first singular behavior entering in the Taylor development of $u$ in the 
neighborhood of $x_0$:
\begin{equation}\label{eq:SingTaylorDevel}
 u(x)=u(x_0) + (x-x_0)u^{(1)}(x_0)+ ... + \frac{(x-x_0)^N}{N!}u^{(N)}(x_0) + C'|x-x_0|^{h(x_0)}\mbox{ ,}
\end{equation}
where $u^{(N)} = d^Nu/dx^N$ and $C'$ a constant. This shows that the velocity increment is only sensitive to exponents $0\le h\le 1$ because it is 
orthogonal to constants only. Thus, we chose for the Eulerian framework $h_{\min}=0$ and $h_{\max}=1$ when dealing with the 
quadratic spectrum (Eq. \ref{eq:DHLN}) and $h_{\min}=1/9$ and $h_{\max}=1$ when dealing with the She-L\'ev\^eque spectrum
(Eq. \ref{eq:DHSL}). Similar choices have been made in the Lagrangian framework, namely $h_{\min}=0$ and $h_{\max}=1$ when 
dealing with the Lagrangian quadratic spectrum (Eq. \ref{eq:DHLNLag}) and the Eulerian quadratic spectrum (Eq. \ref{eq:DHLN}) 
once transformed in the Lagrangian frame using Borgas' relation (Eq. \ref{eq:BorTransf}),  $h_{\min}=1/8$ and $h_{\max}=1$ when 
dealing with the She-L\'ev\^eque spectrum (Eq. \ref{eq:DHSL}) once reexpressed in the Lagrangian frame.

As far as the quadratic spectrum (Eq. \ref{eq:DHLN}) is considered, let us stress that the Eulerian predictions based on 
Eqs. \ref{eq:betah} and \ref{eq:phmeneeul} are similar if one uses $h_{\min}=0$ and $h_{\max}\ge 1$ or 
$h_{\min}=c_1-\sqrt{2c_2}\approx 0.1472$ and $h_{\max}=c_1+\sqrt{2c_2}\approx 0.5944$, the two values of $h$ such that 
$\mathcal D^E(h)\ge 0$. In the Lagrangian framework, considering the quadratic spectrum (Eq. \ref{eq:DHLNLag}), 
the predictions based on  Eqs. \ref{eq:betahlag} and \ref{eq:phmenelag} depend quantitatively on the range of integration. This 
is related to the fact that the decreasing part of the Lagrangian quadratic spectrum (Eq.  \ref{eq:DHLNLag}) reaches 
the zero value at $h\approx 1$. This would strongly suggest the possible existence of $h$-exponents greater than unity, exponents 
that cannot be measured with an increment, as explained at the beginning of this section. This is thus very tempting to study 
Lagrangian velocity fluctuations with multi scale objects orthogonal to polynomials as it may appear in the Taylor 
development (Eq. \ref{eq:SingTaylorDevel}) if singularities $h$ are greater than unity. This is the subject of the following 
section in which higher order increments are considered, as they were in Refs. \cite{CheRou03,ChePhD,FalFau07,FalRou10}.

\subsection{Wavelet analysis of Eulerian and Lagrangian fluctuations.}
\label{sec:HOWavelets}

\begin{figure}[t]
\center{\epsfig{file=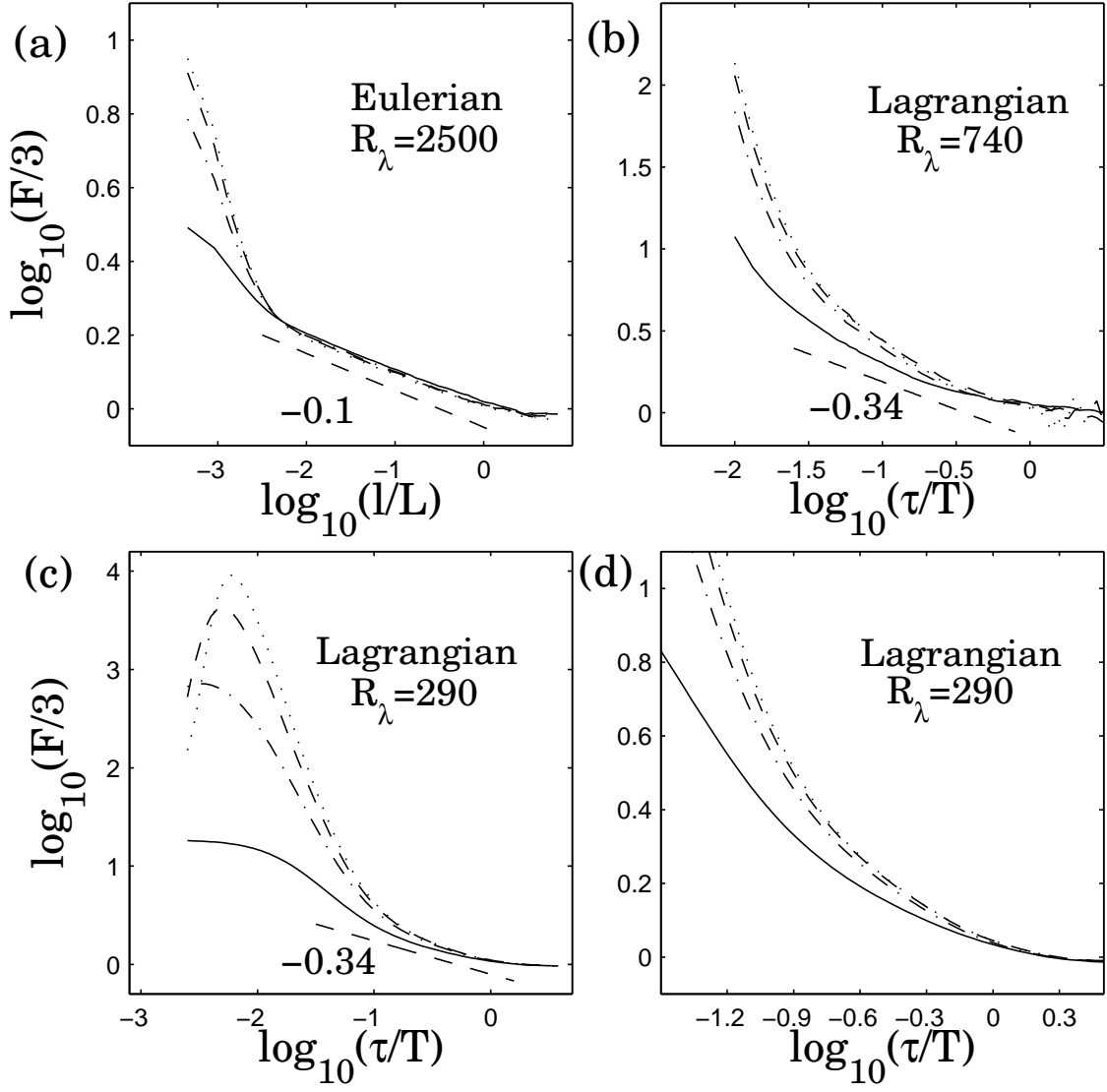,width=15cm}}
\caption{Velocity increment Flatness for various sets of empirical data and increment order. Different lines correspond to different 
velocity increments orders: first (solid), second (dashed-dot), third (dashed) and fourth (dotted). (a) Eulerian velocity flatness 
using the data from Modane's wind tunnel ($\mathcal R_\lambda=2500$). (b) Experimental Lagrangian velocity flatness 
$\mathcal R_\lambda=740$ from Ref. 
\cite{MorMet01}. (c) and (d) Numerical Lagrangian velocity flatness $\mathcal R_\lambda = 290$.} \label{fig:PlotWavelets}
\end{figure}

In order to check whether it exists singularities $h$ greater than unity, we perform a higher order increments study of velocity 
fluctuations. In the Eulerian framework, we will consider $N^{\mbox{th}}$-order velocity increments defined as 
\begin{align}\label{eq:DefEulHOVelIncr}
 \delta_\ell u(x) = \delta_\ell^{(1)}u(x) &= u(x+\ell)-u(x) \notag \\
\forall N\ge2\mbox{ , }\delta_\ell^{(N)}u(x) &= \delta_\ell^{(N-1)}u(x+\ell)-\delta_\ell^{(N-1)}u(x)\mbox{ .}
\end{align}
With such defined $\delta_\ell^{(N)}u$ velocity increments, singularities $0\le h<N$ become accessible (see the discussion 
provided in section \ref{sec:CommentBounds}. 

We display in Fig. \ref{fig:PlotWavelets}(a) the results of the statistical analysis of the first four velocity increments on 
the experimental signal obtained in the Modane's wind tunnel \cite{KahMal98} that provides the highest Reynolds number 
$\mathcal R_\lambda=2500$. We choose to display only the velocity increment flatness $F^{(N)}(\ell)$ defined as 
\begin{equation}\label{eq:FlatNIncr}
 F^{(N)}(\ell) = \frac{\langle (\delta_\ell^{(N)}u)^4\rangle}{\langle (\delta_\ell^{(N)}u)^2\rangle^2}\mbox{ .}
\end{equation}
In the inertial range of scales, namely $\ell \in [10^{-2.5}L;L]$, one can see that the observed power-law is independent on the 
increment order $N$. This means that singularities $h$ greater than unity are not observed on Eulerian fluctuations, as it could 
have been expected from the shape of the well accepted quadratic or She-L\'ev\^eque singularity spectra. This has an important 
implication on the multifractal formalism: the Eulerian singularity spectrum $\mathcal D^E(h)$ is indeed measurable with the 
first order $N=1$ velocity increment. 

At smaller scales $\ell$ 
lying in the dissipative range, one can see a quantitative dependence on the increment order $N$: as $N$ increases, the transition 
from the inertial and dissipative ranges is steeper and steeper. This can be easily understood while generalizing the present 
probabilistic approach (Eqs. \ref{eq:betah} and \ref{eq:phmeneeul}) to any $N^{\mbox{th}}$-order velocity increments, with thus 
the same singularity spectrum $\mathcal D^E(h)$. Indeed, 
the dependence on the order $N$ comes from the behavior in the dissipative range of the $N^{\mbox{th}}$-order velocity increment, 
which, to be consistent with the Taylor development of velocity, reads
\begin{equation}\label{eq:DissDeltaNEul}
 \delta_\ell^{(N)} u(x)\build{\sim}_{\ell\rightarrow 0}^{}\ell^N \partial_x^N u(x)\mbox{ .}
\end{equation}
The fluctuating nature of the dissipative scale $\eta(h)$ (Eq. \ref{eq:etah}) is independent on the way one is looking at 
velocity fluctuations and thus can be parametrized with the exponent $h$, although the constant $\mathcal R^*$ might depend on 
the order $N$. The probabilistic modeling of velocity increments, consistent with the Taylor development \ref{eq:DissDeltaNEul}, 
follows:
\begin{equation}\label{eq:betahN}
 \delta_\ell^{(N)} u \build{=}_{}^{\mbox{law}} \sigma^{(N)}\beta^{(N)}_\ell \delta \mbox{ with } \beta^{(N)}_\ell  =
\frac{\left(\frac{\ell}{L}\right)^h}{\left[
1+\left(\frac{\ell}{\eta(h)}\right)^{-2}\right]^{(N-h)/2}} \mbox{
,}
\end{equation}
the random variable $\delta$ being again a zero average unit variance Gaussian noise, 
$(\sigma^{(N)})^2 = \langle (\delta_L^{(N)}u)^2\rangle = \frac{(2N)!}{(N!)^2}\langle u^2\rangle$ (this can be easily obtained using the 
binomial theorem), and the $h$-distribution being unchanged an given by Eq. \ref{eq:phmeneeul}, but, in this case, $h_{\min}=0$ and 
$h_{\max}=N$. A numerical investigation of the $N^{\mbox{th}}$-order velocity increment 
flatness (Eq. \ref{eq:FlatNIncr}) based on the proposed probabilistic modeling (Eqs. \ref{eq:betahN} and \ref{eq:phmeneeul}) 
shows the same type of behavior as observed on the Modane's wind tunnel velocity data and depicted in Fig. 
\ref{fig:PlotWavelets}(a), namely a $N$-independent power-law in the inertial range and an intermediate dissipative steeper 
and steeper as the order $N$ increases (data not shown, see Ref. \cite{ChePhD}). In the Eulerian framework, the first order 
increment allows to measure fully the singular nature of velocity, the $N^{\mbox{th}}$-order increment appears as a tool that 
highlights unambiguously the dissipative range.

A similar study can be performed on the Lagrangian velocity using the $N^{\mbox{th}}$-order velocity time increments defined in the 
following way:
\begin{align}\label{eq:DefLagHOVelIncr}
 \delta_\tau v(t) = \delta_\tau^{(1)}v(t) &= v(t+\tau)-v(t) \notag \\
\forall N\ge2\mbox{ , }\delta_\tau^{(N)}v(t) &= \delta_\tau^{(N-1)}v(t+\tau)-\delta_\tau^{(N-1)}v(t)\mbox{ .}
\end{align}
We display the statistical analysis of the experimental measurements 
(in Fig. \ref{fig:PlotWavelets}(b), $\mathcal R_\lambda=740$) and of the numerical simulations (in Fig. \ref{fig:PlotWavelets}(c-d),
 $\mathcal R_\lambda=290$) of the Lagrangian velocity using the four first $N^{\mbox{th}}$-order velocity time increments 
(Eq. \ref{eq:DefLagHOVelIncr}). The results and interpretations are very different from the Eulerian case (presented in 
Fig. \ref{fig:PlotWavelets}(a)). We can indeed see that, in the inertial range ($\tau \in [10^{-1.5}T;T]$ for the 
experimental signal and $\tau \in [10^{-0.7}T;T]$ for the numerical simulation), the $N^{\mbox{th}}$-order 
flatness $F^{(N)}(\tau)$ underlies a transition between the $N=1$ and $N=2$ velocity time increment. Recall that in the Lagrangian framework, no clear power-laws are observed (see the discussion of Fig. \ref{fig:FlatLagFit}). We display nevertheless the slope (i.e. $-0.34=-4c_2^L$) that we would be observed if experimental data and simulations were performed at higher Reynolds numbers. As for $N=1$ velocity increments, higher order ($N\ge 2$) velocity increments do not exhibit clear power laws but still do exhibit a steeper behavior. We are thus led to the conclusion that high order velocity increments do behave differently depending on the order $N$: the level of intermittency for $N=1$ is lower than the one observed for $N\ge 2$. This can be 
seen as a proof of the existence of singularities $h$ greater than unity but smaller than 2, since the inertial range is found independent of the order $N$ as long as $N\ge 2$. It also says that the singular nature of Lagrangian 
velocity can be captured by an increment $\delta^{(N)}$ of order $N\ge 2$, the first order increment giving a biased singularity 
spectrum, that underestimates the intermittency coefficient, because of the existence of singularities greater than unity.

In the dissipative range, in a similar fashion than in the Eulerian case, the rapid increase of the flatness is steeper 
and steeper as the order $N$ of the analysing increment increases. This can be obtained  using a similar probabilistic modeling 
of the velocity increment, as it was generalized in the Eulerian frame (i.e. Eq. \ref{eq:betahN}):
\begin{equation}\label{eq:betahlagN}
 \delta_\tau^{(N)} v \build{=}_{}^{\mbox{law}} \sigma^{(N)} \beta^{(N)}_\tau \delta \mbox{ with } \beta^{(N)}_\tau  =
\frac{\left(\frac{\tau}{T}\right)^h}{\left[
1+\left(\frac{\tau}{\tau_{\eta}(h)}\right)^{-\gamma}\right]^{(N-h)/\gamma}} \mbox{
,}
\end{equation}
the expressions of the dissipative time scale $\tau_{\eta}(h)$ (Eq. \ref{eq:TauEtaH}) and the $h$-distribution 
(Eq. \ref{eq:phmenelag}) being unchanged, although the constant $\mathcal R^\dag$ might depends on $N$ and the range of integration 
being $h_{\min}=0$ and $h_{\max}=N$, and $(\sigma^{(N)})^2 = \langle (\delta_T^{(N)}v)^2\rangle = \langle (\delta_L^{(N)}u)^2\rangle 
= \frac{(2N)!}{(N!)^2}\langle u^2\rangle$. 

We have thus shown that some singularities exponent $h$ greater than unity and smaller than 2 exist in the Lagrangian frame, and not in 
the Eulerian frame. This observation is consistent with the Borgas' transformation (Eq. \ref{eq:BorTransf}) that indeed does 
predict these smooth behaviors in the Lagrangian trajectories starting from the well accepted Eulerian singularity 
spectra (Eqs. \ref{eq:DHLN} and \ref{eq:DHSL}). This has a direct implication on the Lagrangian acceleration. Indeed, it suggests 
that acceleration might not be the right quantity to look at from a multi scale perspective. More precisely, it appears to be difficult 
to understand properly acceleration statistics as the dissipative limit of velocity increments statistics in the inertial range 
as it can be done in the Eulerian framework. This is also tempting to interpret the non universal features observed in empirical 
acceleration as a manifestation of the biased behavior of a first order increment and underlines the importance of Lagrangian velocity 
higher order derivatives as unbiased quantities.

\section{Conclusions - Perspectives}

We have shown all along the article that a phenomenological theory of turbulent velocity fluctuations can be written down. Assuming 
the probability laws in the inertial range, we can derive explicit predictions for the statistics of velocity gradients and acceleration 
that can be compared to experimental and numerical data. In the Eulerian framework, assuming a quadratic (Eq. \ref{eq:DHLN}) or a 
log-Poisson (Eq. \ref{eq:DHSL}) singularity spectra, and a universal constant $\mathcal R^*$, we can predict accurately 
the statistics of velocity gradients, including the flatness (Eq. \ref{eq:predflatderiveul}) and skewness (Eq. 
\ref{eq:PredSkewNumEul}). 

In the Lagrangian framework, predictions, when compared to experimental and numerical data, are not as satisfactory as in the Eulerian framework. In particular, the remaing 
constant $\mathcal R^\dag$ is shown to depend strongly on the precise shape of the Lagrangian singularity spectrum $\mathcal D^L(h)$. 
Furthermore, it is shown, using the Borgas' transform (Eq. \ref{eq:BorTransf}), that the two widely accepted Eulerian singularity 
spectra (Eqs. \ref{eq:DHLN} and \ref{eq:DHSL}) lead to quantitative differences when computing the acceleration flatness 
(Eqs. \ref{eq:FaLNBor} and \ref{eq:FaSLBor}), whereas a quadratic approximation for $\mathcal D^L(h)$ (Eq. \ref{eq:DHLNLag}) leads 
to an irrealistic acceleration flatness (Eq. \ref{eq:FaLN}) at high Reynolds number. Nevertheless, these significant differences 
appear as a way to discriminate in the Eulerian framework the two former singularity spectra (Eqs. \ref{eq:DHLN} and \ref{eq:DHSL}). 
We also mention the intrinsic limitations of the (first-order) velocity increment to quantify accurately the singular nature of the 
Lagrangian velocity (section \ref{sec:HOWavelets}). We thus propose to study higher order velocity increments 
(Eqs. \ref{eq:DefEulHOVelIncr} and \ref{eq:DefLagHOVelIncr}) in order to quantify 
precisely the singular nature of the velocity fluctuations. We showed that indeed singularities greater than unity do exist in the 
Lagrangian framework.

As perspectives, we can mention:
\begin{itemize}
 \item To investigate further the multifractal predictions of the velocity derivatives skewness (see section \ref{sec:fullskew}). The present formalism underestimates the skewness of derivatives when compared to data 
(see Fig. \ref{fig:FlatEulTheo}(d)), although the predicted Reynolds number dependence is realistic and a lot scatter is found in 
experimental data. To that regard, the experimental resolution is a key issue. Specially designed numerical experiments could give some hints on the  
influence of the resolution on this quantity, as it has been done in Ref. \cite{Sch07}.
\item A theoretical explanation of the value of the universal constant $\mathcal R^*$ (or similarly of the Kolmogorov constant $c_K$) 
would be welcome. More precisely, we may wonder how this value depends on the shape of the singularity spectrum $\mathcal D^E(h)$. 
This issue is even more crucial in the Lagrangian framework since (see section  \ref{sec:StatAcce}) the constant 
$\mathcal R^\dag$ is very different if we work with different singularity spectra.
\item We have seen that the (first-order) Lagrangian velocity increment is not adapted to study the singular nature of the velocity 
along the trajectories. The acceleration appears to be thus an ill-posed quantity and we could question its 
universality. Indeed, depending on the flow geometry (von Karman flow, wind tunnel, fully periodic DNS, etc.), a lot of scatter is 
found in data (see \cite{Hil02}). To this regard, the second-order time 
derivative of Lagrangian velocity $d^2v/dt^2$ may be more independent on the large scale geometry of the flow than the first-order one. Performing such a measurement (or a numerical simulation) will be much more difficult because of the presence of noise. It 
would be very interesting to quantify precisely the Reynolds number dependence of the variance and flatness of this quantity, as 
it was initiated in Ref. \cite{IshKan07}.
\item The derivation of the Eulerian singularity spectrum from first principles is still missing. Recently, some progresses have 
been made while studying the dynamics of the velocity gradient tensor $A_{ij} = \partial u_j/\partial x_i$ along Lagrangian 
trajectories \cite{Vie84,Ash87,GirPop90,Can92,ChePum99,JeoGir03,CheMen06,CheMen07,Men11}. In particular, in Refs. \cite{CheMen06,CheMen07} is 
proposed a stationary process for the velocity gradient tensor that predicts in a realistic way the longitudinal and transverse intermittencies. This 
theoretical investigation gives an Eulerian singularity spectrum consistent with the one observed on data (Eq. \ref{eq:DHLN}).
\item In this article, we focus on two-points quantities, namely the velocity increments, fully determined by the corresponding 
probability densities $\mathcal P_{\delta_\ell u}(\delta_\ell u)$ and $\mathcal P_{\delta_\tau v}(\delta_\tau v)$. Nothing is said 
on the long-range correlated nature of velocity. Indeed, it is shown in Refs. \cite{MorDel02} and \cite{MorCro04} that the acceleration 
amplitude is correlated over the integral time scale $T$. In a similar fashion, it has been known for a long time that dissipation is 
correlated over the integral length scale $L$ \cite{MonYag75,AntPha81}. Taking into account this peculiar correlation structure 
of velocity allows to propose stochastic processes (one dimensional in the Lagrangian case, three dimensional vectorial for the 
Eulerian case) able to mimic the behavior of velocity in the inertial range \cite{MorDel02,CheRob10}. It remains to propose such 
stochastic processes able to reproduce velocity statistics in the intermediate and dissipative ranges and more generally, to give 
a stochastic representation of the differential action of viscosity as pointed out by the fluctuating nature of the 
dissipative scales (Eqs. \ref{eq:etah} and \ref{eq:TauEtaH}).
\end{itemize}

\section*{Acknowledgements}
We acknowledge for fruitfull discussions P. Abry, P. Borgnat, P. Flandrin. We thank Y. Gagne and the ONERA for providing the velocity signal 
of the wind tunnel of Modane,  P. Tabeling and H. Willaime for the experimental data shown in Fig. \ref{fig:FlatDerivTab}, C. Baudet and A. Naert for the air-jet Eulerian data ($\mathcal R_\lambda=380$), L. Biferale, A. Lanotte, F. Toschi 
for the Lagrangian numerical data ($\mathcal R_\lambda = 400$ shown in Fig. \ref{fig:FlatLagFit}) and N. Mordant for Lagrangian 
experimental data ($\mathcal R_\lambda=740$). 

\appendix

\section{Gaussian approximation and steepest-descent estimation}
\label{ann:SDC}

The aim of this Appendix is to elaborate on a Gaussian approximation that can be made to compute analytically the integrals entering, 
for example in Eqs. \ref{eq:ComputMultiEpsilon} or \ref{eq:predflatderiveul}. Let us consider, without loss of generality,
the integral
\begin{equation}
 \mathcal I(p) = \int_{h_{\min}}^{h_{\max}}\left(\frac{\mathcal R_e}{\mathcal R^*}\right)^{-\frac{p(h-1)+1-\mathcal D^E(h)}{h+1}}dh\mbox{ ,}
\end{equation}
with $\mathcal R_e\rightarrow +\infty$. We first perform a Taylor's development of the exponent of the Reynolds number around the 
exponent $h_p$ that minimizes it, namely
\begin{equation}
 \theta^E(h,p) = \frac{p(h-1)+1-\mathcal D^E(h)}{h+1} = \theta^E(h_p,p)+\left(\frac{\partial^2 \theta^E}{\partial h^2}\right)_{h=h_p}\frac{(h-h_p)^2}{2} + o[(h-h_p)^2]\mbox{ .}
\end{equation}
When assuming that $h_{\min}=-\infty$ and $h_{\max}=+\infty$, we can approximate the integral $\mathcal I(p)$ as a 
Gaussian integral, i.e.
\begin{align}
 \mathcal I(p) &\approx \left(\frac{\mathcal R_e}{\mathcal R^*}\right)^{-\min_h \left[ \frac{p(h-1)+1-\mathcal D^E(h)}{h+1}\right]}
\int_{-\infty}^{+\infty}\left(\frac{\mathcal R_e}{\mathcal R^*}\right)^{-\left(\frac{\partial^2 \theta^E}{\partial h^2}\right)_{h=h_p}\frac{(h-h_p)^2}{2}}dh \notag \\
&=\sqrt{\frac{2\pi}{\left(\frac{\partial^2 \theta^E}{\partial h^2}\right)_{h=h_p}}}
\left(\frac{\mathcal R_e}{\mathcal R^*}\right)^{-\min_h \left[ \frac{p(h-1)+1-\mathcal D^E(h)}{h+1}\right]}\mbox{ .}
\end{align}
Then, the even order moment of velocity gradient (Eq. \ref{eq:predgradeul}) can be obtained in the following 
approximate way (including the normalizing constant $\mathcal Z(0)$):
\begin{equation}\label{eq:PreSDMomGrad}
 \langle (\partial_x u)^{2p}\rangle = \sqrt{\frac{\left(\frac{\partial^2 \theta^E(h,0)}{\partial h^2}\right)_{h=h_0}}{\left(\frac{\partial^2 \theta^E(h,2p)}{\partial h^2}\right)_{h=h_{2p}}}}
\left(\frac{\mathcal R_e}{\mathcal R^*}\right)^{-\min_h \left[ \frac{2p(h-1)+1-\mathcal D^E(h)}{h+1}\right]}\mbox{ .}
\end{equation}
Moreover from Eqs. \ref{eq:PreSDMomGrad} and \ref{eq:ComputMultiEpsilon}, we get a precise estimation of the dissipation:
\begin{equation}
 \langle \epsilon\rangle \build{\approx}_{\mathcal R_e\rightarrow +\infty}^{} 
\sqrt{\frac{\left(\frac{\partial^2 \theta^E(h,0)}{\partial h^2}\right)_{h=h_0}}{\left(\frac{\partial^2 \theta^E(h,2)}{\partial h^2}\right)_{h=h_{2}}}}
\frac{15}{\mathcal R^*}\frac{\sigma^3}{L}\mbox{ .}
\end{equation}
Using the quadratic singularity spectrum Eq. \ref{eq:DHLN} with $c_2=0.025$, we get 
\begin{equation}
  \langle \epsilon\rangle \build{\approx}_{\mathcal R_e\rightarrow +\infty}^{} 0.97\frac{15}{\mathcal R^*}\frac{\sigma^3}{L}\mbox{ ,}
\end{equation}
in very good agreement with the approximation made in Eq. \ref{eq:ComputMultiEpsilonSimp}. Similar calculations can be made for the velocity 
derivative flatness, and we get (for the quadratic singularity spectrum Eq. \ref{eq:DHLN} with $c_2=0.025$):
\begin{equation}
 \frac{ \langle (\partial_x u)^{4}\rangle}{ \langle (\partial_x u)^{2}\rangle^2} 
\build{\approx}_{\mathcal R_e\rightarrow +\infty}^{} 3\times 0.99\times \left(\frac{\mathcal R_e}{\mathcal R^*}\right)^{0.18}\mbox{ ,} 
\end{equation}
once again is very close agreement with the approximation made in Eq. \ref{eq:predflatderiveul} without taking into account these 
corrections.

\section{Velocity increment and gradient probability density functions}\label{ann:PDF}

We follow here the presentation of a classical textbook on random variables \cite{Pap91}. Consider two random variables $\textbf{x}$ 
and  $\textbf{y}$, and a function $g(x,y)$. We form the random variable $\textbf{z}$ as a function of the two random variables  $\textbf{x}$ 
and  $\textbf{y}$, namely
\begin{equation}
 \textbf{z} = g (\textbf{x},\textbf{y})\mbox{ .}
\end{equation}
Consider the domain $D_z$ such that
\begin{equation}
 \{ (\textbf{x},\textbf{y}) \in D_z\} =  \{ g(\textbf{x},\textbf{y}) \le z\}\mbox{ .}
\end{equation}
The distribution function $F_z(z)$ defined as the probability $P$ to have $\textbf{z}\le z$ is given by \cite{Pap91}
\begin{equation}
 F_z(z) = P\{\textbf{z}\le z\} = P\{ (\textbf{x},\textbf{y}) \in D_z\}=\iint_{D_z}\mathcal P_{x,y}(x,y)dxdy\mbox{ ,}
\end{equation}
where $\mathcal P_{x,y}$ is the joint-density of the random variables $\textbf{x}$ and  $\textbf{y}$. Then, the probability density 
function $\mathcal P_z(z) = dF_z(z)/dz$ is given by the derivative, with respect to $z$, of the distribution function $F_z(z)$.

Consider now the random variable $\delta_\ell u$ formed as the product of two random variables written as 
$\delta_\ell u = \sigma\beta_\ell(h)\delta$, with the function $\beta_\ell (h)>0$ definite positive (Eqs. \ref{eq:betah}, 
\ref{eq:betahlag} or \ref{eq:DistrDeltalu}). In this case, the domain $D_{\delta_\ell u}$ is easily obtained
\begin{equation}
  \{ (\beta_\ell(h),\delta) \in D_{\delta_\ell u}\} = \{ h\in[h_{\min},h_{\max}],\delta\in]-\infty,\frac{\delta_\ell u}{\beta_\ell(h)}]  \}\mbox{ .}
\end{equation}
We get for the distribution function $F_{\delta_\ell u}$ of the velocity increments
\begin{equation}\label{eq:DistributionIncrement}
 F_{\delta_\ell u} (\delta_\ell u) = \int_{h=h_{\min}}^{h_{\max}} \int_{\delta=-\infty}^{\frac{\delta_\ell u}{\sigma\beta_\ell (h)}}
\mathcal P_{\delta,h}(\delta,h)d\delta dh\mbox{ .}
\end{equation}
The probability density function of the random variable $\delta_\ell u$ is readily obtained from Eq. \ref{eq:DistributionIncrement} 
taking a derivative with respect to $\delta_\ell u$, and we get
\begin{equation}
 \mathcal P_{\delta_\ell u} (\delta_\ell u) = \frac{dF_{\delta_\ell u} (\delta_\ell u)}{d\delta_\ell u} = 
\int_{h=h_{\min}}^{h_{\max}}\frac{1}{\sigma \beta_\ell (h)}\mathcal P_{\delta,h}\left(\frac{\delta_\ell u}{\sigma\beta_\ell (h)},h\right)dh\mbox{ ,}
\end{equation}
which justifies Eq. \ref{eq:DistrDeltalu}. If furthermore, the random variables $h$ and $\delta$ are assumed independent, as in 
Eqs. \ref{eq:betah} and \ref{eq:betahlag}, then the joint density can be factorized, i.e. 
$\mathcal P_{\delta,h}(\delta,h) = \mathcal P_{\delta}(\delta) \mathcal P_{h}(h)$ and the distribution Eq. \ref{eq:PredPDFEulEdgeworth} is justified.


\begin{thebibliography}{00}

\bibitem{Ric22}
L. F. Richardson. Weather Prediction by Numerical Process, Cambridge University Press, Cambridge, 1922.

\bibitem{Kol41}
A. N. Kolmogorov, The local structure of turbulence in incompressible
viscous fluid for very large Reynolds numbers, Dokl. Akad. Nauk SSSR 30
(1941) 301 [in Russian]. English translation: Proc. R. Soc. London,
Ser. A 434 (1991), 9.

\bibitem{Bat53}
G. K. Batchelor, The Theory of Homogeneous Turbulence, Cambridge University Press, Cambridge, 1953.

\bibitem{TenLum72}
H. Tennekes and J. L. Lumley, A First Course in Turbulence, MIT Press,
Cambridge, MA, 1972.

\bibitem{Kra74}
R. H. Kraichnan, On Kolmogorov’s inertial-range theories, J. Fluid
Mech. 62 (1974), 305.

\bibitem{MonYag75}
A. S. Monin and A. M. Yaglom, Statistical Fluid Mechanics, MIT Press,
Cambridge, MA, 1975.

\bibitem{Fri95}
U. Frisch, Turbulence, Cambridge University Press, Cambridge, 1995.

\bibitem{Pop00}
S. B. Pope, Turbulent Flows, Cambridge University Press, Cambridge,
2000.
\bibitem{Tsi01}
A. Tsinober, An Informal Introduction to Turbulence, Kluwer Academic,
Dordrecht, 2001.


\bibitem{YeuZho97}
P. K. Yeung and Y. Zhou, Universality of the Kolmogorov constant in numerical simulations
of turbulence. Phys. Rev. E 56 (1997), 1746.
\bibitem{DonSre10}
D. A. Donzis and K. R. Sreenivasan, The bottleneck effect and the Kolmogorov
constant in isotropic turbulence, J. Fluid Mech. 657 (2010), 171.


\bibitem{CasGag90}
B. Castaing, Y. Gagne and E. Hopfinger, Velocity probability density functions of high Reynolds number turbulence, 
Physica D 46 (1990), 177.
\bibitem{BenBif91}
R. Benzi, L. Biferale, G. Paladin, A. Vulpiani, and M. Vergassola, 
Multifractality in the statistics of the velocity gradients in turbulence, Phys. Rev. Lett. 67 (1991), 2299.
\bibitem{KaiSre92}
P. Kailasnath, K. R. Sreenivasan, and G. Stolovitzky, Phys. Rev. Lett. 68 (1992), 2766.


\bibitem{Obo62}
A. M. Oboukhov, Some specific features of atmospheric turbulence, J. Fluid Mech. 13 (1962), 77. 

\bibitem{Kol62}
A. N.  Kolmogorov, A refinement of previous hypotheses concerning the local structure of turbulence in a viscous incompressible fluid at high Reynolds number, 
J. Fluid Mech. 13 (1962), 77. 


\bibitem{KahMal98}
H. Kahalerras, Y. Mal\'ecot, Y. Gagne, and B. Castaing, Intermittency and Reynolds number, Phys. Fluids 10 (1998), 910.


\bibitem{OttMan00}
S. Ott and J. Mann, An experimental investigation of the relative diffusion of particle pairs in three-dimensional turbulent flow,
J. Fluid Mech. 422 (2000), 207. 

\bibitem{PorVot01}
A. La Porta, G. A. Voth, A. M. Crawford, J. Alexander and  E. Bodenschatz, 
Fluid particle accelerations in fully developed turbulence, Nature 409 (2001), 1017.

\bibitem{VotPor02}
G. A. Voth. A. La Porta, A. Crawford, J. Alexander and E. Bodenschatz, Measurement of particle accelerations in 
fully developed turbulence, J. Fluid Mech. 469 (2002) 121.

\bibitem{MorMet01}
N. Mordant, P. Metz, O. Michel, and J.-F. Pinton, Measurement of Lagrangian Velocity in Fully Developed Turbulence,
Phys. Rev. Lett. 87 (2001), 214501.

\bibitem{MorDel02}
N. Mordant, J. Delour, E. L\'ev\^eque, A. Arneodo, and J.-F. Pinton, 
Long Time Correlations in Lagrangian Dynamics: A Key to Intermittency in Turbulence, Phys. Rev. Lett. 89 (2002), 254502.

\bibitem{MorDel03}
N. Mordant, J. Delour, E. L\'ev\^eque, O. Michel, A. Arneodo and J.-F. Pinton, 
Lagrangian Velocity Fluctuations in Fully Developed Turbulence: Scaling, Intermittency, and Dynamics,  
J. Stat. Phys. 113 (2003), 701.

\bibitem{CheRou03}
L. Chevillard, S. G. Roux, E. L\'ev\^eque, N. Mordant, J.-F. Pinton, and A. Arneodo,  
Lagrangian Velocity Statistics in Turbulent Flows: Effects of Dissipation, Phys. Rev. Lett. 91 (2003), 214502. 

\bibitem{MorCro04}
N. Mordant, A. M. Crawford and E. Bodenschatz, Experimental Lagrangian acceleration probability density function measurement,
Physica D 193 (2004), 245.

\bibitem{XuBou06}
H. Xu, M. Bourgoin, N. T. Ouellette, and E. Bodenschatz, 
High order Lagrangian velocity statistics in turbulence, Phys. Rev. Lett. 96 (2006), 024503.

\bibitem{BerOtt09}
J. Berg, S. Ott, J. Mann, and B. L\"{u}thi,   Experimental investigation of Lagrangian structure functions in turbulence, 
Phys. Rev. E 80 (2009), 026316. 


\bibitem{YeuPop89}
P. K. Yeung and S. B. Pope,  Lagrangian statistics from direct numerical simulations of isotropic turbulence, 
J. Fluid Mech. 207 (1989), 531.

\bibitem{Yeu01}
P. K. Yeung, Lagrangian characteristics of turbulence and scalar transport in direct numerical simulations, J. Fluid Mech. 427 (2001), 241.

\bibitem{BifBof04}
L. Biferale, G. Boffetta, A. Celani, B. J. Devenish, A. Lanotte, and F. Toschi,
Multifractal statistics of Lagrangian velocity and acceleration in turbulence, Phys. Rev. Lett. 93 (2004), 064502. 


\bibitem{Yeu02}
P. K. Yeung, Lagrangian investigations of turbulence, Ann. Rev. Fluid Mech. 34 (2002), 115.

\bibitem{BifBod08}
L. Biferale, E. Bodenschatz, M. Cencini, A. S. Lanotte, N. T. Ouellette, F. Toschi, and H. Xu, Lagrangian structure functions in turbulence: A quantitative comparison between experiment and direct numerical simulation, Phys. Fluids \textbf{20}, 065103 (2008). 

 
\bibitem{ArnICTR08}
A. Arneodo et al., Universal Intermittent Properties of Particle Trajectories in Highly Turbulent Flows, 
Phys. Rev. Lett.  100 (2008), 254504.
\bibitem{TosBod09}
F. Toschi and E. Bodenschatz, Lagrangian properties of particles in turbulence, Ann. Rev. Fluid Mech. 41 (2009), 375.
\bibitem{FriVer91}
U. Frisch and M. Vergassola, A prediction of the multifractal model: the intermediate dissipation range, Europhys. Lett. 14 (1991), 439.
\bibitem{BeckSuper}
C. Beck and E.G.D. Cohen, Superstatistics, Physics A \textbf{322}, 267 (2003).
\bibitem{Fried03}
R. Friedrich, Statistics of Lagrangian Velocities in Turbulent Flows, Phys. Rev. Lett. \textbf{90}, 084501 (2003).
\bibitem{ZybSir10}
K. P. Zybin and V. A. Sirota, Lagrangian and Eulerian Velocity Structure Functions in Hydrodynamic Turbulence, Phys. Rev. Lett. \textbf{104}, 154501 (2010).
\bibitem{DaiFri12}
A. Daitche, R. Friedrich, O. Kamps, J. L$\ddot{\mbox{u}}$lff, M. Voskuhle and M. Wilczek, The Lundgren-Monin-Novikov Hierarchy: Kinetic Equations for Turbulence, This specal issue of the Comptes Rendus de l'Acad\'emie des Sciences (2012).
\bibitem{DuSaw95}
S. Du, B. L. Sawford, J. D. Wilson, and D. J. Wilson, 
Estimation of the Kolmogorov constant (C0) for the Lagrangian structure function, using a second-order Lagrangian model of grid turbulence,
Phys. Fluids 7 (1995), 3083.
\bibitem{LieAsa02}
R. Lien and E. D'Asaro, The Kolmogorov constant for the Lagrangian velocity spectrum and structure function, Phys. Fluids 14 (2002), 4456.


\bibitem{CheCas05}
L. Chevillard, B. Castaing, and E. L\'ev\^eque,
On the rapid increase of intermittency in the near-dissipation range of fully developed turbulence, Eur. Phys. J. B 45 (2005), 561.

\bibitem{CheCas06}
L. Chevillard, B. Castaing, E. L\'ev\^eque, and A. Arneodo,
Unified multifractal description of velocity increments statistics
in turbulence: Intermittency and skewness, Physica D 218 (2006),
77.


\bibitem{TabZoc96}
P. Tabeling, G. Zocchi, F. Belin, J. Maurer, and H. Willaime, Probability density functions, skewness, and flatness in large Reynolds 
number turbulence, Phys. Rev. E 53 (1996), 1613.
\bibitem{TabWil02}
P. Tabeling and H. Willaime, Transition at dissipative scales in large-Reynolds-number turbulence,  Phys. Rev. E 65 (2002), 066301.

\bibitem{ChaCha00}
O. Chanal, B. Chabaud, B. Castaing and B. H\'ebral, Intermittency in a turbulent low temperature gaseous helium jet, 
Eur. Phys. J. B 17 (2000), 309.



\bibitem{WilFri09}
M. Wilczek and R. Friedrich, Dynamical origins for non-Gaussian vorticity distributions in turbulent flows, 
Phys. Rev. E 80 (2009), 0160316.

\bibitem{WilFri10}
M. Wilczek, A. Daitche and R. Friedrich, On the velocity distribution in
homogeneous isotropic turbulence: correlations and deviations from Gaussianity, J. Fluid Mech. \textbf{676}, 191 (2011). 




\bibitem{Arneodoetal96}
A. Arneodo \textit{et al.}, Structure functions in turbulence, in various flow configurations, at Reynolds number between 30 and 5000, using extended self-similarity,
 Europhys. Lett. 34 (1996), 411.

\bibitem{SheLev94}
Z.-S. She and E. L\'ev\^eque,  Universal scaling laws in fully developed turbulence, Phys. Rev. Lett. 72 (1994), 336.

\bibitem{Dub94}
B. Dubrulle, Intermittency in fully developed turbulence: Log-Poisson statistics and generalized scale covariance,
Phys. Rev. Lett. 73 (1994) 959. 



\bibitem{MuzBac93}
J. F. Muzy, E. Bacry and A. Arneodo, Multifractal formalism for fractal signals: The structure-function approach versus the wavelet-transform modulus-maxima method,
 Phys. Rev. E 47 (1993), 875.

\bibitem{WenAbr07}
H. Wendt, P. Abry, and S. Jaffard, Bootstrap for empirical multifractal analysis, 
IEEE Signal Proc. Mag. 24 (2007) 38. 



\bibitem{PalVul87}
G. Paladin and A. Vulpiani, Degrees of freedom of turbulence,
Phys. Rev. A \textbf{35} (1987), 1971.



\bibitem{MalAur00}
Y. Mal\'ecot, C. Auriault, H. Kahalerras, Y. Gagne, O. Chanal, B. Chabaud, and B. Castaing, A statistical estimator of turbulence intermittency in physical
and numerical experiments, Eur. Phys. J. B 16 (2000), 549.

\bibitem{GagCas04}
Y. Gagne, B. Castaing, C. Baudet, and Y. Mal\'ecot, Reynolds dependence of third-order velocity structure functions,
Phys. Fluids 16 (2004), 482.





\bibitem{Nel90}
M. Nelkin, Multifractal scaling of velocity derivatives in
turbulence, Phys. Rev. A 42 (1990), 7226.



\bibitem{VanAnt80}
C. W. Van Atta and R. A. Antonia, Reynolds number dependence of skewness and flatness factors of turbulent velocity derivatives,
Phys. Fluids 23 (1980), 252.

\bibitem{SreAnt97}
K. R. Sreenivasan and R. A. Antonia, The phenomenology of small-scale turbulence, Ann. Rev. Fluid Mech. 29 (1997) 435.

\bibitem{GylAyy04}
A. Gylfason, S. Ayyalasomayajula and Z. Warhaft, Intermittency, pressure and acceleration statistics from hot-wire measurements 
in wind-tunnel turbulence, J. Fluid Mech. 501 (2004), 213.



\bibitem{IshKan07}
T. Ishihara, Y. Kaneda, M. Yokokawa, K. Itakura and A. Uno, Small-scale statistics in high-resolution direct numerical simulation of turbulence: Reynolds number dependence of one-point velocity gradient statistics,
J. Fluid Mech. 592 (2007), 335.

\bibitem{AntCha81}
R. Antonia, A. Chambers and B. Satyaprakash, Reynolds number dependence of high-order moments of the streamwise turbulent velocity derivative, 
Boundary Layer Met. 21 (1981) 159.


\bibitem{ChePhD}
L. Chevillard, Unified Multifractal Description of the Intermittency Phenomenon in Eulerian and Lagrangian Turbulence, PhD Thesis, 
University of Bordeaux (2004), unpublished, can be found online at http://tel.archives-ouvertes.fr/.

\bibitem{Men96}
C. Meneveau, Transition between viscous and inertial-range scaling of turbulence structure functions, Phys. Rev. E 54 (1996) 3657.

\bibitem{Bat51}
G. K. Batchelor, Pressure fluctuations in isotropic turbulence, Proc. Cambridge Philos. Soc. 47 (1951), 359.


\bibitem{BosChe10}
W. Bos, L. Chevillard, J. Scott and R. Rubinstein,  Reynolds number effect on the velocity increment skewness in isotropic turbulence,  Phys. Fluids \textbf{24}, 015108 (2012).


\bibitem{Pap91}
A. Papoulis, Probability, Random Variables and Stochastic Processes, McGraw-Hill Inc., New York, 1991.

\bibitem{BenBif10}
R. Benzi, L. Biferale, R. Fisher, D.Q. Lamb and F. Toschi, Inertial range Eulerian and Lagrangian statistics from numerical simulations of isotropic turbulence, J. Fluid Mech. 653 (2010) 221. 

\bibitem{DelMuz01}
J. Delour, J.-F. Muzy and A. Arneodo, Intermittency of 1D velocity spatial profiles in turbulence: a magnitude cumulant analysis, Eur. Phys. J. B  23 (2001), 243.


\bibitem{BofLil02}
G. Boffetta, F. De Lillo and S. Musacchio, Lagrangian statistics and temporal intermittency in a shell model of turbulence,
Phys. Rev. E 66 (2002), 066307.



\bibitem{HomKam09}
H. Homann, O. Kamps, R. Friedrich and R. Grauer, Bridging from Eulerian to Lagrangian statistics in 3D hydro- and magnetohydrodynamic turbulent flows, 
New Journ. Phys. 11 (2009), 73020.

\bibitem{KamFri09}
O. Kamps, R. Friedrich, and R. Grauer, Exact relation between Eulerian and Lagrangian velocity increment statistics, 
Phys. Rev. E 79 (2009), 066301.



\bibitem{BenBifCal10}
R. Benzi, L. Biferale, E. Calzavarini, D. Lohse and F. Toschi, 
Velocity-gradient statistics along particle trajectories in turbulent flows: The refined similarity
hypothesis in the Lagrangian frame, Phys. Rev. E 80 (2009), 066318.

\bibitem{YuMen10}
H. Yu and C. Meneveau, Lagrangian refined Kolmogorov similarity hypothesis for gradient time evolution and 
correlation in Turbulent Flows, Phys. Rev. Lett. 104 (2010), 084502.


\bibitem{Hil02}
R.J. Hill, Scaling of acceleration in locally isotropic turbulence, J. Fluid Mech. 452 (2002), 361.



\bibitem{Meneveauetal90}
C. Meneveau, K.R. Sreenivasan, P. Kailasnath, and M. S. Fan, Joint
multifractal measures: Theory and applications to turbulence,
Phys. Rev. A 41 (1990), 894.
\bibitem{MenSre91}
C. Meneveau and K. R. Sreenivasan, The multifractal nature of
turbulent energy dissipation, J. Fluid Mech. 224 (1991), 429.

\bibitem{SreMen88}
K. R. Sreenivasan and C. Meneveau, Singularities of the equations
of fluid motion, Phys. Rev. A 38 (1988), 6287.

\bibitem{Bor93}
M. S. Borgas, The multifractal lagrangian nature of turbulence,
Phil. Trans. R. Soc. Lond. A 342 (1993), 379.

\bibitem{BifCen99}
L. Biferale, M. Cencini, D. Vergni, and A. Vulpiani, Exit time of turbulent signals: A way to detect the 
intermediate dissipative range, Phys. Rev. E 60 (1999), R6295.

\bibitem{FalFau07}
E. Falcon, S. Fauve and C. Laroche, Observation of intermittency in wave turbulence, Phys. Rev. Lett. 98 (2007), 154501.

\bibitem{FalRou10}
E. Falcon, S. Roux and C. Laroche, On the origin of intermittency in wave turbulence, Europhys. Lett. 90 (2010), 34005.

\bibitem{Sch07}
J. Schumacher, Sub-Kolmogorov-scale fluctuations in fluid turbulence, Europhys. Lett. 80 (2007), 54001.

\bibitem{Vie84}
P. Vieillefosse, Internal motion of a small element of fluid in an inviscid
flow, Physica A 125 (1984), 150.
\bibitem{Can92}
B. J. Cantwell, Exact solution of a restricted Euler equation for the velocity
gradient tensor, Phys. Fluids A 4 (1992), 782.
\bibitem{Ash87}
W. T. Ashurst, A. R. Kerstein, R. M. Kerr, and C. H. Gibson, Alignment
of vorticity and scalar gradient with strain rate in simulated Navier–Stokes
turbulence, Phys. Fluids 30 (1987), 2343.
\bibitem{GirPop90}
S. S. Girimaji and S. B. Pope, A diffusion model for velocity gradients in
turbulence, Phys. Fluids A 2 (1990), 242.
\bibitem{ChePum99}
M. Chertkov, A. Pumir, and B. I. Shraiman, Lagrangian tetrad dynamics
and the phenomenology of turbulence, Phys. Fluids 11 (1999), 2394.
\bibitem{JeoGir03}
E. Jeong and S. S. Girimaji, Velocity-gradient dynamics in turbulence:
Effect of viscosity and forcing, Theor. Comput. Fluid Dyn. 16 (2003), 421.
\bibitem{CheMen06}
L. Chevillard and C. Meneveau, Lagrangian dynamics and statistical geometric
structure of turbulence, Phys. Rev. Lett. 97 (2006), 174501.
\bibitem{CheMen07}
L. Chevillard and C. Meneveau, Intermittency and universality in a Lagrangian
model of velocity gradients in three-dimensional turbulence, C.
R. Mec. 335 (2007), 187.
\bibitem{Men11}
C. Meneveau, Lagrangian dynamics and models of the velocity gradient tensor in turbulent flows, Annual Rev. Fluid Mech. 43, 219 (2011).
\bibitem{AntPha81}
R. A. Antonia, N. Phan‐Thien, and B. R. Satyaprakash, Autocorrelation and spectrum of dissipation fluctuations in a turbulent jet, 
Phys. Fluids 24 (1981), 554. 

\bibitem{CheRob10}
L. Chevillard, R. Robert and V. Vargas, A stochastic representation of the local structure of turbulence, Europhys. Lett. 89 (2010), 54002. 

\end{thebibliography}
\end{document}